\documentclass[%
aip,
 amsmath,amssymb,
 reprint,%
]{revtex4-1}

\usepackage{graphicx}
\usepackage{dcolumn}
\usepackage{bm}
\usepackage{threeparttable}
\usepackage{adjustbox}
\usepackage{color}

\DeclareRobustCommand{\katoerase}{\bgroup\markoverwith{\textcolor[rgb]{0,0.6,0}{\rule[.5ex]{2pt}{0.8pt}}}\ULon}

\newcommand{\argmin}{\mathop{\mathrm{argmin}}\limits}

\usepackage[utf8]{inputenc}
\usepackage[T1]{fontenc}
\usepackage{mathptmx}

\begin{document}

\preprint{manuscript for {\it Journal of Rheology}}

\title[]{Rheo-SINDy: Finding a constitutive model from rheological data for complex fluids \\using sparse identification for nonlinear dynamics}

\author{Takeshi Sato}
\email{takeshis@se.kanazawa-u.ac.jp}
 \affiliation{
Advanced Manufacturing Technology Institute, Kanazawa University, Kanazawa 920-1192, Japan
 }

\author{Souta Miyamoto}%
\affiliation{Department of Chemical Engineering, Graduate School of Engineering, Kyoto University, Kyoto 615-8510, Japan}

\author{Shota Kato}
\affiliation{Graduate School of Informatics, Kyoto University, Kyoto 606-8501, Japan}

\date{\today}

\begin{abstract}
Rheology plays a pivotal role in understanding the flow behavior of fluids by discovering governing equations that relate deformation and stress, known as constitutive equations. Despite the importance of these equations,  current methods for deriving them lack a systematic methodology, often relying on sense of physics and incurring substantial costs. To overcome this problem, we propose a novel method named {\it Rheo}-SINDy, which employs the sparse identification of nonlinear dynamics (SINDy) algorithm for discovering constitutive models from rheological data. {\it Rheo}-SINDy was applied to five distinct scenarios, four with well-established constitutive equations and one without predefined equations. Our results demonstrate that {\it Rheo}-SINDy successfully identified accurate models for the known constitutive equations and derived physically plausible approximate models for the scenario without established equations. Notably, the identified approximate models can accurately reproduce nonlinear shear rheological properties, especially at steady state, including shear thinning. These findings validate the robustness of {\it Rheo}-SINDy in handling data complexities and underscore its efficacy as a tool for advancing the development of data-driven approaches in rheology.
\end{abstract}
\maketitle
%
\section{\label{sec:Intro}Introduction}
Mathematical models grounded in physical laws offer profound insights into the behavior of complex systems across science and engineering.
These models clarify the underlying mechanisms governing system dynamics and empower predictions and innovations in technology and natural science.
Traditionally, model derivation has leaned heavily on theoretical and empirical knowledge, often requiring expert intuition.
Data-driven methods have become capable of assisting in developing mathematical models and constructing models that provide advanced predictions~\cite{Brunton2022}.
These data-driven methods involve the sparse identification~\cite{Brunton2016,De_Silva2020-da,Kaheman2022-jt,Fasel2022-ae}, symbolic regression~\cite{Schmidt2009-zv,Bongard2007-oc,Reinbold2021-fa,Udrescu2020-ny,Cranmer2020,Lemos2023}, and physics-informed machine learning methods~\cite{Karniadakis2021-rk,Raissi2019-bu,Jia2021-mn,Rosofsky2023-ae}.
These methods have emerged as powerful tools for deriving governing equations directly from data.

{\it Rheology} is one of the scientific fields that address the properties of flowing matter, which plays a crucial role in many industries, such as designs of chemical processes, by providing insights into the flow behavior of complex fluids.
One of the roles of rheology is to discover or derive governing equations that relate deformation and stress, referred to as {\it constitutive equations}~\cite{Larson1988}.
Accurate constitutive equations are necessary to predict the flows of complex fluids under complex boundary conditions. 
Derivations of constitutive equations starting from the principles of continuum mechanics have achieved significant success in the field of rheology, yielding several practical constitutive equations~\cite{Larson1988}.
Additionally, Ilg and Kr\"{o}ger proposed the derivation of constitutive equations following the guidelines of nonequilibrium thermodynamics~\cite{Ilg2011}.
Nevertheless, it is generally difficult to theoretically obtain constitutive equations for complex fluids from molecular models.
Such cases usually explore mesoscopic coarse-grained models, which are based on molecular theories and suitable for numerical simulations.
For example, for polymeric liquids, standard molecular theories have been proposed~\cite{Bird1987, Rouse1953, Doi1986} and refined mesoscopic models have been constructed based on them~\cite{Masubuchi2001,Doi2003,Likhtman2005}.
In these models, the motion of individual (coarse-grained) molecules is numerically tracked.
Although these models can reproduce rheological data with high accuracy, they require significantly more computational time compared to constitutive equations. Thus, a clear methodology is desired for obtaining constitutive equations from available {\it data} with the assistance of the rheological knowledge.

Data-driven methods have addressed the aforementioned challenges and advanced rheological studies such as constitutive modeling, flow predictions of complex fluids, and model selection~\cite{Jamali2023-review,Miyamoto2024-review}.
Some applications have successfully identified constitutive relations of complex fluids or governing equations to predict the dynamics of fluids with knowledge of rheology.
These studies have employed neural networks (NNs), including deep NN~\cite{Fang2022-DNN}, graph NN~\cite{Mohammadamin2022-GNN}, recurrent NN~\cite{Jin2023-RNN}, physics-informed NN~\cite{Mahmoudabad2021-RhINN,Jamali2021-PINN,Sadat2022-RhINN}, multi-fidelity NN~\cite{Mohammadamin2021-MFNN}, and tensor basis NN~\cite{Lennon2023-RUDE}. Gaussian process regressions (GPRs) have also been employed, for example, for strain-rate dependent viscosity~\cite{Zhao2018} or for viscoelastic properties~\cite{Zhao2021,Seryo2020-hp,Seryo2021-ug,Miyamoto2023-rq}.

Despite the success of NNs and GPRs, with a few exceptions such as the recent work of Lennon and coworkers~\cite{Lennon2023-RUDE}, their black-box nature often obscures the underlying physics, making symbolic regression techniques more appealing for transparency and interoperability~\cite{Brunton2022}.
For example, this technique can successfully identify the governing equations for fluid flows~\cite{Reinbold2021-fa}. Moreover, the sparse identification of nonlinear dynamics (SINDy)~\cite{Brunton2016}, which is one of such methods, has been utilized to track (reduced order) dynamics in the field of fluid mechanics~\cite{Fukami2021-zx}. 
Inspired by these successes, symbolic regression methods have recently started to be used in the field of rheology as well.
For example, Mohammadamin and coworkers relied on SINDy for flexibly identifying the constitutive equations for an elasto-visco-plastic fluid~\cite{Mohammadamin2024-SINDy}.
Generally, the performance of SINDy is significantly influenced by the training data collection method, the candidate terms selected, and the optimization method. However, a comprehensive study to test SINDy for rheological data has not yet been conducted. Before applying SINDy to real-world rheological data, it is highly desirable to investigate fundamental learning strategies such as how to collect training data among various rheological tests and which optimization method to implement.


In this study, we employ SINDy to find constitutive models from rheological data, which we call as {\it Rheo}-SINDy.
For the {\it Rheo}-SINDy regressions, we prepare a training dataset including stress trajectories under simple and oscillatory shear flows and choose the candidate terms based on rheological knowledge of fundamental constitutive equations. Furthermore, multiple optimization methods are compared to find the effective ones for obtaining constitutive equations.
This paper demonstrates five case studies. The first four cases verify the performance of {\it Rheo}-SINDy to identify the known constitutive equations, while the last one attempts to find the unknown constitutive equation for a coarse-grained mesoscopic model. 
Through the case studies, we validate the effectiveness of {\it Rheo}-SINDy and propose a strategy to find constitutive equations from rheological data. The details are shown below.

\section{\label{sec:Methods}Methods}
\begin{figure*}[t]
 \centering
  \includegraphics[width=6.5in]{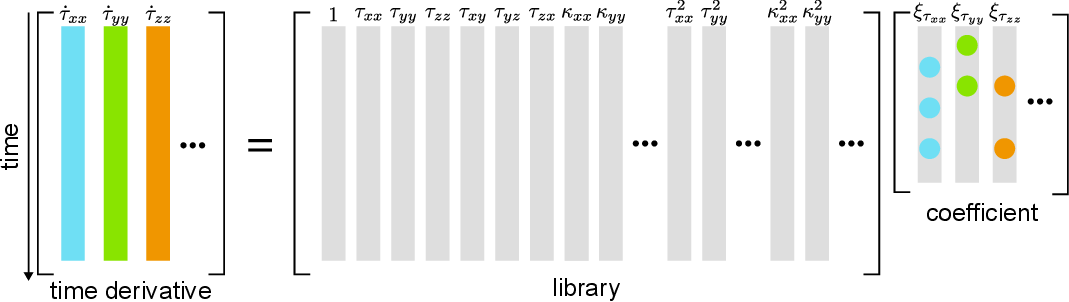}
  \caption{Schematic illustration of {\it Rheo}-SINDy.}
  \label{Fig_SS}
\end{figure*}
We use a data-driven method known as a sparse identification of nonlinear dynamics (SINDy), which was originally developed by Brunton and coworkers~\cite{Brunton2016}, to obtain constitutive equations of complex fluid dynamics.
The SINDy framework considers dynamical systems generally expressed by the following differential equation:
\begin{equation}
\frac{{\rm d} {\bm x}(t)}{{\rm d}t} = \dot {\bm x} (t) = {\bm f} [\bm x(t)],
\label{dynamical_eq}
\end{equation}
where the vector ${\bm x}(t)$ represents the state of a system at time $t$ and the function $\bm f [\bm x (t)]$ determines the dynamics of the state ${\bm x}(t)$.
The basic idea of SINDy is to find dominant terms for describing the dynamics out of numerous candidates using a sparse identification method.
One can determine the (sparse) representation of $\bm f$ by a dataset including a collection of $\bm x (t)$ and $\dot {\bm x} (t)$.
The regression to points of $\bm x (t)$ and $\dot {\bm x} (t)$ is computed with sparsity-promoting techniques, such as $\ell_1$-regularization.

In the rheological community, it is of great importance to determine a relation between stress and strain rate.
This relation is a so-called constitutive model or constitutive equation.
Most constitutive equations are differential equations that depend on the (extra) stress tensor $\bm \tau$ and velocity gradient tensor $\bm \kappa$.
We prefer to use the so-called extra stress tensor $\bm \tau$ as the stress tensor because this tensor satisfies $\bm \tau = \bm 0$ at equilibrium~\cite{Larson1988}, which is convenient for the SINDy regression.
The total stress tensor $\bm \sigma$ can be obtained by the relation $\bm \sigma = \bm \tau + G \bm I$, where $G$ is the modulus and $\bm I$ is the unit tensor. 
A general form for such constitutive equations can be written as
\begin{equation}
\frac{{\rm d} {\bm \tau}(t)}{{\rm d}t} = {\dot {\bm \tau}}(t) = {\bm f} [\bm \tau (t), \bm \kappa (t)].
\label{CE}
\end{equation}
The velocity gradient tensor $\bm \kappa (t)$ is manipulated during rheological measurements.
In formal constitutive equations, the time derivative should be frame-invariant, such as the upper-convected derivative~\cite{Larson1988}.
As shown later in Sec.~\ref{sec:res_and_dis}, the terms that appeared in the upper-convected derivative are recovered by the SINDy regressions.

We use the SINDy algorithm to find constitutive equations for complex fluids from data, and we refer to this technique as {\it Rheo}-SINDy.
{\it Rheo}-SINDy requires three types of training data, two of which are transient stress data $\bm T$ and those time derivatives $\dot {\bm T}$ summarized as the following matrices:
\begin{equation}
{\bm T}
= \left [ \begin{array}{cccc}
   {\bm t}_{xx}  & {\bm t}_{yy} & \cdots & {\bm t}_{zx} \\
\end{array}
\right ]
= \left [ \begin{array}{cccc}
   \tau_{xx} (t_1) & \tau_{yy} (t_1) & \cdots & \tau_{zx} (t_1) \\
   \tau_{xx} (t_2) & \tau_{yy} (t_2) & \cdots & \tau_{zx} (t_2) \\
   \vdots & \vdots & \ddots & \vdots          \\
   \tau_{xx} (t_n) & \tau_{yy} (t_n) & \cdots & \tau_{zx} (t_n)
\end{array}
\right ]
\label{T}
\end{equation}
and
\begin{equation}
{\dot {\bm T}}
= \left [ \begin{array}{cccc}
   {\dot {\bm t}}_{xx}  & {\dot {\bm t}}_{yy} & \cdots & {\dot {\bm t}}_{zx} \\
\end{array}
\right ]
= \left [\begin{array}{cccc}
   {\dot {\tau}}_{xx} (t_1) & {\dot {\tau}}_{yy} (t_1) & \cdots & {\dot {\tau}}_{zx} (t_1) \\
   {\dot {\tau}}_{xx} (t_2) & {\dot {\tau}}_{yy} (t_2) & \cdots & {\dot {\tau}}_{zx} (t_2) \\
   \vdots & \vdots & \ddots & \vdots          \\
   {\dot {\tau}}_{xx} (t_n) & {\dot {\tau}}_{yy} (t_n) & \cdots & {\dot {\tau}}_{zx} (t_n)
\end{array}
\right ],
\label{dot_T}
\end{equation}
where ${\bm t}_{\mu\nu}$ ($\mu\nu \in \{xx,yy,zz,xy,yz,zx\}$) is the stress data for $n$ sequential times. 
In this study, the stress data are collected by applying $\bm \kappa (t)$ to systems of prescribed constitutive equations or mesoscopic models of viscoelastic fluids.
The time derivatives of the stress data $\dot {\bm T}$ are computed by a numerical differentiation method. 
When applying {\it Rheo}-SINDy with experimental data, we need to address the errors in time derivative data due to limited experimental time resolutions. To address this, it is effective to use a high-accuracy numerical differentiation scheme, such as the differentiation scheme with total variation regularization~\cite{Chartrand2011}, as reported by Brunton and coworkers~\cite{Brunton2016}. Nevertheless, in this study, we use a simple finite difference method and demonstrate that the correct equations can be regressed with this method. 
The remaining required data is the velocity gradient data $\bm K$ summarized as
\begin{equation}
{\bm K}
= \left [ \begin{array}{cccc}
   {\bm k}_{xx}  & {\bm k}_{yy} & \cdots & {\bm k}_{zx} \\
\end{array}
\right ]
= \left [ \begin{array}{cccc}
   \kappa_{xx} (t_1) & \kappa_{yy} (t_1) & \cdots & \kappa_{zx} (t_1) \\
   \kappa_{xx} (t_2) & \kappa_{yy} (t_2) & \cdots & \kappa_{zx} (t_2) \\
   \vdots & \vdots & \ddots & \vdots \\
   \kappa_{xx} (t_n) & \kappa_{yy} (t_n) & \cdots & \kappa_{zx} (t_n)
\end{array}
\right ],
\label{K}
\end{equation}
where ${\bm k}_{\mu\nu}$ ($\mu,\nu \in \{x,y,z\}$) is the velocity gradient data for $n$ time steps. 

In {\it Rheo}-SINDy, we construct a library matrix of functions $\bm \Theta$, which includes various nonlinear functions, expressed as
\begin{align}
{\bm \Theta}
&= \left [ \begin{array}{cccc}
   {\bm \theta}_{1}  & {\bm \theta}_{2} & \cdots & {\bm \theta}_{N_{\bm \Theta}} \\
\end{array}
\right ] \nonumber \\
&= \left [ \begin{array}{ccccccc}
   {\bm 1} & {\bm T} & {\bm K} & ({\bm T} \otimes {\bm T}) & ({\bm T} \otimes {\bm K}) & ({\bm K} \otimes {\bm K}) & \cdots
\end{array}
\right ],
\end{align}
where $N_{\bm \Theta}$ is the total number of library functions and ${\bm T} \otimes {\bm K}$, for example, denotes all possible combinations of the products of the row components in ${\bm T}$ and ${\bm K}$ for each time $t_i$ ($1 \le i \le n$).
The library ${\bm \Theta}$ can incorporate not only polynomials but also other functions, such as sinusoidal functions.
If the library does not contain necessary functions, the correct expression cannot be obtained by {\it Rheo}-SINDy. Thus, functions to be included in the library must be selected carefully by utilizing expertise in rheology and knowledge of the target data. The detailed procedure varies from case to case and is provided in Sec.~\ref{sec:res_and_dis}.

Using these expressions, we can rewrite Eq.~\eqref{CE} as
\begin{equation}
\dot {\bm T} = {\bm \Theta} \bm \Xi,
\label{SINDy_Matrix}
\end{equation}
where $\bm \Xi$ is the sparse coefficient matrix written as
\begin{equation}
\bm \Xi
= \left [ \begin{array}{cccc}
   {\bm \xi}_{xx}  & {\bm \xi}_{yy} & \cdots & {\bm \xi}_{zx} \\
\end{array}
\right ]
= \left [ \begin{array}{cccc}
   \xi_{xx,1} & \xi_{yy,1} & \cdots & \xi_{zx,1} \\
   \xi_{xx,2} & \xi_{yy,2} & \cdots & \xi_{zx,2} \\
   \vdots & \vdots & \ddots & \vdots          \\
   \xi_{xx,N_{\bm \Theta}} & \xi_{yy,N_{\bm \Theta}} & \cdots & \xi_{zx,N_{\bm \Theta}}
\end{array}
\right ].
\end{equation}
To determine the coefficients $\bm \Xi$, we solve the following optimization problem for each row:
\begin{equation}
  \hat{\bm \xi}_{\mu\nu} = {\argmin_{{\bm \xi}_{\mu\nu}}} \| \dot {\bm t}_{\mu\nu} - {\bm \Theta} {\bm \xi}_{\mu\nu} \|_2^2 + R ({\bm \xi}_{\mu\nu}),
\label{regression}
\end{equation}
where $\hat{\bm \xi}_{\mu\nu}$ is the optimized sparse vector, $|| \cdots ||_2$ is the $\ell_2$-norm defined as
\begin{equation}
|| \bm x ||_2 = \left ( \sum_{i} x_i^2 \right )^{1/2},
\label{ell_2}
\end{equation}
and $R({\bm \xi}_{\mu\nu})$ is the regularization term. 
In SINDy, the appropriate optimization method generally depends on the specific problem, as has already been demonstrated by Fukami and coworkers~\cite{Fukami2021-zx}. 
To obtain a sparse solution of $\bm\Xi$ from rheological data, we apply the following five methods~\cite{Fukami2021-zx}:
(i) the sequentially thresholded least square algorithm (STLSQ),
(ii) sequentially thresholded Ridge regression (STRidge),
(iii) least absolute shrinkage and selection operator (Lasso),
(iv) Elastic-Net (E-Net),
and (v) adaptive-Lasso (a-Lasso). 

These methods employ different regularization terms as shown in Table~\ref{Table01}.
\begin{table}[b]
  \caption{\label{Table01}The regularization term $R({\bm \xi}_{\mu\nu})$ for the sparse regression methods.}
  \centering
  \begin{tabular}{cl}
  \hline
  \hline
  method & regularization term $R({\bm \xi}_{\mu\nu})$ \\
  \hline
  STLSQ   & $\lambda_0 || \bm \xi_{\mu\nu} ||_0$ \\
  STRidge & $\lambda_0 || \bm \xi_{\mu\nu} ||_0 + \lambda_2 || \bm \xi_{\mu\nu} ||_2^2$ \\
  Lasso   & $\lambda_1 || \bm \xi_{\mu\nu} ||_1$ \\
  E-Net   & $\lambda_1 || \bm \xi_{\mu\nu} ||_1 + \lambda_2 || \bm \xi_{\mu\nu} ||_2^2$ \\
  a-Lasso & $\lambda_1 || \bm \xi_{\mu\nu}' ||_1$ \\
  \hline
  \hline
  \end{tabular}
\end{table}
The hyperparameters of $\ell_i$ norm ($i=0, 1, 2$) are denoted as $\lambda_i$ ($>0$).
The $\ell_0$ and $\ell_1$ norms are defined as
\begin{equation}
  || \bm \xi_{\mu\nu} ||_0 = \sum_{j} \delta(\xi_{\mu\nu,j})
\end{equation}
and
\begin{equation}
  || \bm \xi_{\mu\nu} ||_1 = \sum_{j} |\xi_{\mu\nu,j}|,
\end{equation}
where $\delta(\xi_{\mu\nu,j})$ is the Kronecker delta function,
which is equal to $1$ if $\xi_{\mu\nu,j} \neq 0$ and $0$ otherwise.
The vector $\bm \xi'_{\mu\nu}$ in the a-Lasso is defined as $\bm \xi'_{\mu\nu} = \bm w_{\mu\nu} \otimes \bm \xi_{\mu\nu}$, where $\otimes$ is the element-wise product and $\bm w_{\mu\nu}$ is the adaptive weight vector and its $j$-th element is defined as $w_{\mu\nu,j} = |\xi_{\mu\nu,j}|^{-\delta}$ with $\delta$ being the positive constant.

The STLSQ and STRidge were implemented by iteratively conducting the least square regression and the Ridge regression, respectively, while setting the coefficients with smaller absolute values than a certain threshold $\alpha$ ($>0$) to zero based on the original papers~\cite{Brunton2016, Rudy2017}.
Since this coefficient selection by $\alpha$ replaces the role of the $\ell_0$ norm, $\lambda_0$ is set to zero in this implementation.
In the STRidge, the hyperparameter $\lambda_2$ was set to $0.05$.
The Lasso, E-Net, and a-Lasso were implemented using the scikit-learn library~\cite{Pedregosa2011}.
In this library,
the loss functions $L ({\bm \xi}_{\mu\nu})$ for the Lasso and E-Net are respectively defined as
\begin{equation}
  L ({\bm \xi}_{\mu\nu}) = \frac{1}{2n} || \dot {\bm t}_{\mu\nu} - {\bm \Theta} \bm \xi_{\mu\nu} ||_2^2 + \alpha || \bm \xi_{\mu\nu} ||_1
  \label{Lasso}
\end{equation}
and
\begin{equation}
  L({\bm \xi}_{\mu\nu}) = \frac{1}{2n} || \dot {\bm t}_{\mu\nu} - {\bm \Theta} \bm \xi_{\mu\nu} ||_2^2 + \alpha \beta || \bm \xi_{\mu\nu} ||_1 + \frac{\alpha (1 - \beta)}{2} || \bm \xi_{\mu\nu} ||_2^2,
  \label{ENet}
\end{equation}
where $\beta$ is the $\ell_1$ ratio and $\alpha$ and $\beta$ are the hyperparameters.
These two loss functions have the same form when $\beta = 1$. 
In this study, $\beta$ for the E-Net was set to $0.5$.
According to the original paper of the a-Lasso~\cite{Zou2006}, it can be implemented as the Lasso problem as the following steps:
\begin{enumerate}
  \item Define ${\bm \Theta}'= \left [{\bm \theta}_{1}', \cdots, {\bm \theta}_{N_{\bm \Theta}}' \right ]$, where ${\theta}_{j}' = {\theta}_{j} / w_{\mu\nu,j}$ ($j=1, \ldots, N_{\bm \Theta}$).
  \item Solve the Lasso problem to obtain $\hat {\bm \xi}'_{\mu\nu}$ using Eq.~\eqref{Lasso} with ${\bm \Theta}'$.
  \item Output $\hat {\xi}_{\mu\nu,j} = \hat {\xi}'_{\mu\nu,j} / w_{\mu\nu,j}$ ($j=1, \ldots, N_{\bm \Theta}$)
\end{enumerate}
The adaptive weight $w_{\mu\nu,j}$ depends on the coefficients,
and thereby the output coefficients can be varied in each iteration.
To obtain the converged solution, we initialized the weights as unit vectors $\bm w = \bm 1$ and repeated the above steps until the coefficients $\hat {\xi}_{\mu\nu,j}$ no longer change~\cite{Fukami2021-zx}.
Here, the hyperparameter $\delta$ was set to $3$ (see Sec.~S1 in the supplementary material for the effect of $\delta$).
As shown above, each method has a hyperparameter $\alpha$ to penalize the solution complexity, which is to be tuned for obtaining good predictive yet parsimonious representations.
For this purpose, we test various $\alpha$ values and pick the one with the smallest number of terms among the results whose error has the same order as the minimum error (when $\alpha$ is sufficiently small).

In this study, we focus on {\it shear} rheological measurements that give fundamental rheological properties because they are well-studied and suitable for discussing the applicability of our method to rheology data.
Under shear flow, among the components of $\bm \kappa$, only $\kappa_{xy}$ has non-zero values.
Here, $x$ is the velocity direction, and $y$ is the velocity gradient direction.
Since the major stress components are $\tau_{xx}$, $\tau_{yy}$, $\tau_{zz}$, and $\tau_{xy}$ under shear flow, we only use these components to conduct {\it Rheo}-SINDy.

\section{\label{sec:CS}Case Studies}
For case studies, we first test whether {\it Rheo}-SINDy can find appropriate constitutive equations from training data generated by phenomenological constitutive equations, namely the Upper Convected Maxwell (UCM) model and the Giesekus model. 
Subsequently, we apply {\it Rheo}-SINDy to data generated by several dumbbell-based models. This section provides a brief overview of the UCM, Giesekus, and dumbbell models used in this study and the conditions for creating the training datasets. Table~\ref{Table02} summarizes the conditions for generating training datasets.

\begin{table*}
\caption{\label{Table02}Conditions for generating training datasets.}
\begin{tabular}{lcccccc}
\hline
\hline
model & parameter & shear type\footnote{``S'' means the steady shear flow and ``O'' means the oscillatory shear flow.}  & shear parameters & $\Delta t$ & $\Delta t_\mathrm{train}$ & simulation time \\ \hline
UCM      & -- & S &$\dot \gamma \in \{ 1, 1.7, \ldots, 100\}$\footnote{Ten equally spaced values on a logarithmic scale between $\dot \gamma=1$ and $\dot \gamma = 100$ were used.} & $10^{-4}$ &$10^{-2}$ & $10$\\
UCM      & -- & O &$\gamma_0=2$, $\omega=1$ & $10^{-4}$ & $10^{-2}$ & $100$ \\
Giesekus & $\alpha_{\rm G} = 0.5$ & O & $\gamma_0 = 2$, $\omega \in \{0.1, 0.2, \ldots, 1\}$& $10^{-4}$ & $10^{-2}$ & $100$ \\
Hookean dumbbell\footnote{Three different numbers of dumbbells ($N_{\rm p}\in\{10^3,10^4,10^5\}$) were addressed.} & $n_\mathrm{K}=10$ & O &$\gamma_0 = 2$, $\omega=0.5$ & $10^{-3}$ & $10^{-2}$ & $100$ \\
FENE-P dumbbell\footnote{The closed form expression ($N_{\rm p} \to \infty$) was examined.} & $n_\mathrm{K}=10$ & O &$\gamma_0 = 2$ or $8$, $\omega \in \{0.1, 0.2, \ldots, 1\}$ & $10^{-4}$ & $10^{-2}$ & $100$\\
FENE dumbbell\footnote{The number of dumbbells was set to $N_{\rm p}=10^4$.} & $n_\mathrm{K}=10$ & O & $\gamma_0=2$ or $8$, $\omega \in \{0.1, 0.2, \ldots, 1\}$ & $10^{-4}$ & $10^{-2}$ & $100$ \\ 
\hline
\hline
\end{tabular}
\end{table*}

\subsection{\label{sec:UCM_eqs}Upper Convected Maxwell (UCM) Model}
The simplest constitutive equation for viscoelastic fluids is the UCM model~\cite{Larson1988} shown as
\begin{equation}
\frac{{\rm d} {\bm \tau}}{{\rm d} t} - {\bm \tau} \cdot {\bm \kappa}^{\rm T} - {\bm \kappa} \cdot {\bm \tau} = - \frac{1}{\lambda} \bm \tau + 2 G \bm D.
\label{UCM}
\end{equation}
Here, the left-hand side of Eq.~\eqref{UCM} is the upper-convected time derivative of $\bm \tau$, $\bm \kappa^{\rm T}$ is the transposed $\bm \kappa$, $\lambda$ is the relaxation time, $G$ is the modulus, and $\bm D$ is the deformation rate tensor defined as $\bm D = (\bm \kappa + {\bm \kappa}^{\rm T})/2$.
Using $\lambda$ as the unit time and $G$ as the unit stress (i.e., $\lambda = G = 1$), we can obtain dimensionless expressions for time $\tilde t = t/\lambda$, velocity gradient tensor $\tilde{\bm \kappa} = \lambda \bm \kappa$, and stress $\tilde{\bm \tau} = {\bm \tau}/G$. In what follows, we omit the tilde in dimensionless variables for simplicity.
The dimensionless form of the UCM model under shear flow is thus written as
\begin{equation}
{\dot \tau}_{xx} = -\tau_{xx} + 2\tau_{xy} \kappa_{xy}, \label{UCM_shear_xx}
\end{equation}
\begin{equation}
{\dot \tau}_{yy/zz} = -{\tau}_{yy/zz} = 0, \label{UCM_shear_yy_zz}
\end{equation}
\begin{equation}
{\dot \tau}_{xy} = - \tau_{xy} + \kappa_{xy} + \tau_{yy} \kappa_{xy} = - \tau_{xy} + \kappa_{xy}. 
\label{UCM_shear_xy}
\end{equation}
Here, since the initial conditions for $\bm \tau$ are set to the values of $\bm \tau$ at equilibrium, namely $\bm \tau = \bm 0$,
${\tau}_{yy/zz}$ of the UCM model is zero under shear flow.

For the UCM model, we generate training data by numerically solving Eqs.~\eqref{UCM_shear_xx}--\eqref{UCM_shear_xy} under two shear flow scenarios: steady shear and oscillatory shear tests.
For the steady shear test, the shear rate is kept constant ($\kappa_{xy} = \dot \gamma$) across various values ($\dot \gamma \in \{ 1, 1.7, 2.8, 4.6, 7.7, 13, 22, 36, 60, 100\}$) with simulations running from $t=0$ to $t=10$ using a time step of $\Delta t = 1.0\times 10^{-4}$.
The oscillatory shear test introduces a time-dependent oscillatory shear strain, $\gamma (t) = \gamma_0 \sin (\omega t)$ (i.e., $\kappa_{xy} (t) = \dot \gamma (t) = \gamma_0 \omega \cos (\omega t)$), with $\gamma_0 = 2$ and $\omega = 1$, over a period from $t=0$ to $t=100$, employing the same time step.
In both tests, data are collected at intervals of $\Delta t_{\rm train} = 1 \times 10^{-2}$, resulting in a total of $10^4$ data points for the training data.

\subsection{\label{sec:Giesekus_eqs}Giesekus Model}
The Giesekus model, which is one of the most popular phenomenological constitutive equations~\cite{Giesekus1982}, shows typical shear rheological properties and is used to fit various complex fluids, including polymer solutions and wormlike micellar solutions.
The tensorial form of the Giesekus constitutive equation can be written as
\begin{equation}
\frac{{\rm d} {\bm \tau}}{{\rm d} t} - {\bm \tau} \cdot {\bm \kappa}^{\rm T} - {\bm \kappa} \cdot {\bm \tau} = - \frac{1}{\lambda} \bm \tau - \frac{\alpha_{\rm G}}{G\lambda} \bm \tau \cdot \bm \tau + 2 G \bm D,
\label{Giesekus}
\end{equation}
where $\alpha_{\rm G}$ is the parameter governing the nonlinear response of the Giesekus model.
The Giesekus equation under shear flow is thus given by
\begin{eqnarray}
{\dot \tau}_{xx} &=& -\tau_{xx} - \alpha_{\rm G} ( \tau_{xx}^2 + \tau_{xy}^2)  + 2 \tau_{xy} \kappa_{xy}, \label{G_shear_xx}\\
{\dot \tau}_{yy} &=& -\tau_{yy} - \alpha_{\rm G} ( \tau_{yy}^2 + \tau_{xy}^2), \label{G_shear_yy} \\
{\dot \tau}_{zz} &=& 0, \label{G_shear_zz} \\
{\dot \tau}_{xy} &=& -\tau_{xy} - \alpha_{\rm G} ( \tau_{xx} + \tau_{yy} ) \tau_{xy} + \tau_{yy}\kappa_{xy} + \kappa_{xy}. \label{G_shear_xy}
\end{eqnarray}
Here, all quantities are non-dimensionalized by using $\lambda$ as the unit time and $G$ as the unit stress.
From Eqs.~\eqref{G_shear_xx}--\eqref{G_shear_xy},
the total number of collect terms in the Giesekus model is $12$.

We generate the training data by solving Eqs.~\eqref{G_shear_xx}--\eqref{G_shear_xy} numerically with $\alpha_{\rm G} = 0.5$ and $\Delta t = 1 \times 10^{-4}$.
We note that the Giesekus model with $\alpha_{\rm G} = 0.5$ gives sufficient nonlinear features under shear flow.
We applied the oscillatory shear flow with $\gamma_0 = 2$ and various $\omega$ values ($\omega \in \{0.1, 0.2, \ldots, 1\}$) for $0\le t \le 100$. From the computed stress data, we collected data at the interval of $\Delta t_{\rm train} = 1 \times 10^{-2}$.

\subsection{\label{sec:dumbbell_models}Dumbbell Models}
\begin{figure}[t]
 \centering
  \includegraphics[width=2in]{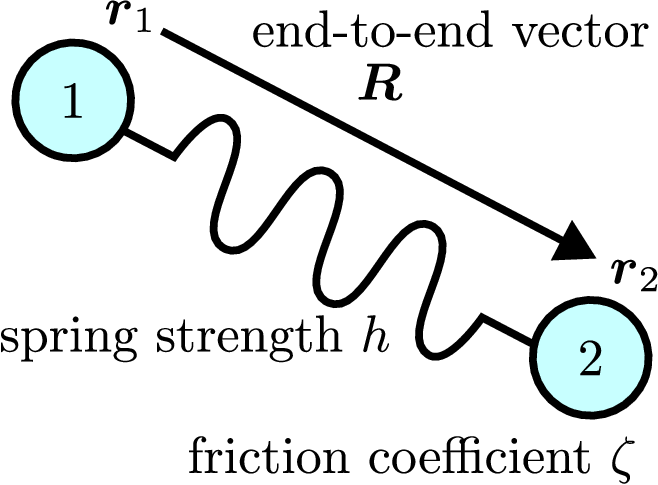}
  \caption{Schematic illustration of the dumbbell model.}
  \label{Fig02}
\end{figure}

The dumbbell-based models have been widely utilized in numerous previous studies for the computation of viscoelastic fluids and are regarded as the standard mesoscopic model for viscoelastic fluids~\cite{Bird1987}. As a stepping stone to investigating more complex models, we chose this model, which can be considered as a fundamental model in rheology.
As illustrated in Fig.~\ref{Fig02}, a dumbbell consists of two beads (indexed as $1$ or $2$) and a spring that connects them.
The Langevin equations for the positions of the two beads ${\bm r}_{1/2} (t)$ can be written as
\begin{equation}
  \zeta \left [ \frac{{\rm d} {\bm r}_i(t)}{{\rm d} t}  - {\bm \kappa} \cdot {\bm r}_i(t) \right ]
  = - h (t) \left \{ {\bm r}_i (t) - {\bm r}_j (t) \right \} + {\bm F}_{{\rm B}i} (t),
\label{dumbbell}
\end{equation}
with $(i,j) = (1,2)$ or $(2,1)$. Here, $\zeta$ is the friction coefficient, $h(t)$ is the spring strength,
and ${\bm F}_{{\rm B}i} (t)$ is the Brownian force acting on the bead $i$.
The time evolution equation for the end-to-end vector $\bm R (t)$ ($={\bm r}_2(t) - {\bm r}_1(t)$)  of the beads is thus obtained as
\begin{equation}
\zeta \left [ \frac{{\rm d} {\bm R} (t)}{{\rm d} t}  -  {\bm \kappa} \cdot {\bm R}(t) \right ] = -2 h (t) {\bm R} (t) + \left \{ {\bm F}_{{\rm B}2}(t) - {\bm F}_{{\rm B}1}(t) \right \}.
\label{dumbbell_ete}
\end{equation}
The Brownian force is characterized by the first and second-moment averages as
\begin{equation}
\langle {\bm F}_{{\rm B}i}(t) \rangle = {\bm 0}
\label{F_B_1st}
\end{equation}
and
\begin{equation}
\langle {\bm F}_{{\rm B}i}(t) {\bm F}_{{\rm B}j}(t') \rangle = 2 \zeta k_{\rm B} T \delta_{ij}\delta (t-t') {\bm I},  \label{F_B_2nd}
\end{equation}
where $k_{\rm B}$ is the Boltzmann constant and $T$ is the temperature.
From the end-to-end vector $\bm R (t)$, the stress tensor can be expressed as
\begin{equation}
{\bm \tau} (t) = \rho \langle h(t) {\bm R}(t){\bm R}(t) \rangle - \rho k_{\rm B} T {\bm I},
\label{stress}
\end{equation}
where $\rho$ is the density of dumbbells.

There are several expressions for the spring strength $h(t)$.
The most basic one is the Hookean spring, defined as
\begin{equation}
h (t) = h_{\rm eq} = \frac{3k_{\rm B}T}{n_{\rm K}b_{\rm K}^2},
\label{Hookean}
\end{equation}
where $n_{\rm K}$ is the number of Kuhn segments per spring and $b_{\rm K}$ is the Kuhn length.
Reproduction of some nonlinear rheological properties, such as shear thinning under shear flow, necessitates dealing with finite extensible nonlinear elastic (FENE) effects.
Although the exact expression for FENE springs is given by the inverse Langevin function, the following empirical expression is widely used~\cite{Bird1987}:
\begin{equation}
h (t) = h_{\rm eq}\frac{1 - \langle {R}_{\rm eq}^2 \rangle/R_{\rm max}^2}{1 - {\bm R}^2(t)/R_{\rm max}^2},
\label{FENE}
\end{equation}
where $\langle R_{\rm eq}^2 \rangle^{1/2} = (n_{\rm K})^{1/2}b_{\rm K}$ is the equilibrium length of the springs,
and $R_{\rm max} = n_{\rm K}b_{\rm K}$ is the maximum length of the springs.
As shown later in Sec.~\ref{sec_FENE_equations},
a constitutive equation cannot be analytically obtained for the FENE dumbbell model.
To address the FENE spring more analytically,
the following approximate expression of the FENE spring has been proposed~\cite{Bird1987}:
\begin{equation}
  h (t) = h_{\rm eq}\frac{1 - \langle {R}_{\rm eq}^2 \rangle/R_{\rm max}^2}{1 - \langle {\bm R}^2(t) \rangle/R_{\rm max}^2 } = h_{\rm eq} f_{\rm FENE} (t).
  \label{FENE_P}
\end{equation}
This spring is referred to as the FENE-P spring.
Here, ``P'' means Peterlin, who proposed the approximate form of the FENE spring law.
The average appearing in Eq.~\eqref{FENE_P} makes it possible to obtain the analytical constitutive equation.

We use $\lambda = \zeta / 4h_{\rm eq}$ as the unit time and $G = \rho k_{\rm B}T$ as the unit stress for the dumbbell models.
To simplify the expressions, we omit the tilde representing dimensionless quantities in what follows.

\subsubsection{Hookean Dumbbell Model}
The most basic dumbbell model is the Hookean dumbbell model, where Hookean springs are employed (cf. Eq.~\eqref{Hookean}).
From Eqs.~\eqref{dumbbell_ete}, \eqref{stress}, and \eqref{Hookean}, the Hookean dumbbell model reduces to the constitutive equation for the UCM model (cf. Eq.~\eqref{UCM}) in the limit of $N_{\rm p} \rightarrow \infty$ with $N_{\rm p}$ being the number of dumbbells.

For the Hookean dumbbell model, we generate training data by Brownian dynamics (BD) simulations with the finite numbers of dumbbells ($N_{\rm p} \in \{10^3, 10^4, 10^5\}$). 
In the BD simulations of this study, to integrate Eq.~\eqref{dumbbell_ete}, we use the explicit Euler method with a small time step. 
We apply the oscillatory shear flow, $\gamma (t) = \gamma_0 \sin (\omega t)$, with $\gamma_0 = 2$ and $\omega = 0.5$, over a period from $t=0$ to $t=100$.
The simulations are run with $\Delta t = 1 \times 10^{-3}$ for $0\le t \le 100$ and data are collected at the interval of $\Delta t_{\rm train} = 1 \times 10^{-2}$. Each simulation is conducted with five different random seeds, and their average data is used for training.
Due to the characteristics of the BD simulation, the training data inherently include noise originating from the finite $N_{\rm p}$.
We here test whether {\it Rheo}-SINDy can find from the noisy data the constitutive equations for the UCM model shown in Eqs.~\eqref{UCM_shear_xx}--\eqref{UCM_shear_xy}. 

\begin{table*}
\caption{\label{Table03}Settings of {\it Rheo}-SINDy used in case studies.}
\begin{tabular}{lcccc}
\hline
\hline
model & exact equations & optimization\footnote{The value of $\alpha$ is picked from sparser solutions whose error is the same order of magnitude as the minimum error.} & library & $N_{\bm \Theta}$ \\ \hline
UCM & Eqs.~\eqref{UCM_shear_xx}--\eqref{UCM_shear_xy}    & all five methods\footnote{All five methods are the STLSQ, STRidge, Lasso, E-Net, and a-Lasso explained in Sec.~\ref{sec:Methods}.} & polynomial (up to 3rd order) & 35 \\
Giesekus & Eqs.~\eqref{G_shear_xx}--\eqref{G_shear_xy}    & STLSQ, STRidge, a-Lasso & polynomial (up to 2nd order) & 15 \\
Hookean dumbbell & Eqs.~\eqref{UCM_shear_xx}--\eqref{UCM_shear_xy} & STRidge, a-Lasso & polynomial (up to 2nd order) & 15 \\
FENE-P dumbbell & Eqs.~\eqref{FENE_P_xx}--\eqref{FENE_P_xy} & STRidge, a-Lasso & Eq.~\eqref{Theta_FENE_P_C} & 26 \\
FENE-P dumbbell & Eqs.~\eqref{FENE_P_s_xx}--\eqref{FENE_P_s_xy} & STRidge, a-Lasso & Eq.~\eqref{Theta_FENE_P_stress} &
 29 \\
FENE dumbbell & N/A & STRidge\footnote{The results are shown in the supplementary material.}, a-Lasso & Eq~.\eqref{Theta_FENE_P_stress} & 29 \\ 
\hline
\hline
\end{tabular}
\end{table*}

\begin{figure*}[t]
 \centering
  \includegraphics[width=6.5in]{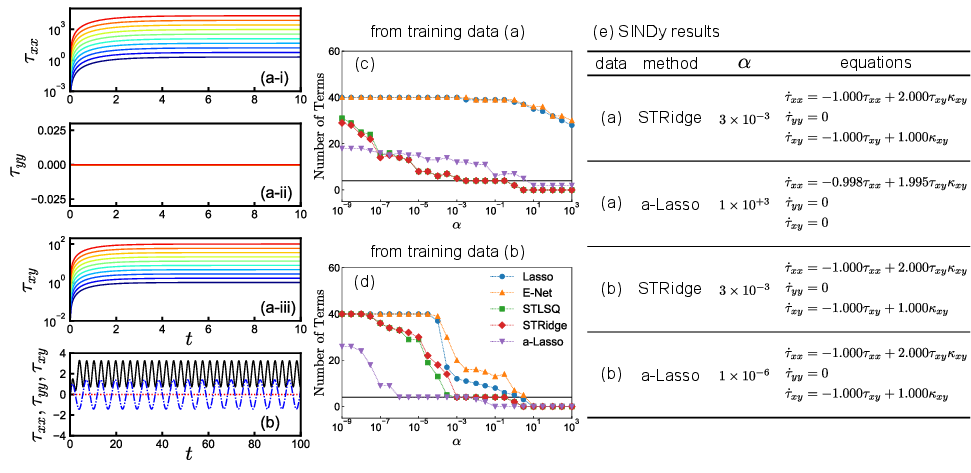}
  \caption{Training data obtained by the UCM model (cf. Eqs.~\eqref{UCM_shear_xx}--\eqref{UCM_shear_xy}) (a) under steady shear flow ($\kappa_{xy} = {\dot \gamma}$) and (b) under oscillatory shear flow ($\kappa_{xy} = \gamma_0 \omega\cos(\omega t)$).
  The numbers of total terms obtained (c) from the training data (a) and (d) from the training data (b). (e) The constitutive equations obtained by {\it Rheo}-SINDy.
  In (b), $xx$-, $yy$-, and $xy$-components of the stress tensor are plotted with the black solid, red dotted, and blue dash-dotted lines, respectively. In (c) and (d), the numbers of total terms for five different optimization methods are plotted against the hyperparameter $\alpha$. The black horizontal lines in (c) and (d) indicate the correct number of the terms in the UCM model.}
  \label{Fig03}
\end{figure*}

\subsubsection{\label{sec:FENE_P_equations}FENE-P Dumbbell Model}
We next address the so-called FENE-P dumbbell model,
where Eq.~\eqref{FENE_P} is utilized as the spring strength.
As shown below,
the FENE-P dumbbell model has an analytical solution and is utilized for various flow problems~\cite{Laso1993,Graham2014}.

Due to the assumption shown in Eq.~\eqref{FENE_P}, a simple representation of the time evolution for the conformation tensor ${\bm C} = \langle {\bm R} (t) {\bm R} (t) \rangle$ can be obtained as
\begin{equation}
\frac{{\rm d} {\bm C}}{{\rm d} t} - {\bm C} \cdot {\bm \kappa}^{\rm T} - {\bm \kappa} \cdot {\bm C} = - f_{\rm FENE}(t) {\bm C} + \frac{n_{\rm K}}{3}{\bm I}.
\label{conformation_tensor}
\end{equation}
The stress tensor is thus obtained by
\begin{equation}
{\bm \tau} (t) = \rho h(t) {\bm C} (t) - \rho k_{\rm B} T {\bm I}.
\label{stress_FENE_P}
\end{equation}
Under shear flow, Eq.~\eqref{conformation_tensor} reduces to the following expressions:
\begin{equation}
{\dot C}_{xx} = -f_{\rm FENE} C_{xx} + 2C_{xy} \kappa_{xy} + \frac{n_{\rm K}}{3}, \label{FENE_P_xx}
\end{equation}
\begin{equation}
{\dot C}_{yy/zz} = -f_{\rm FENE} C_{yy/zz} + \frac{n_{\rm K}}{3}, \label{FENE_P_yy_zz}
\end{equation}
\begin{equation}
{\dot C}_{xy} = -f_{\rm FENE} C_{xy} + C_{yy} \kappa_{xy}. \label{FENE_P_xy}
\end{equation}
Using {\it Rheo}-SINDy, we test whether or not Eqs.~\eqref{FENE_P_xx}--\eqref{FENE_P_xy} can be discovered from the data.

Although it has not been as widely recognized due to its complexity,
the FENE-P dumbbell model can also be expressed in the form of the constitutive equation (i.e., the stress expression)~\cite{Mochimaru1983}.
From the textbook of Bird and coworkers~\cite{Bird1987}, the constitutive equation for the FENE-P model is
\begin{equation}
\frac{{\rm d}  {\bm \tau}}{{\rm d}t} -  {\bm \tau} \cdot  {\bm \kappa}^{\rm T} -  {\bm \kappa} \cdot  {\bm \tau} = - f_{\rm FENE} (t) {\bm \tau}  + 2{\bm D} + \frac{{\rm D} \ln Z}{{\rm D}t} ( {\bm \tau} + {\bm I}),
\label{FENE_P_CE}
\end{equation}
where ${\rm D}(\cdots)/{\rm D}t$ is the substantial derivative and $Z$ is the function expressed as
\begin{equation}
Z = \frac{1}{1 - \langle {\bm R}^2(t)/R_{\rm max}^2 \rangle} = 1 + \frac{1}{3 n_{\rm K} Z^{-1}_{\rm eq}} ( {{\rm tr}} {\bm \tau} + 3).
\label{Z_CE}
\end{equation}
Here, $Z_{\rm eq}$ indicates $Z$ at equilibrium.
From Eq.~\eqref{Z_CE}, we can see that ${{\rm tr}} {\bm \tau}$ is tightly related to the (squared) length of dumbbells.
Since we do not address the spatial gradient in rheological calculations,
${\rm D}(\cdots)/{\rm D}t$ simply reduces to ${\rm d}(\cdots)/{\rm d}t$.
Using Eqs.~\eqref{conformation_tensor}, \eqref{FENE_P_CE}, and \eqref{Z_CE},
the constitutive equations for the FENE-P dumbbell model under shear flow can be expressed as
\begin{eqnarray}
\dot \tau_{xx} &=& - \left \{ 1 + \frac{1}{3 (n_{\rm K} - 1)} \right \} \tau_{xx} - \frac{1}{3 (n_{\rm K} - 1)} (\tau_{yy} + \tau_{zz}) \nonumber \\
& & - \frac{1}{9n_{\rm K}(n_{\rm K}-1)} ({\rm tr} \bm \tau)^2 - \frac{1}{3n_{\rm K}} \left ( 2 + \frac{1}{n_{\rm K} - 1} \right ) {\rm tr} \bm \tau \tau_{xx} \nonumber \\
& & + 2 \left \{ 1 + \frac{1}{3(n_{\rm K} - 1)} \right \} \tau_{xy}\kappa_{xy} - \frac{1}{9n_{\rm K}(n_{\rm K} - 1)} ({\rm tr} \bm \tau)^2 \tau_{xx} \nonumber \\
& & + \frac{2}{3(n_{\rm K} - 1)} \tau_{xx} \tau_{xy}\kappa_{xy}, \label{FENE_P_s_xx}
\end{eqnarray}
\begin{eqnarray}
\dot \tau_{yy/zz} &=&  - \left \{ 1 + \frac{1}{3 (n_{\rm K} - 1)} \right \} \tau_{yy/zz} - \frac{1}{3 (n_{\rm K} - 1)} (\tau_{xx} + \tau_{zz/yy}) \nonumber \\
& &  - \frac{1}{9n_{\rm K}(n_{\rm K} - 1)} ({\rm tr} \bm \tau)^2 - \frac{1}{3n_{\rm K}} \left ( 2 + \frac{1}{n_{\rm K} - 1} \right ) {\rm tr} \bm \tau \tau_{yy/zz} \nonumber \\
& & + \frac{2}{3(n_{\rm K} - 1)} \tau_{xy}\kappa_{xy} - \frac{1}{9n_{\rm K}(n_{\rm K} - 1)} ({\rm tr} \bm \tau)^2 \tau_{yy/zz} \nonumber \\
& & + \frac{2}{3(n_{\rm K} - 1)} \tau_{yy/zz} \tau_{xy}\kappa_{xy}, \label{FENE_P_s_yy_zz}
\end{eqnarray}
\begin{eqnarray}
\dot \tau_{xy} &=& - \tau_{xy} + \kappa_{xy} + \tau_{yy}\kappa_{xy} - \frac{1}{3n_{\rm K}} \left ( 2 + \frac{1}{n_{\rm K} - 1} \right ) {\rm tr} \bm \tau \tau_{xy} \nonumber \\
& & - \frac{1}{9n_{\rm K}(n_{\rm K} - 1)} ({\rm tr} \bm \tau)^2 \tau_{xy} + \frac{2}{3(n_{\rm K} - 1)} \tau_{xy}^2\kappa_{xy}. \label{FENE_P_s_xy}
\end{eqnarray}
For the derivation, please refer to Sec.~S2 in the supplementary material.
From Eqs.~\eqref{FENE_P_s_xx}--\eqref{FENE_P_s_xy},
the constitutive equation for the FENE-P model can be expressed by a polynomial of up to a third degree in $\bm \tau$ and $\bm \kappa$.
Here, we note that Eqs.~\eqref{FENE_P_s_xx}--\eqref{FENE_P_s_xy} become equivalent to the UCM model shown in Eqs.~\eqref{UCM_shear_xx}--\eqref{UCM_shear_xy} in the limit of $n_{\rm K} \rightarrow \infty$.

To generate noise-free training data, we solve Eqs.~\eqref{stress_FENE_P}--\eqref{FENE_P_xy} with $n_{\rm K} = 10$ and $\Delta t = 1 \times 10^{-4}$ for $0\le t \le 100$.
We apply the oscillatory shear flow with $\gamma_0 = 2$ or $8$ and various $\omega$ values ($\omega \in \{0.1, 0.2, \ldots, 1\}$).
From the computed stress data, we collect data at the interval of $\Delta t_{\rm train} = 1 \times 10^{-2}$.

\subsubsection{\label{sec_FENE_equations}FENE Dumbbell Model}
We finally address the FENE dumbbell model, where the spring strength is represented by Eq.~\eqref{FENE}.
Since the FENE dumbbell model does not use any simplification for the spring strength (e.g., Peterlin approximation shown in Eq.~\eqref{FENE_P}), its analytical constitutive equation has not been obtained.
We apply {\it Rheo}-SINDy to this case to see if an {\textit {approximate}} constitutive equation can be obtained.
The obtained equations are validated by comparing the data obtained by numerically solving them with the data obtained by BD simulations.

The training data are generated by the BD simulations using Eqs.~\eqref{dumbbell_ete}--\eqref{stress} and \eqref{FENE} with $n_{\rm K} = 10$, $N_{\rm p} = 10^4$, and $\Delta t = 1 \times 10^{-4}$ for $0\le t \le 100$.
We apply the oscillatory shear flows with the same parameters as those in the FENE-P dumbbell model.
The BD simulation results with five different random seeds are averaged for each condition.
Since we do not use any approximation for the spring strength, the values of $h(t)$ differ for each individual dumbbell.
From the computed stress data, we collected data at the interval of $\Delta t_{\rm train} = 1 \times 10^{-2}$.

\section{\label{sec:res_and_dis}Results and Discussions}
In this section, we present the results of the case studies for five models explained in Sec.~\ref{sec:CS}. Through case studies on phenomenological constitutive equations (i.e., the UCM and Giesekus models), we first investigate two key questions for deriving accurate equations: (i) whether oscillatory shear tests or simple (steady) shear tests are more appropriate, and (ii) which of the five optimization methods presented in Sec.~\ref{sec:Methods} is most suitable. Subsequently, using the experimental and optimization methods identified as appropriate by this investigation, we attempt to obtain the constitutive equations of the dumbbell models. The methods of identifying constitutive models are summarized in Table~\ref{Table03}. 
In the subsequent sections, we examine how the hyperparameter $\alpha$ affects the {\it Rheo}-SINDy results. 
To discuss the effect of the hyperparameter $\alpha$, we also examine the results with various $\alpha$ ($10^{-9} \le \alpha \le 10^3$).
Among the $\alpha$ values examined, we pick $\alpha$ with the smallest number of terms among the results whose error has the same order as the minimum error. For comparison, {\it Rheo}-SINDy results with $\alpha$ other than the one chosen by this criterion are also shown. 

\subsection{\label{sec:UCM_res}Upper Convected Maxwell Model}
Through this case study, we first check the appropriate methods to take the shear rheological data for {\it Rheo}-SINDy.
Figure~\ref{Fig03} shows the training data and results for the UCM model.
Figure~\ref{Fig03}(a) and (b) are the stress data under the steady shear flows with the various shear rates and those under the oscillatory shear flow.

We conducted the {\it Rheo}-SINDy regressions by using the polynomial library that includes up to third order terms of $\tau_{xx}$, $\tau_{yy}$, $\tau_{xy}$, and $\kappa_{xy}$. Thus, there were $35$ candidate terms for each component of the constitutive equation.
The terms related to $\tau_{zz}$ were excluded because they do not contribute to the UCM dynamics.
Figures~\ref{Fig03}(c) and (d) present the numbers of total terms varying with the hyperparameter $\alpha$ obtained by {\it Rheo}-SINDy using the training data (a) and (b), respectively.
Figure~\ref{Fig03}(c) indicates that sparse solutions can be obtained by the STLSQ, STRidge, and a-Lasso, but not by the Lasso and E-Net.
Moreover, regarding the number of terms, the STLSQ and STRdge exhibit similar behavior.
Specifically, the STLSQ and STRidge with $3\times 10^{-3} \le \alpha \le 3\times 10^{-1}$ successfully discovered equations including the correct number of terms.
Figure~\ref{Fig03}(d) indicates that the STLSQ, STRidge, and a-Lasso yielded the correct number of terms, though all five methods gave sparse solutions.
In most of the cases where the number of terms obtained was correct, the obtained coefficients were also correct. These results suggest that the oscillatory shear test is more appropriate than the steady shear test to obtain the correct constitutive equations for the UCM model. 
Figure~\ref{Fig03}(e) lists the constitutive equations obtained by the STRidge and a-Lasso.
We can see that the STRidge and a-Lasso can give the correct constitutive equations, except for the a-Lasso in the steady shear test.
Furthermore, we confirmed that the correct equations were obtained even for $\alpha$ values not shown in Fig.~\ref{Fig03}(e) in the case of the UCM model.
These findings show the basic validity of finding the constitutive equations from the rheological data by {\it Rheo}-SINDy.
Figure~\ref{Fig03} indicates that the STLSQ, STRidge, and a-Lasso demonstrate better performance in discovering the correct constitutive equations compared to the Lasso and E-Net; thus, we use the former three methods in the following discussion.

\subsection{\label{sec:Giesekus_res}Giesekus Model}
\begin{figure}[t]
 \centering
  \includegraphics[width=3in]{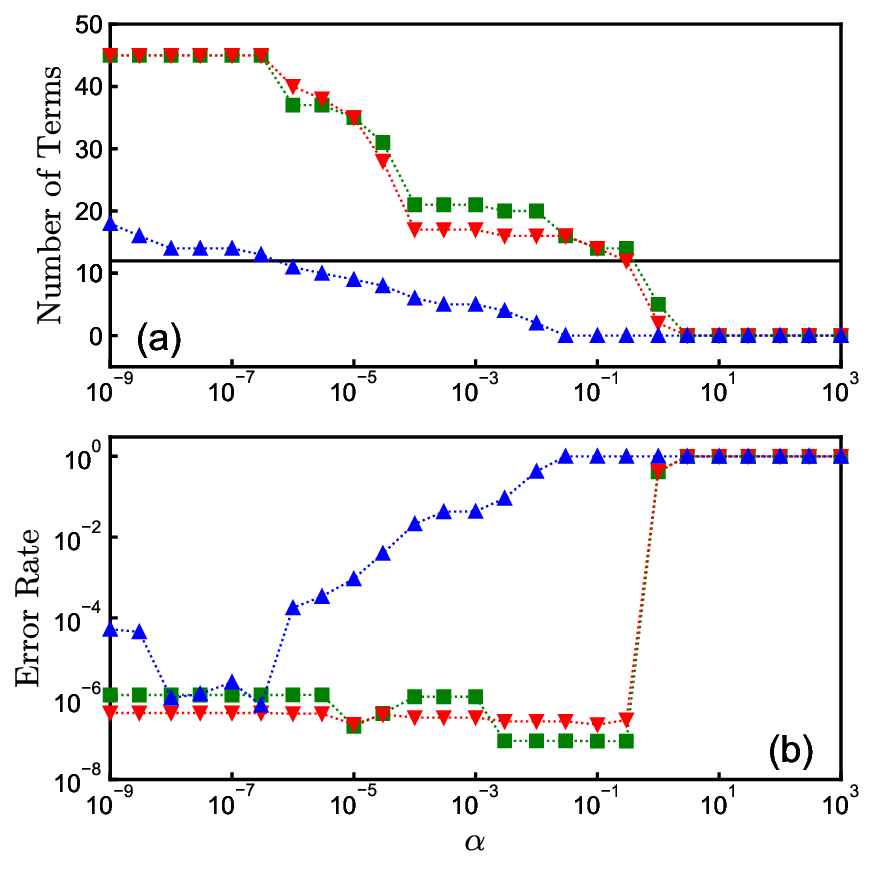}
  \caption{(a) The number of total terms and (b) the error rate of the constitutive equations obtained by {\it Rheo}-SINDy for the Giesekus model. The used optimization methods are the STLSQ (green squares), STRidge (red reverse triangles), and a-Lasso (blue triangles). The regressions were conducted with the multiple ($10$) data trajectories of $\kappa_{xy} = \gamma_0 \omega \cos(\omega t)$ with $\gamma_0 = 2$ and $\omega \in \{0.1, 0.2, \ldots, 1\}$ for $0\le t \le 100$, respectively.}
  \label{Fig04}
\end{figure}
\begin{figure*}[t]
 \centering
  \includegraphics[width=6.5in]{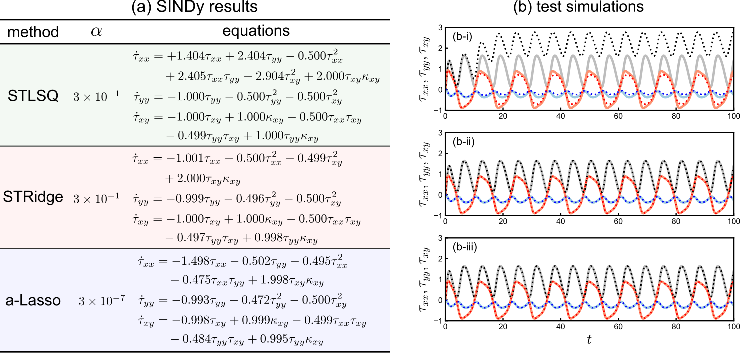}
  \caption{(a) The constitutive equations obtained by {\it Rheo}-SINDy with (i) the STLSQ, (ii) STRidge, and (iii) a-Lasso from the multiple data trajectories, and (b) their test simulation results under the oscillatory shear flow with $\gamma_0 = 4$ and $\omega = 0.5$. The training data are the same as those in Fig.~\ref{Fig04}. The exact equations for the Giesekus model under shear flow are shown in Eqs.~\eqref{G_shear_xx}--\eqref{G_shear_xy}. 
  In (b), the $xx$-, $yy$-, and $xy$-components of the stress tensor are shown with black, blue, and red lines, respectively. The dotted and solid lines in (b) denote the predictions by the equations shown in (a) and those by the exact Giesekus model, respectively.}
  \label{Fig05}
\end{figure*}
\begin{figure*}[t]
 \centering
  \includegraphics[width=6.5in]{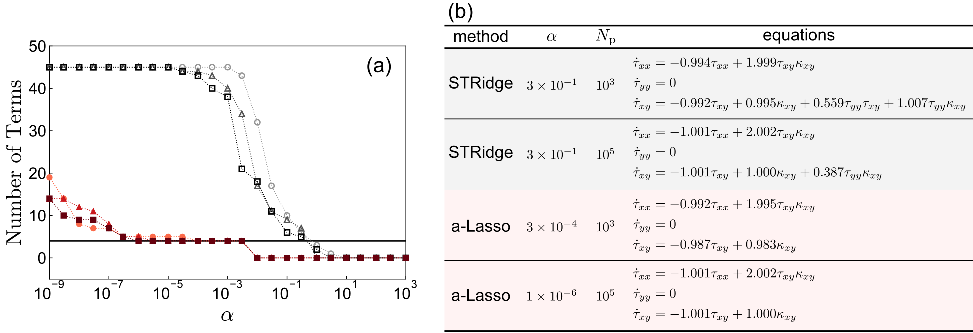}
  \caption{(a) The total number of terms and (b) the constitutive equations obtained by {\it Rheo}-SINDy with the STRidge (black open symbols) and a-Lasso (red closed symbols) for the Hookean dumbbell model (Eqs.~\eqref{UCM_shear_xx}--\eqref{UCM_shear_xy}).
  In (a), the horizontal line indicates the correct number of terms, and circle, triangle, and square symbols represent the results for $N_{\rm p} = 10^3$, $10^4$, and $10^5$, respectively.
  }
  \label{Fig06}
\end{figure*}
We here explain the results of {\it Rheo}-SINDy for the Giesekus model.
This case used the polynomial library consisting of up to second-order terms of $\tau_{xx}$, $\tau_{yy}$, $\tau_{xy}$, and $\kappa_{xy}$, which contain sufficient candidate terms to obtain the exact equations.
Figure~\ref{Fig04} shows (a) the total number of terms and (b) the error rate obtained by {\it Rheo}-SINDy for the training data of the Giesekus model.
The error rate is defined as the sum of the mean squared errors (MSEs) of $\dot {\bm t}_{\mu\nu} - {\bm \Theta} \hat{\bm \xi}_{\mu\nu}$.
The MSEs were scaled so that the maximum value of each method was 1.
Figure~\ref{Fig04}(a) and (b) indicates that the a-Lasso evidently provides a sparser solution compared to the other two methods when the error rates are comparable.
We note that, similar to the number of terms, coefficient values generally depend on $\alpha$.

Figures~\ref{Fig05}(a) and (b) show the constitutive equations found by {\it Rheo}-SINDy and the test simulation results, respectively.
For test simulations shown in Fig.~\ref{Fig05}(b), we employed the oscillatory shear flow with $\gamma_0 = 4$ and $\omega = 0.5$, which is outside of the parameters in the training data described in Sec.~\ref{sec:Giesekus_eqs}.
Figure~\ref{Fig05}(a) reveals that the STRidge with $\alpha = 3 \times 10^{-1}$ gave almost exact constitutive equations, including the value of $\alpha_{\rm G}$ (cf. Eqs.~\eqref{G_shear_xx}--\eqref{G_shear_xy}).
As inferred from this, the predictions based on the constitutive equations obtained by the STRidge demonstrate a good agreement with the test data, as shown in Fig.~\ref{Fig05}(b-ii).
In contrast to the success of the STRidge, the STLSQ and a-Lasso failed to identify the correct solution, as indicated in Fig.~\ref{Fig05}(a).
The constitutive equations obtained by the STLSQ with $\alpha = 3 \times 10^{-1}$ has a low error rate as shown in Fig.~\ref{Fig04}(b), but its predicted $\tau_{xx}$ significantly deviates from the test data as seen in Fig.~\ref{Fig05}(b-i).
In contrast, although the a-Lasso did not provide the correct solution for $\tau_{xx}$, the test simulations with the obtained constitutive equations exhibit a good agreement with the test data.
These test simulations demonstrate that the STRidge and a-Lasso are promising approaches for {\it Rheo}-SINDy.

\subsection{Hookean Dumbbell Model}
We next explain the results for the Hookean dumbbell model.
We here used the polynomial library that includes up to second order terms of $\tau_{xx}$, $\tau_{yy}$, $\tau_{xy}$, and $\kappa_{xy}$. Thus, the total number of candidate terms was $N_{\bm \Theta} = 15$ for each component.

Figure~\ref{Fig06} shows the {\it Rheo}-SINDy results for the Hookean dumbbell model with the different numbers of dumbbells.
We note that the standard deviation of $\bm \tau$ in the training data decreases proportionally with $N_{\rm p}^{-1/2}$.
From Fig.~\ref{Fig06}(a), as the value of $N_{\rm p}$ increases, sparser solutions are obtained, especially for {\it Rheo}-SINDy with the STRidge.
Unlike the case of the UCM model (cf. Fig.~\ref{Fig03}), which can be considered as the ``noise-free'' case of the Hookean dumbbell model, the STRidge provides the correct number of terms only within a narrow range of $\alpha$ values.
Nevertheless, if we choose the appropriate $\alpha$ value, the nearly correct constitutive equations can be found by the STRidge, as shown in the upper part of Fig.~\ref{Fig06}(b).
We note that the terms containing $\tau_{yy}$ appear in the time evolution equation for $\tau_{xy}$ obtained by the STRidge. Although these terms do not affect the predictions because ${\tau}_{yy} = 0$, we speculate that the appearance of these terms is due to the correlation effects of the noise in $R_x$ and $R_y$ on the stress (cf. Eq.~\eqref{stress}).
When comparing the STRidge and a-Lasso, it is evident that the a-Lasso provides stable and sparse solutions across a broader range of $\alpha$ values,
regardless of the $N_{\rm p}$ value.
Furthermore, we confirm that the correct equations can be obtained using the a-Lasso, as shown in the lower part of Fig.~\ref{Fig06}(b). 
Comparing the results obtained by the a-Lasso with different $N_{\rm p}$, we found that the correct number of terms is obtained in almost the same $\alpha$ range for $N_{\rm p} \ge 10^4$. This indicates that using $N_{\rm p} = 10^4$ provides sufficient results in this case. 
This partially suggests the effectiveness of the a-Lasso in discovering essential terms from noisy data.

\subsection{\label{res:FENE_P}FENE-P Dumbbell Model}
We next examine whether {\it Rheo}-SINDy can find more complex differential equations (i.e., the FENE-P dumbbell model) than the UCM model and the Giesekus model.
We utilized {\it Rheo}-SINDy with $\bm T$ replaced by $\bm C$ to discover the differential equations for the conformation tensor $\bm C$ of the FENE-P dumbbell model explained in Sec.~\ref{sec:FENE_P_equations}. In this case,
we prepared the following custom library:
\begin{equation}
{\bm \Theta}
= \left [ \begin{array}{cccc}
   {\bm 1} & {\bm \Omega}(t_1) & {\bm \Omega}^2(t_1) & f_{\rm FENE}(t_1) {\bm \Omega}(t_1) \\
   {\bm 1} & {\bm \Omega}(t_2) & {\bm \Omega}^2(t_2) & f_{\rm FENE}(t_2) {\bm \Omega}(t_2) \\
   \vdots  & \vdots & \vdots & \vdots \\
   {\bm 1} & {\bm \Omega} (t_n) & {\bm \Omega}^2 (t_n) & f_{\rm FENE} (t_n) {\bm \Omega} (t_n)
\end{array}
\right ], \label{Theta_FENE_P_C}
\end{equation}
where ${\bm \Omega}$ consists of non-zero components of $\bm C$ under shear flow ($C_{xx}$, $C_{yy}$, $C_{zz}$, and $C_{xy}$) and $\kappa_{xy}$, and ${\bm \Omega}^2$ is the vector composed of all the multiplied combinations of the ${\bm \Omega}$ components.
The total number of library functions was thus $N_{\bm \Theta} = 26$.

\begin{figure}[t]
   \centering
    \includegraphics[width=2.75in]{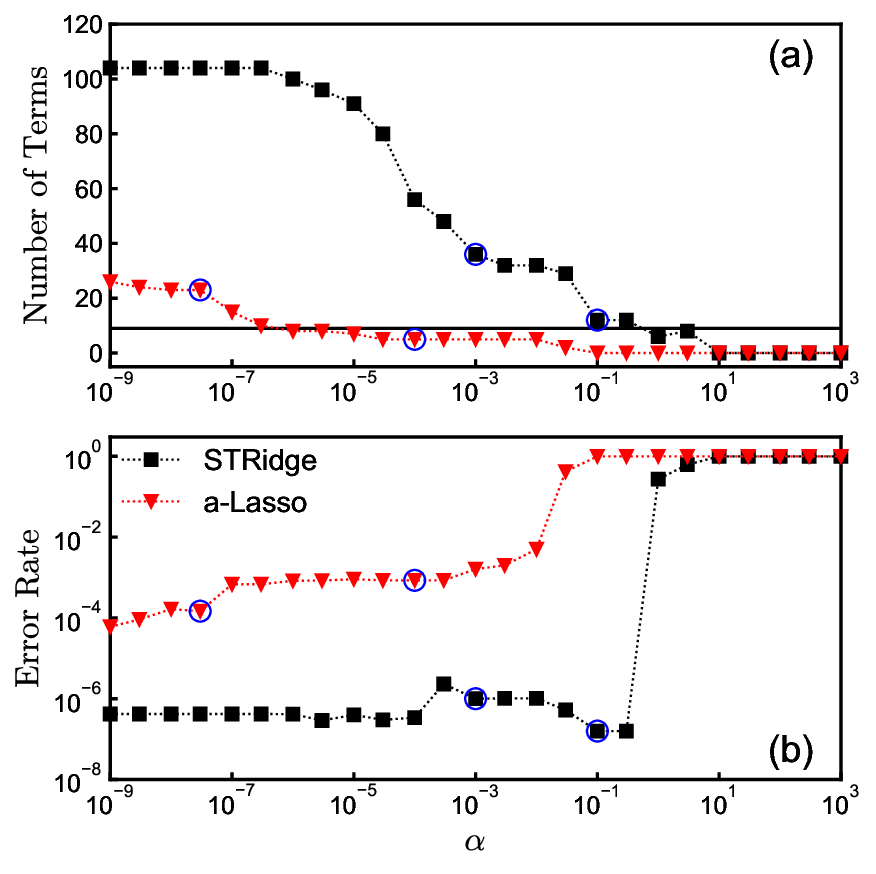}
    \caption{(a) The total number of terms and (b) the error rate of the constitutive equations obtained by {\it Rheo}-SINDy with the STRidge (black squares) and a-Lasso (red reverse triangles) for the FENE-P dumbbell model in the conformation expression (Eqs.~\eqref{FENE_P_xx}--\eqref{FENE_P_xy} with $n_{\rm K} = 10$).
    The horizontal line in (a) indicates the correct number of terms. 
    The blue circles represent the $\alpha$ values selected for test simulations. Based on our criterion, $\alpha=1\times10^{-1}$ for the STRidge and $\alpha=3\times10^{-8}$ for the a-Lasso are picked.}
    \label{Fig07}
\end{figure}
\begin{figure*}[t]
   \centering
    \includegraphics[width=5.0in]{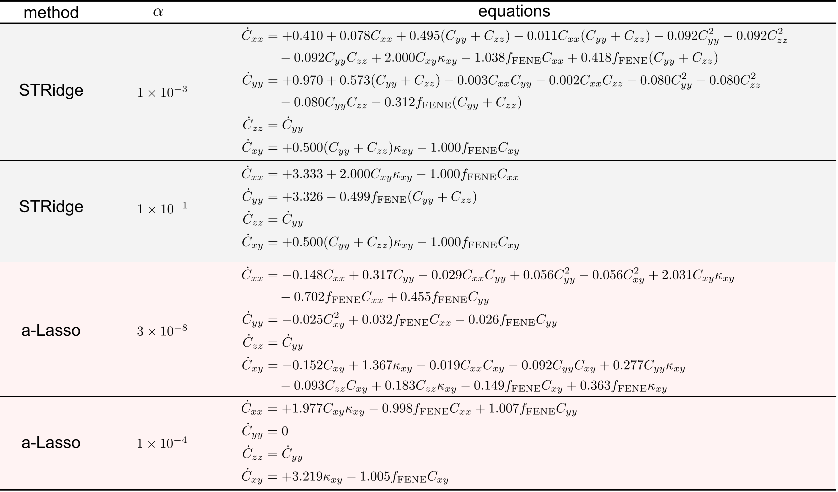}
    \caption{The constitutive equations obtained by {\it Rheo}-SINDy with the STRidge and a-Lasso for the FENE-P dumbbell model in the conformation expression (Eqs.~\eqref{FENE_P_xx}--\eqref{FENE_P_xy} with $n_{\rm K} = 10$). Based on our criterion, the appropriate $\alpha$ values were determined as $\alpha=1\times10^{-1}$ for the STRidge and $\alpha=3\times10^{-8}$ for the a-Lasso.}
    \label{Fig08}
\end{figure*}
\begin{figure}[t]
   \centering
    \includegraphics[width=2.75in]{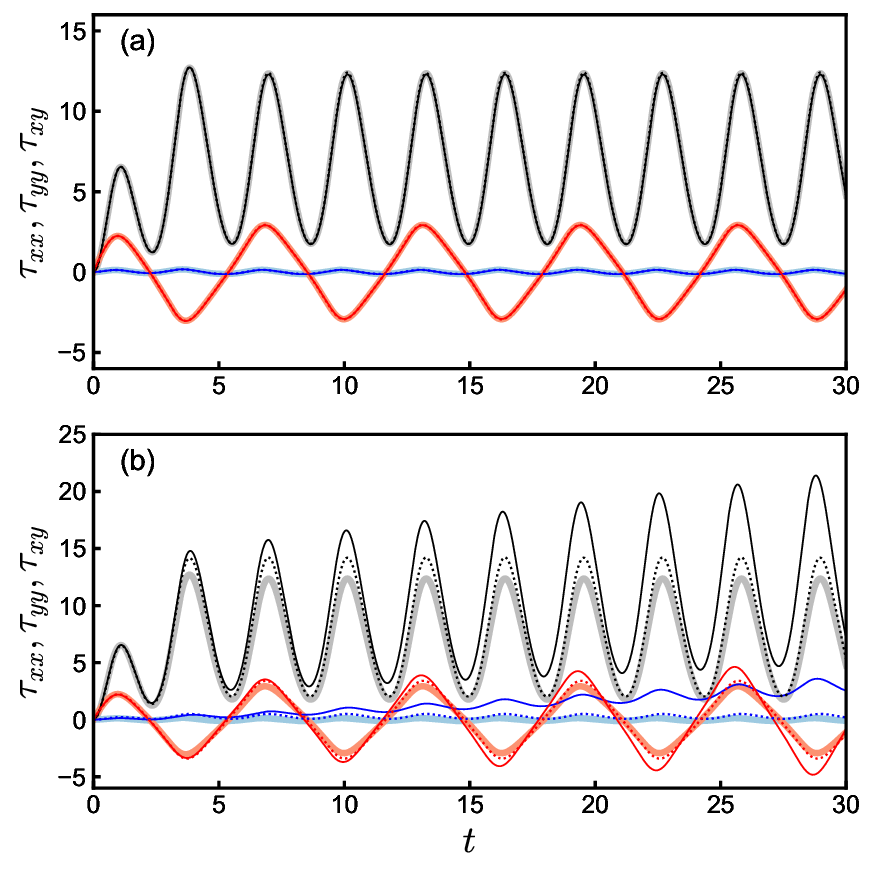}
    \caption{Test simulation results under the oscillatory shear flow with $\gamma_0 = 4$ and $\omega = 1$ using the equations obtained by {\it Rheo}-SINDy with (a) the STRidge and (b) a-Lasso for the FENE-P dumbbell model in the conformation expression. The obtained data from the equations were converted to the stress using the dimensionless form of Eq.~\eqref{stress_FENE_P}.
    The black, blue, and red lines show $\tau_{xx}$, $\tau_{yy}$, and $\tau_{xy}$. The bold, thin dotted, and thin solid lines indicate the exact solutions (Eqs.~\eqref{stress_FENE_P}--\eqref{FENE_P_xy} with $n_{\rm K} = 10$), predictions with smaller $\alpha$ values ($\alpha = 1 \times 10^{-3}$ for the STRidge and $\alpha = 3 \times 10^{-8}$ for the a-Lasso), and predictions with larger $\alpha$ values ($\alpha = 1 \times 10^{-1}$ for the STRidge and $\alpha = 1 \times 10^{-4}$ for the a-Lasso), respectively.
    }
    \label{Fig09}
\end{figure}
\begin{figure}[t]
 \centering
  \includegraphics[width=2.75in]{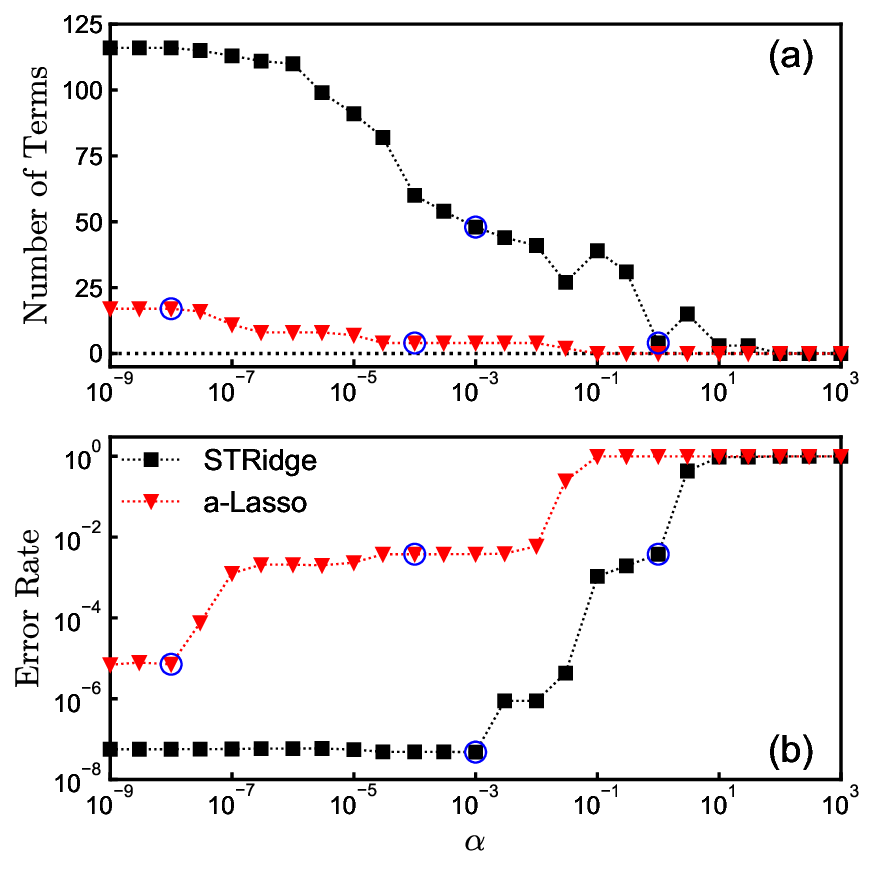}
  \caption{(a) The total number of terms and (b) the error rate of the constitutive equation obtained by {\it Rheo}-SINDy with the STRidge (black squares) and the a-Lasso (red reverse triangles) for the FENE-P dumbbell model in the stress expression (Eqs.~\eqref{FENE_P_s_xx}--\eqref{FENE_P_s_xy} with $n_{\rm K} = 10$). The horizontal short-dashed line in (a) indicates that the number of terms is zero. The blue circles represent the $\alpha$ values selected for test simulations. Based on our criterion, the appropriate $\alpha$ values were determined as $\alpha=1\times10^{-3}$ for the STRidge and $\alpha=1\times10^{-8}$ for the a-Lasso.}
  \label{Fig10}
\end{figure}
\begin{figure*}[t]
 \centering
  \includegraphics[width=5.0in]{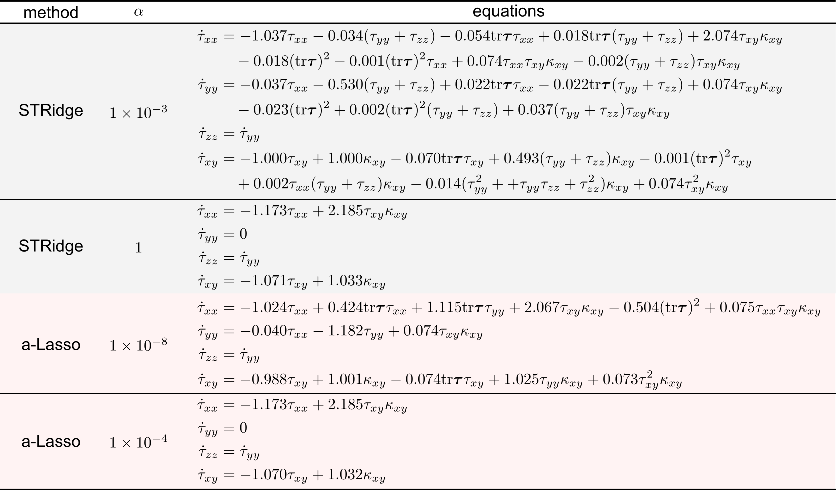}
  \caption{The constitutive equations obtained by {\it Rheo}-SINDy with the STRidge and a-Lasso for the FENE-P dumbbell model in the stress expression (Eqs.~\eqref{FENE_P_s_xx}--\eqref{FENE_P_s_xy} with $n_{\rm K} = 10$). Based on our criterion, the appropriate $\alpha$ values were determined as $\alpha=1\times10^{-3}$ for the STRidge and $\alpha=1\times10^{-8}$ for the a-Lasso.}
  \label{Fig11}
\end{figure*}
Figure~\ref{Fig07} displays the results: (a) the total number of predicted terms and (b) the error rate as a function of the hyperparameter $\alpha$.
The error rate is defined as the sum of the mean squared errors (MSEs) of $\dot {\bm t}_{\mu\nu} - {\bm \Theta} \hat{\bm \xi}_{\mu\nu}$ scaled so that the maximum value of each method is 1.
Similar to the results for the phenomenological constitutive equations shown in Secs.~\ref{sec:UCM_res} and \ref{sec:Giesekus_res}, the a-Lasso provides sparser solutions than the STRidge, and the STRidge gives lower error rates than the a-Lasso.
Figure~\ref{Fig08} presents the differential equations obtained by the STRidge and a-Lasso for two $\alpha$ values: the one selected by our proposed criterion (i.e., with the smallest number of terms, while the corresponding error is of the same order as the minimum error), and the other has a larger $\alpha$ value (i.e., with a smaller number of terms).
From the lower part of Fig.~\ref{Fig08}, while the a-Lasso provides sparser solutions, they are not correct (cf. Eqs.~\eqref{FENE_P_xx}--\eqref{FENE_P_xy}).
Specifically, in all cases for $C_{xx}$, $C_{yy}$, and $C_{zz}$, the a-Lasso failed to identify the constant term in Eqs.~\eqref{FENE_P_xx} and \eqref{FENE_P_yy_zz}, which is a possible source of larger errors compared to the STRidge.
In the case of the STRidge, we confirmed that by choosing $\alpha$ based on our criterion ($\alpha = 1 \times 10^{-1}$),
nearly correct differential equations can be obtained, as shown in the upper part of Fig.~\ref{Fig08}.
Since the $yy$-component and $zz$-component of the stress are equivalent, the exact equations can be recovered by setting $C_{yy} = C_{zz}$.
Thus, we conclude that the correct differential equations for the FENE-P dumbbell model can be obtained by {\it Rheo}-SINDy.

To validate the obtained equations in Fig.~\ref{Fig08}, we generated data using the equations under conditions different from those used to generate the training data (namely, the oscillatory shear flow with $\gamma_0=4$ and $\omega=1$), and converted these data to the stress data using the dimensionless form of Eq.~\eqref{stress_FENE_P}. Figure~\ref{Fig09} compares the generated data with the correct data generated from Eqs.~\eqref{stress_FENE_P}--\eqref{FENE_P_xy}.
As shown in this figure,  the equations obtained by the STRidge can reproduce the exact solutions even when the equations are not exactly correct ($\alpha = 1 \times 10^{-3}$).
In contrast, the test simulations with the differential equations obtained by the a-Lasso show the deviations from the test data, especially for $\tau_{xx}$. 
In particular, although the equations with $\alpha = 3 \times 10^{-8}$ have the small error rate for the training data (cf. Fig.~\ref{Fig07}(b)), those predictions deviate from the test data, and this deviation increases over time. 
These results emphasize the need to choose an appropriate optimization method to obtain reasonable solutions.

We then examine whether the stress expression of the constitutive equation for the FENE-P dumbbell model (cf. Eqs.~\eqref{FENE_P_s_xx}--\eqref{FENE_P_s_xy}) can be found by {\it Rheo}-SINDy.
For such a purpose, we prepared the following custom library:
\begin{equation}
{\bm \Theta}
= \left [ \begin{array}{cccc}
   1 & \{ {\rm tr} {\bm \tau} (t_1) \}^{p}{\bm T}_{\rm s}(t_1) & \{ {\rm tr} {\bm \tau} (t_1) \}^{2} & \{ {\bm T}_{\rm s} (t_1) \}^{p} \kappa_{xy} (t_1) \\
   1 & \{ {\rm tr} {\bm \tau} (t_2) \}^{p}{\bm T}_{\rm s}(t_2) & \{ {\rm tr} {\bm \tau} (t_2) \}^{2} & \{ {\bm T}_{\rm s} (t_2) \}^{p} \kappa_{xy} (t_2) \\
   \vdots & \vdots & \vdots & \vdots \\
   1 & \{ {\rm tr} {\bm \tau} (t_n) \}^{p}{\bm T}_{\rm s}(t_n) & \{ {\rm tr} {\bm \tau} (t_n) \}^{2} & \{ {\bm T}_{\rm s} (t_n) \}^{p} \kappa_{xy} (t_n) \\
\end{array}
\right ], \label{Theta_FENE_P_stress}
\end{equation}
where ${\bm T}_{\rm s}$ includes $\{\tau_{xx}, \tau_{yy}, \tau_{zz}, \tau_{xy} \}$ and $p$ ($=0, 1, 2$) is the polynomial order. Thus, the total number of library functions was $N_{\bm \Theta} = 29$.
We designed the library to include all terms present in Eqs.~\eqref{FENE_P_s_xx}--\eqref{FENE_P_s_xy}. Moreover, we excluded terms that could potentially become large, such as higher-order terms involving $\kappa_{xy}$.
When such terms are included in the solutions, the differential equations may be unstable, and in worse cases, they may diverge. Furthermore, as proposed by Lennon and coworkers, the principle of frame-invariance can also be used to constrain the candidate terms in $\bm \Theta$~\cite{Lennon2023-RUDE}, which will be considered in our future work.

\begin{figure}[t]
 \centering
  \includegraphics[width=2.75in]{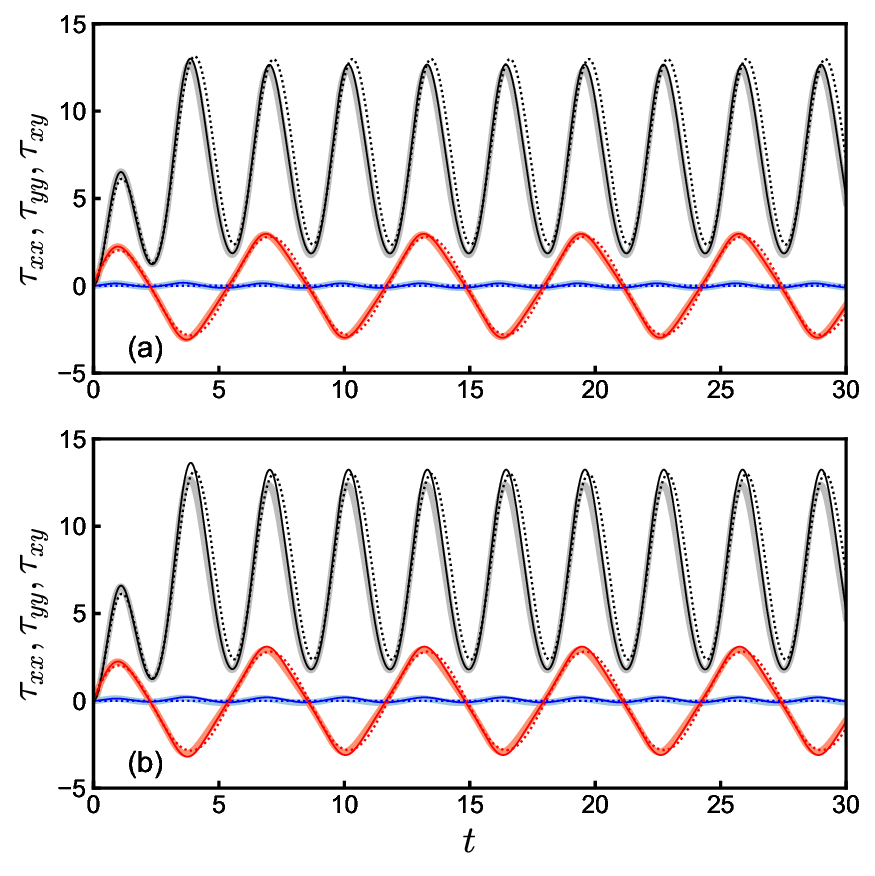}
  \caption{Test simulation results under the oscillatory shear flow with $\gamma_0 = 4$ and $\omega = 1$ using the constitutive equations obtained by {\it Rheo}-SINDy with (a) the STRidge and (b) a-Lasso for the FENE-P dumbbell model in the stress expression. The black, blue, and red lines represent the $xx$-, $yy$-, and $xy$-components of the stress tensor, respectively.
  The bold lines show the exact solutions. The thin solid and short-dashed lines indicate the results with smaller $\alpha$ values ($\alpha = 1 \times 10^{-3}$ for the STRidge and $\alpha = 1 \times 10^{-8}$ for the a-Lasso) and with larger $\alpha$ values ($\alpha = 1$ for the STRidge and $\alpha = 1 \times 10^{-4}$ for the a-Lasso), respectively.}
  \label{Fig12}
\end{figure}
Figure~\ref{Fig10} shows (a) the total number of terms and (b) the error rate of the constitutive equation obtained by {\it Rheo}-SINDy with the STRidge and a-Lasso for the FENE-P dumbbell model.
Similar to the previous case, the a-Lasso yields sparser solutions than the STRidge.
Based on the number of terms and the error rates shown in Fig.~\ref{Fig10}, we chose two $\alpha$, one is based on the proposed criterion and the other has a smaller number of terms. Figure~\ref{Fig11} presents the equations obtained using the chosen $\alpha$.
From Fig.~\ref{Fig11}, the equations predicted by the STRidge with $\alpha = 1$ and the a-Lasso with $\alpha = 1 \times 10^{-4}$ are almost the same; conversely, the expressions for small $\alpha$ values significantly differ between the two methods.
For the STRidge with $\alpha = 1\times 10^{-3}$, the identified equations are close to the correct equations (cf. Eqs.~\eqref{FENE_P_s_xx}--\eqref{FENE_P_s_xy}).
Furthermore, the coefficient values for the terms obtained correctly are close to the correct values.
For the a-Lasso with $\alpha = 1\times 10^{-8}$, several coefficients for the correctly obtained terms, such as $\tau_{xx}$, $\tau_{xy}\kappa_{xy}$, and $\tau_{xx}\tau_{xy}\kappa_{xy}$ in the equation for $\dot\tau_{xx}$, are close to the exact values, but for other several terms, such as ${\rm tr} \bm \tau \tau_{xx}$ in the equation for $\dot\tau_{xx}$, the correct coefficient values are not obtained.
Nevertheless, from Fig.~\ref{Fig12}, which shows the test simulation results, the equations obtained by the STRidge with $\alpha = 1\times 10^{-3}$ and the a-Lasso with $\alpha = 1\times 10^{-8}$ can well reproduce the exact solutions including the small oscillation of $\tau_{yy}$.
Although the equations obtained by the STRidge and a-Lasso demonstrate similar performance in the test simulations shown in Fig.~\ref{Fig12}, the difference in predictions is quantified by their MSEs shown in Table~\ref{Table04}.
When $\alpha$ is small, the STRidge outperforms the a-Lasso.
The STRidge, however, generally provides sparse solutions within a narrow range of $\alpha$ values, requiring careful selection of $\alpha$. 
In the case of the FENE-P dumbbell model, as shown in Figs.~\ref{Fig07} and \ref{Fig10}, the error rate obtained by the STRidge is smaller than that obtained by the a-Lasso over the investigated $\alpha$ region. This indicates that the STRidge is superior to the a-Lasso in representing the training data. However, as shown in the next section, even with a low error rate, the performance in test simulations for the extrapolation region is not guaranteed. This indicates that the optimization method needs to be chosen based on the specific problem. 

\begin{table}[t]
\caption{\label{Table04}The mean squared errors (MSEs) between the predicted and correct constitutive equations for the FENE-P dumbbell model in the stress expression (Eqs.~\eqref{FENE_P_s_xx}--\eqref{FENE_P_s_xy} with $n_{\rm K} = 10$)}.
\begin{tabular}{ccccc}
\hline
\hline
method & $\alpha$ & MSE ($\tau_{xx}$)  & MSE ($\tau_{yy}$) & MSE ($\tau_{xy}$) \\
\hline
STRidge & $1\times 10^{-3}$ & $8.1\times10^{-3}$ & $3.1\times10^{-4}$ & $1.1\times10^{-3}$\\
STRidge & $1$ & $2.3$ & $8.2\times10^{-3}$ & $8.9\times10^{-2}$\\
a-Lasso & $1\times 10^{-8}$ & $3.2\times10^{-1}$ & $3.4\times10^{-3}$ & $2.1\times10^{-2}$\\
a-Lasso & $1\times 10^{-4}$ & $2.3$ & $8.2\times10^{-3}$ & $9.0\times10^{-2}$\\
\hline
\hline
\end{tabular}
\end{table}
\subsection{FENE Dumbbell Model}
\begin{figure}[t]
 \centering
  \includegraphics[width=2.75in]{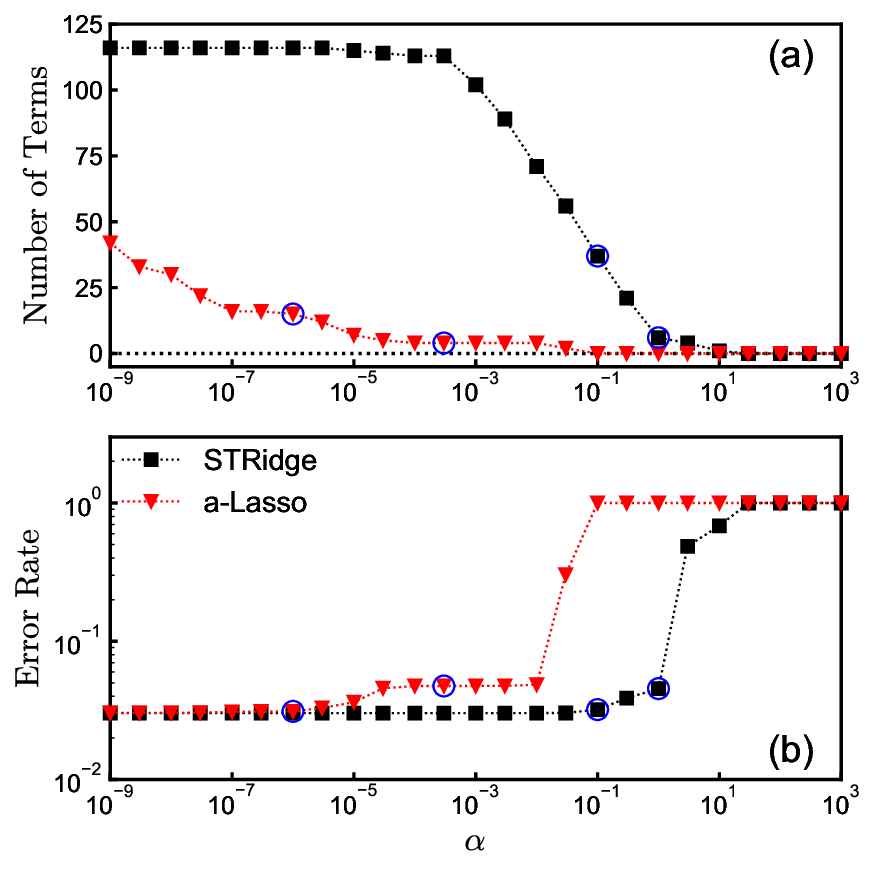}
  \caption{(a) The total number of terms and (b) the error rate of the constitutive equation obtained by {\it Rheo}-SINDy with the STRidge (black squares) and the a-Lasso (red reverse triangles) for the FENE dumbbell model. The horizontal short-dashed line in (a) indicates that the number of terms is zero. The blue circles represent the $\alpha$ values selected for test simulations. Based on our criterion, the appropriate $\alpha$ values were determined as $\alpha=1\times10^{-1}$ for the STRidge and $\alpha=1\times10^{-6}$ for the a-Lasso.}
  \label{Fig13}
\end{figure}
\begin{figure*}[t]
 \centering
  \includegraphics[width=6.0in]{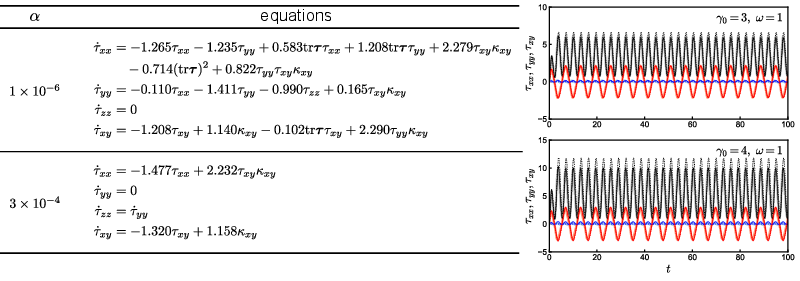}
  \caption{The constitutive equations obtained by {\it Rheo}-SINDy with the a-Lasso for the FENE dumbbell model (left) and the test simulation results under the oscillatory shear flows with $\gamma_0 = 3$ and $\omega = 1$ (right upper panel) and $\gamma_0 = 4$ and $\omega = 1$ (right lower panel). The black, blue, and red lines represent the $xx$-, $yy$-, and $xy$-components of the stress tensor, respectively. The bold lines show the exact solutions, and the thin solid and short-dashed lines show the results with the smaller $\alpha$ value ($\alpha = 1 \times 10^{-6}$), which is chosen based on our criterion, and the larger $\alpha$ value ($\alpha = 3 \times 10^{-4}$).}
  \label{Fig14}
\end{figure*}

We next address the FENE dumbbell model.
As explained in Sec.~\ref{sec_FENE_equations}, the FENE dumbbell model does not have an analytical expression of the constitutive equation. Thus, we here aim to identify an {\textit {approximate}} constitutive equation using {\it Rheo}-SINDy.

To prepare the library $\bm \Theta$ for the FENE dumbbell model, we utilize the physical insights obtained from the analytical expression of the FENE-P dumbbell model.
We assume the constitutive equation for the FENE-P dumbbell model is {\it similar} to that for the FENE dumbbell model.
Since the FENE-P dumbbell model is a simplified version of the FENE dumbbell model, we believe that this is a reasonable assumption.
Here, we note that the stress expression shown in Eq.~\eqref{stress_FENE_P} is no longer applicable to the FENE dumbbell model since the values of $h(t)$ differ for each individual dumbbell.
Thus, it is invalid to obtain stress through the conformation tensor $\bm C$ using Eq.~\eqref{stress_FENE_P}.
Based on the above considerations, we decided to use the custom library presented in Eq.~\eqref{Theta_FENE_P_stress}, which was also used to discover the constitutive equation for the FENE-P dumbbell model.

Figure~\ref{Fig13} compares (a) the total number of terms and (b) the error rate predicted by the STRidge and a-Lasso.
Similar to the previous discussions, we obtain sparse solutions over a wide range of $\alpha$ values with the a-Lasso, whereas the STRidge gives sparse solutions only within a limited range of $\alpha$.
The left table in Fig.~\ref{Fig14} shows the equations obtained by the a-Lasso with two $\alpha$ values chosen in the same way as Fig.~\ref{Fig10}.
The predictions obtained by the STRidge are inferior to those obtained by the a-Lasso shown in Fig.~\ref{Fig14}; therefore, we focus on the a-Lasso result below (the STRidge results are discussed in Sec.~S3 in the supplementary material).
From the left table in Fig.~\ref{Fig14}, if $\alpha$ is appropriately chosen, the a-Lasso can give sparse equations with coefficients of reasonable (not excessively large) magnitudes.
Comparing the equations for the FENE-P model obtained by the a-Lasso with $\alpha = 1 \times 10^{-8}$ (Fig.~\ref{Fig11}) and those for the FENE model obtained by the a-Lasso with $\alpha = 1\times 10^{-6}$ (Fig.~\ref{Fig14}), the appearing terms are almost identical, which demonstrates the similarity between these models.
The difference in the coefficients thus represents the difference between these models.
The right panels in Fig.~\ref{Fig14} show the test simulation results obtained by the equations shown in the left table.
With the large $\alpha$ ($\alpha = 3 \times 10^{-4}$),
the identified equation for $\tau_{yy}$ becomes ${\dot \tau}_{yy} = 0$,
which fails to reproduce the oscillatory behavior of $\tau_{yy}$. 
On the contrary, the equations obtained with $\alpha=1 \times 10^{-6}$ can reproduce well the BD simulation results outside the range of the training data, including the oscillatory behavior of $\tau_{yy}$.

This success suggests that {\it Rheo}-SINDy with the a-Lasso is effective for discovering {\it unknown} constitutive equations.
Nevertheless, we note that the equations presented in Fig.~\ref{Fig14} may fail to predict test data significantly outside the range of the training data.
Reproducing such highly nonlinear data would require the nonlinear terms dropped in Fig.~\ref{Fig14}.
In this sense, the constitutive equations for the FENE dumbbell model obtained here are appropriately referred to as the {\textit {approximate}} constitutive equations.

\begin{figure}[t]
 \centering
  \includegraphics[width=2.75in]{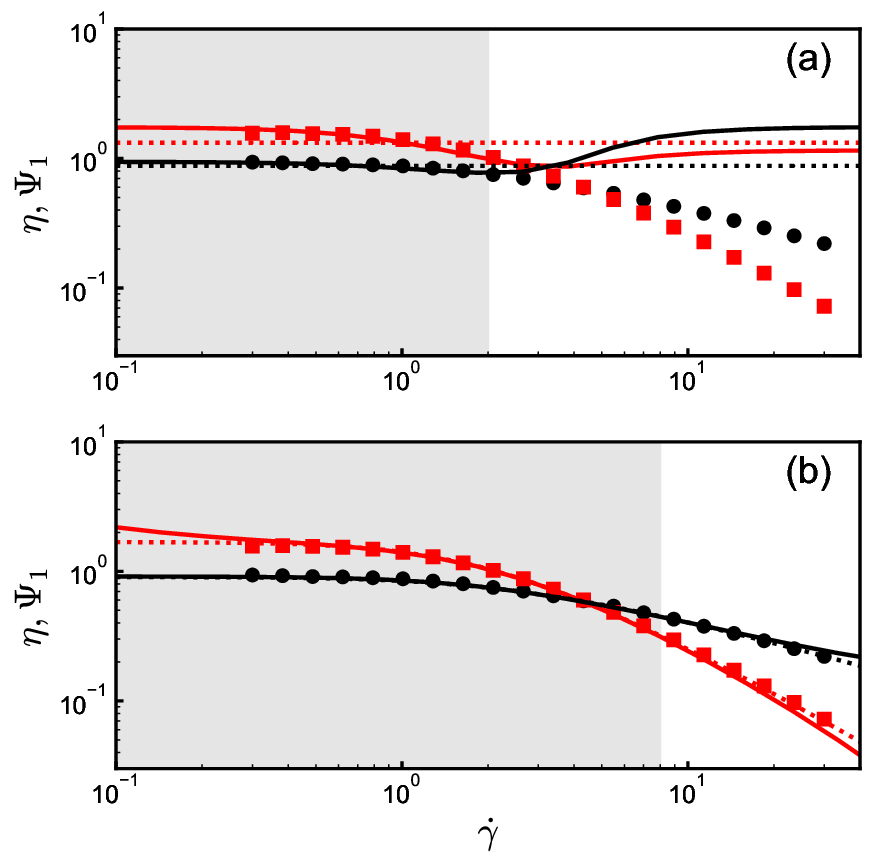}
  \caption{Steady state shear viscosity $\eta$ (black) and normal stress coefficient $\Psi_1$ (red) for the training data (a) with $\gamma_0 = 2$ and (b) with $\gamma_0 = 8$. The symbols are the BD simulation data, and the lines are the {\it Rheo}-SINDy predictions. The solid and short-dashed lines show the {\it Rheo}-SINDy results with the smaller $\alpha$ value ($\alpha = 1 \times 10^{-6}$ for both (a) and (b)) and the larger $\alpha$ value ($\alpha = 3 \times 10^{-4}$ for (a) and $\alpha = 1 \times 10^{-5}$ for (b)). The shaded regions show the ranges of shear rates in the training data.}
  \label{Fig15}
\end{figure}

To further investigate rheological quantities under shear flow, we conducted shear simulations under constant shear rates. 
Figure~\ref{Fig15} compares the steady state shear viscosity $\eta \equiv \sigma_{xy}/\dot \gamma$ and first normal stress coefficient $\Psi_1 \equiv (\sigma_{xx} - \sigma_{yy})/\dot \gamma^2 $ obtained from BD simulations with the approximate constitutive equations obtained by {\it Rheo}-SINDy. 
Although the constant shear flows considered in Fig.~\ref{Fig15} differ from the oscillatory shear flows used to generate the training data, {\it Rheo}-SINDy can reproduce the results of BD simulations, including shear thinning, within the shear rate region of the training data and slightly beyond it, as shown in Fig.~\ref{Fig15}(a). 
However, in the high shear rate region, the {\it Rheo}-SINDy predictions with $\alpha = 1\times10^{-6}$ exhibit plateaus for $\eta$ and $\Psi_1$ that differ from the results of BD simulations.
To improve the {\it Rheo}-SINDy predictions in the high shear rate region, we tested the training data with larger $\gamma_0$ ($\gamma_0 = 8$) while keeping the same $\omega$ values as those used to obtain Figs.~\ref{Fig13} and \ref{Fig14}. 
We note that the generated data with $\gamma_0 = 8$ show nonlinear stress responses, as shown in Sec.~S4 in the supplementary material. 
Figure~\ref{Fig15}(b) indicates that the model found by {\it Rheo}-SINDy with the training data that includes higher shear rates successfully extrapolates continuous shear thinning to the high shear rate regime beyond the training data. 

\begin{figure}[t]
 \centering
  \includegraphics[width=2.75in]{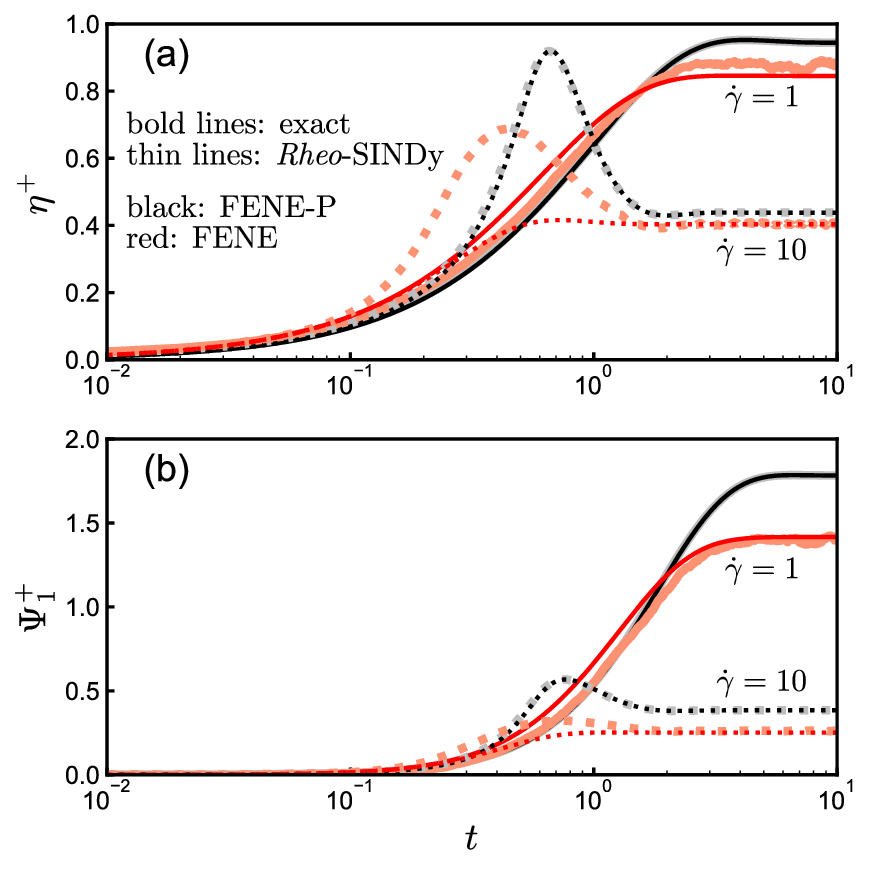}
  \caption{(a) Transient state shear viscosity $\eta^{+}$ and (b) first normal stress coefficient $\Psi_1^{+}$ under the steady shear flow with $\dot \gamma = 1$ (solid lines) and $10$ (dotted lines) for the FENE-P (black) and FENE (red) dumbbell models. The thin and bold lines were obtained by {\it Rheo}-SINDy and by the equations for the FENE-P and FENE dumbbell models, respectively. The {\it Rheo}-SINDy regressions were conducted for the training data using the same parameters ($\gamma_0 = 8$ and $\omega \in \{0.1, 0.2, \ldots, 1\}$) for both the FENE-P and FENE dumbbell models. While the {\it Rheo}-SINDy results for the FENE-P dumbbell model were obtained by the STRidge with $\alpha = 1 \times 10^{-4}$, those for the FENE dumbbell model were obtained by the a-Lasso with $\alpha = 1 \times 10^{-5}$, which is the same parameter as that in Fig.~\ref{Fig15}(b).}
  \label{Fig16}
\end{figure}

It would be interesting to compare the results obtained by {\it Rheo}-SINDy for the FENE-P and FENE dumbbell models.
Figure~\ref{Fig16} indicates the transient shear viscosity $\eta^{+}$ and first normal stress coefficient $\Psi_1^{+}$ obtained by {\it Rheo}-SINDy and those exact solutions for the FENE-P and FENE dumbbell models. We note that, for comparison, the {\it Rheo}-SINDy regressions were conducted using the training data with the same parameters ($\gamma_0 = 8$ and $\omega \in \{0.1, 0.2, \ldots, 1\}$) for both models.
For the FENE-P dumbbell model, the regression for the stress expression (cf. Eqs.~\eqref{FENE_P_s_xx}--\eqref{FENE_P_s_xy}) was performed. From the success of the STRidge in this case (cf. Fig.~\ref{Fig10}), we used the STRidge with $\alpha = 1 \times 10^{-4}$. The {\it Rheo}-SINDy results for the FENE dumbbell model were obtained using the a-Lasso with $\alpha = 1 \times 10^{-5}$, which is the same parameter as that determined in Fig.~\ref{Fig15}(b). 
Figure~\ref{Fig16} shows that {\it Rheo}-SINDy for the FENE-P dumbbell model accurately reproduces the transient behavior shown in the exact solutions. However, while {\it Rheo}-SINDy for the FENE dumbbell model successfully reproduces the steady state values of $\eta^{+}$ and $\Psi_1^{+}$, as expected from Fig.~\ref{Fig15}(b), it fails to describe the stress overshoot, especially for the large $\dot \gamma$. 
Although the objective of this study is not to develop a sophisticated constitutive model that approximates the FENE dumbbell model, this discrepancy can be addressed through three possible strategies: (1) adding training data obtained by steady shear flow with high shear rate, (2) using the conformation tensor $\bm C$ as a latent variable, and (3) incorporating higher-order terms into the library for regression. 
For point (1), it has been reported that including training data obtained from constant shear flow with high shear rate is effective~\cite{Miyamoto2023-rq}. This enables the development of regression models that can predict transient behavior, even for more complex models. For point (2), the conformation tensor often allows for simpler descriptions, as shown for the FENE-P dumbbell model. It might be effective to use $\bm C$ to obtain regression equations in the FENE dumbbell model. This case will require an additional regression to relate $\bm C$ and $\bm \tau$, replacing Eq.~\eqref{stress_FENE_P} for the FENE-P dumbbell model. For point (3), Du and coworkers have reported that the approximated model with high-order terms achieves high predictive performance~\cite{Du2005}. Following this study, including higher-order terms in the library can enhance the representational ability of {\it Rheo}-SINDy. These considerations are interesting subjects for future studies. 

Thanks to the equations obtained using {\it Rheo}-SINDy, it is possible to provide a physical interpretation with the assistance of rheological knowledge. For example, from the comparison of the equations obtained for the FENE-P dumbbell model (cf. Fig.~\ref{Fig11}) and those for the FENE dumbbell model (cf. Fig.~\ref{Fig14}), the equations for larger $\alpha$ value ($\alpha = 1\times 10^{-4}$ for the FENE-P dumbbell model and $\alpha = 3\times 10^{-4}$ for the FENE dumbbell model) are similar except for the coefficient values. 
We note that these $\alpha$ values are not selected based on our criteria. However, to simply analyze the representation obtained by {\it Rheo}-SINDy, we consider the sparser solution here. The terms in these equations are the same as those for the UCM model (and thus the Hookean dumbbell model).
This indicates that all of these models share the same origin based on the dumbbell model.
We can also discuss the linear term of stress, which represents the relaxation of stress (see Eq.~\eqref{UCM}).
Since the relaxation time at equilibrium ($\lambda = \zeta/4h_{\rm eq}$) is taken as the unit time in this study,
the coefficient of this term should be $-1$ at equilibrium (and thus for the UCM model, see Eqs.~\eqref{UCM_shear_xx}--\eqref{UCM_shear_xy}).
From Figs.~\ref{Fig11} and \ref{Fig14}, the coefficient of the linear term of stress with the larger $\alpha$ values is smaller than $-1$,
which indicates $\lambda_{\rm sf} < \lambda_{\rm eq}$ with the subscript ``sf'' and ``eq'' standing for ``shear flow'' and ``equiliblium'', respectively.
This indicates that under shear flow, the values of spring strength for the FENE-P and FENE dumbbell models become larger than $h_{\rm eq}$, which implies the appearance of the FENE effects under flow.
From this discussion, it is evident that {\it Rheo}-SINDy can provide physically interpretable constitutive equations.

\begin{figure*}[t]
 \centering
  \includegraphics[width=5.0in]{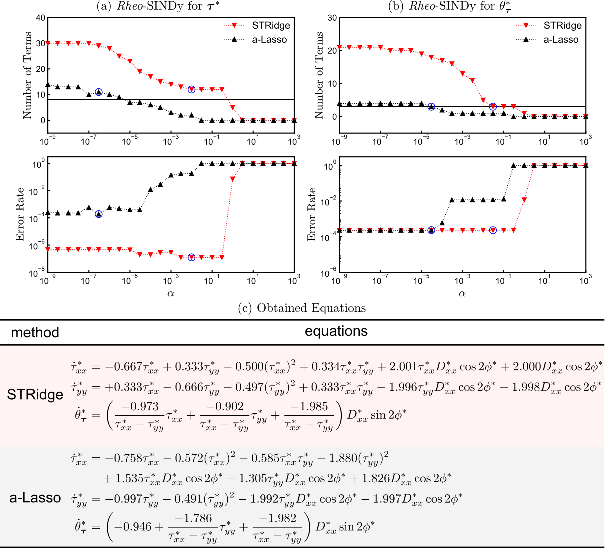}
  \caption{The total number of terms and the error rate of the constitutive equations obtained by {\it Rheo}-SINDy with the STRidge (black squares) and the a-Lasso (red reverse triangles) for (a) ${\bm \tau}^{\ast}$ and for (b) $\theta^{\ast}_{\bm \tau}$. The blue circles represent the appropriate $\alpha$ values. The horizontal lines in the top figures indicate the correct number of terms. (c) The constitutive equations obtained by {\it Rheo}-SINDy with the STRidge and the a-Lasso.}
  \label{Fig17}
\end{figure*}

\subsection{Material Frame Indifference}
Finally, we show a method to obtain constitutive models that address the principle of material objectivity. This principle requires that constitutive equations are independent of rotation. However, the results obtained by the {\it Rheo}-SINDy regression method presented thus far do not explicitly satisfy the rotational indifference.

To account for this indifference, Young and coworkers demonstrated that it is effective to separate the rotational components within the velocity gradient tensor from constitutive equations~\cite{Young2023}. In this approach, the velocity gradient tensor $\bm \kappa$ in the laboratory frame (LF) is decomposed into the deformation rate tensor $\bm D$ and the rotation rate tensor $\bm W$ as $\bm \kappa = \bm D + \bm W$, where $\bm D = (\bm \kappa + \bm \kappa^{\rm T})/2$ and $\bm W = (\bm \kappa - \bm \kappa^{\rm T})/2$. We then consider a constitutive model in the reference frame (RF) rotated by $\bm W$. Here, we note that the constitutive model in the RF should depend only on $\bm D$. We can compute the rotation matrix $\bm R(\theta_{\bm W})$ around an axis that relates the LF and RF as 
\begin{equation}
\frac{{\rm d} {\bm R} (\theta_{\bm W})}{{\rm d} t} = {\bm W} \cdot {\bm R} (\theta_{\bm W}),
\end{equation}
where $\theta_{\bm W}$ is the rotation angle. Using the rotation matrix $\bm R(\theta_{\bm W})$, the stress tensor $\bm \tau$ and the strain rate tensor $\bm D$ are respectively transformed into $\bm \tau' = \bm R^{\rm T} (\theta_{\bm W}) \bm \tau \bm R (\theta_{\bm W})$ and $\bm D' = \bm R^{\rm T} (\theta_{\bm W}) \bm D \bm R (\theta_{\bm W})$, where prime denotes quantities in the RF. We note that Young and coworkers demonstrated that the constitutive equation of the UCM model in the RF can be written similarly to that in the LF using $\bm \tau'$ and $\bm D'$~\cite{Young2023}.

Based on the study of Young and coworkers~\cite{Young2023}, we here propose a more rigorous approach to account for material objectivity. 
Note that Lennon and coworkers proposed another method to employ prescribed tensor terms satisfying the frame indifference~\cite{Lennon2023-RUDE}.
Since the current {\it Rheo}-SINDy method does not address constraints of inter-coefficients for tensorial terms, the identified equations should be constructed using invariants of the stress and strain rate tensors.
In this approach, $\bm \tau'$ and $\bm D'$ in the RF are diagonalized using rotation matrices around an axis $\bm P(\theta^{\ast}_{\bm \tau})$ and $\bm P(\theta^{\ast}_{\bm D})$ as $\bm \tau^{\ast} = \bm P^{\rm T} (\theta^{\ast}_{\bm \tau}) \bm \tau' \bm P (\theta^{\ast}_{\bm \tau})$ and $\bm D^{\ast} = \bm P^{\rm T} (\theta^{\ast}_{\bm D}) \bm \tau' \bm P (\theta^{\ast}_{\bm D})$, respectively. Here, $\theta^{\ast}_{\bm \tau}$ and $\theta^{\ast}_{\bm D}$ are the rotation angles for $\bm \tau^{\ast}$ and $\bm D^{\ast}$, respectively.  
With $\bm \tau^{\ast}_{\bm \tau}$, $\bm \tau^{\ast}_{\bm D}$, and $\phi^{\ast} = \theta_{\bm \tau}^{\ast} - \theta_{\bm D}^{\ast}$, 
the constitutive equations are expressed as follows:
\begin{align}
\dot {\bm \tau}^{\ast} = \bm f_{{\bm \tau}^{\ast}} ({\bm \tau}^{\ast}, {\bm D}^{\ast}, \phi^{\ast}) \label{dot_tau_star} \\
\dot {\theta}_{\bm \tau}^{\ast} = f_{{\theta}_{\bm \tau}^{\ast}} ({\bm \tau}^{\ast}, {\bm D}^{\ast}, \phi^{\ast}) \label{dot_theta_star}
\end{align}
Here, $\phi^{\ast}$ indicates the adjustment for using rotation matrices with different rotation angles to obtain $\bm \tau^{\ast}$ and $\bm D^{\ast}$. 
This type of constitutive equations can cover constitutive relations under deformations in a plane.
In the supplementary material (Sec.~S5), we demonstrate the derivation of constitutive equations for $\bm \tau^{\ast}$ and $\bm \theta^{\ast}_{\bm \tau}$ in the UCM and Giesekus models. 
Hereafter, we show that {\it Rheo}-SINDy can give Eqs.~\eqref{dot_tau_star} and \eqref{dot_theta_star} for the Giesekus model. 
From $\bm \tau^{\ast}$ and $\theta^{\ast}_{\bm \tau}$, $\bm \tau$ in LF can be determined by the relationship $\bm \tau = \bm P(\theta_{\bm W}) \bm P(\theta^{\ast}_{\bm \tau}) \bm \tau^{\ast} \bm P^{\rm T}(\theta^{\ast}_{\bm \tau}) \bm P^{\rm T}(\theta_{\bm W})$.

Using the same training data shown in Sec.~\ref{sec:Giesekus_eqs}, we here conduct the {\it Rheo}-SINDy regressions. 
In the Giesekus model, the constitutive equations for $\bm \tau^{\ast}$ and $\theta^{\ast}_{\bm \tau}$ are as follows (see the supplementary material for the detailed derivations):
\begin{align}
\dot{\tau}^{\ast}_{xx} &= +2\tau^{\ast}_{xx} D_{xx}^{\ast} \cos 2\phi^{\ast} - {\tau}^{\ast}_{xx} - \alpha_{\rm G} ({\tau}^{\ast}_{xx})^2 + 2 D_{xx}^{\ast} \cos 2\phi^{\ast} \label{G_eq_diag_01} \\
\dot{\tau}^{\ast}_{yy} &= -2\tau^{\ast}_{yy} D_{xx}^{\ast} \cos 2\phi^{\ast} - {\tau}^{\ast}_{yy} - \alpha_{\rm G} ({\tau}^{\ast}_{yy})^2 - 2 D_{xx}^{\ast} \cos 2\phi^{\ast} \label{G_eq_diag_02} \\
\dot{\tau}^{\ast}_{zz} &= -{\tau}^{\ast}_{zz} - \alpha_{\rm G} ({\tau}^{\ast}_{zz})^2 = 0\label{G_eq_diag_03} \\
\dot {\theta}_{\bm \tau}^{\ast} &= - \left ( \frac{\tau^{\ast}_{xx}}{\tau^{\ast}_{xx} - \tau^{\ast}_{yy}} + \frac{\tau^{\ast}_{yy}}{\tau^{\ast}_{xx} - \tau^{\ast}_{yy}} + \frac{2}{\tau^{\ast}_{xx} - \tau^{\ast}_{yy}}  \right ) D_{xx}^{\ast} \sin 2\phi^{\ast} \label{G_eq_diag_04}
\end{align}
Similarly to Eqs.~\eqref{G_shear_xx}--\eqref{G_shear_xy}, Eqs.~\eqref{G_eq_diag_01}--\eqref{G_eq_diag_04} are nondimensionalized using $\lambda$ and $G$. 
We prepared polynomial libraries up to second order consisting of $\{\tau^{\ast}_{xx}, \tau^{\ast}_{yy}, \tau^{\ast}_{zz}, D^{\ast}_{xx}\sin (2\phi^{\ast}), D^{\ast}_{xx}\cos (2\phi^{\ast})\}$ for the regression of $\bm \tau^{\ast}$, and $\{1/\Delta, \tau^{\ast}_{xx}/\Delta, \tau^{\ast}_{yy}/\Delta, \tau^{\ast}_{zz}/\Delta, D^{\ast}_{xx}\sin (2\phi^{\ast}), D^{\ast}_{xx}\cos (2\phi^{\ast})\}$ with the shorthand $\Delta = \tau^{\ast}_{xx} - \tau^{\ast}_{yy}$ for the regression of $\theta^{\ast}_{\bm \tau}$. Thus, there were $N_{\bm \Theta} = 21$ and $N_{\bm \Theta} = 28$ candidate terms for each component of $\bm \tau^{\ast}$ and $\bm \theta^{\ast}_{\bm \tau}$, respectively. 
Based on dimensional consideration, it is natural to create the library for the regression of $\theta^{\ast}_{\bm \tau}$ using terms of $\bm \tau^{\ast}/\Delta$.

Figure~\ref{Fig17} indicates the {\it Rheo}-SINDy results for $\bm \tau^{\ast}$ and $\theta^{\ast}_{\bm \tau}$. 
The trends in the number of terms and error rate shown in Figs.~\ref{Fig17}(a) and (b) are the same as those in the regressions with {\it Rheo}-SINDy at the LF (cf. Fig.~\ref{Fig04}). 
The lower panel in Fig.~\ref{Fig17}(a) shows that the a-Lasso has a larger error rate than the STRidge, reflecting the inaccuracies in the regression, especially in $\tau^{\ast}_{xx}$, as shown in Fig.~\ref{Fig17}(c). 
Nevertheless, in the regressions of $\theta^{\ast}_{\bm \tau}$, the STRidge and a-Lasso have similar error rates, which indicates that these methods have comparable performance in this case. 
We note that the expression of $\dot \theta^{\ast}_{\bm \tau}$ obtained by the a-Lasso, $\dot \theta^{\ast}_{\bm \tau} \simeq -(1 + 2\tau^{\ast}_{yy}/\Delta + 2/\Delta) D^{\ast}_{xx} \sin 2\phi^{\ast}$, reduces the correct equation shown in Eq.~\eqref{G_eq_diag_04}. 
The {\it Rheo}-SINDy regressions for $\bm \tau^{\ast}$ using STRidge have small error rates, but the obtained expressions shown in Fig.~\ref{Fig17}(c) differ from Eqs.~\eqref{G_eq_diag_01} and \eqref{G_eq_diag_02}.
This is because the Giesekus model with $\alpha_{\rm G}=0.5$ is equivalent to the Leonov model~\cite{Larson1988}. Interestingly, by conducting the diagonalization, the equivalent mathematical expression becomes evident through the {\it Rheo}-SINDy regression. 

Through the discussion in this section, we show that a constitutive equation that takes into account material objectivity can be obtained from {\it Rheo}-SINDy. This indicates that the technical preparations are now in place to precisely handle more complex models using {\it Rheo}-SINDy.

\section{\label{sec:Concluding_Remarks}Concluding Remarks}
We tested that the sparse identification for nonlinear dynamics (SINDy) modified for nonlinear rheological data, which we call {\it Rheo}-SINDy, is effective in finding constitutive equations of complex fluid dynamics.
We found that {\it Rheo}-SINDy can successfully identify correct equations from training data generated from {\it known} constitutive equations, as well as provide approximate constitutive equations (or reduced order models) from training data generated by mesoscopic models when constitutive equations are analytically {\it unknown}.

{\it Rheo}-SINDy was applied for two phenomenological constitutive equations (i.e., the upper convected Maxwell model and Giesekus model), which revealed the following two things.
First, compared to constant shear tests, oscillatory shear tests are appropriate for generating training data.
Second, the sequentially thresholded Ridge regression (STRidge) and adaptive Lasso (a-Lasso) are effective in finding appropriate constitutive equations.
We then examined the commonly used mesoscopic model,
namely the dumbbell model with three different representations of spring strength: the Hookean, FENE-P, and FENE springs.
Although the Hookean and FENE-P dumbbell models have analytical constitutive equations, for the FENE dumbbell model, there is no analytical expression of the constitutive equation.
We confirmed through the Hookean dumbbell model that even in the presence of noise, the a-Lasso provides the correct solution over a wide range of the hyperparameter $\alpha$.
{\it Rheo}-SINDy was also effective in discovering the complex constitutive equations for the FENE-P dumbbell model.
This case study revealed that the identification of complex equations requires the preparation of an appropriate custom library based on prior physical knowledge.
Using physical insights obtained from the Hookean and FENE-P dumbbell models, we attempted to find {\it approximate} constitutive equations for the FENE dumbbell model.
We found that the a-Lasso can successfully give the approximate constitutive equations as combinations of known elemental terms given in a function library, which can be used in predictions beyond the range of the training data. Finally, we present a regression method that yields a constitutive model satisfying rotational indifference. 
These results indicate that {\it Rheo}-SINDy is ready to be applied to data obtained from more complex models or experimental data. 

From our investigation, {\it Rheo}-SINDy with the STRidge or a-Lasso is effective for discovering constitutive equations from nonlinear rheological data.
We found that the STRidge is generally superior in terms of retaining correct terms, while the a-Lasso is more robust to the selection of $\alpha$ than the STRidge.
To obtain correct constitutive equations, in addition to selecting the appropriate optimization method, we are required to design an appropriate library by using physical insights, namely {\textit {domain knowledge}}.
Designing such a proper library necessitates not only including necessary terms but also excluding unnecessary terms. 
For such a purpose, candidate terms can be chosen to satisfy the principle of frame invariance~\cite{Lennon2023-RUDE}.
Furthermore, we may conduct regression with physics-informed constraints on the coefficients~\cite{Loiseau2018}.

This research is expected to have an impact on fields such as rheology and fluid mechanics.
From a rheological perspective,
for several systems such as entangled polymers~\cite{Sato2019, Miyamoto2023} and wormlike micellar solutions~\cite{Sato2020, Sato2022}, sophisticated mesoscopic models suitable for numerical simulations under flow have been proposed.
These mesoscopic models can generate reasonable training data not only under shear flow but also under extensional flow.
Finding new approximate models from the data obtained by these mesoscopic simulations would be an interesting research subject.
Furthermore, it would be desirable to conduct {\it Rheo}-SINDy for experimental data obtained by Large Amplitude Oscillatory Shear (LAOS) experiments~\cite{Hyun2011}. Since the LAOS measurements do not provide all the major stress components under shear flow, exploring methods for discovering the constitutive equations from experimental data would be a future challenge.
When approximate constitutive models are identified, those models can be employed for predictions of complex flows, which would deepen our understanding of complex fluids.
We will continue our research in these directions.

\section*{Data Availability Statement}
The data that support the findings of this study are available upon reasonable request from the authors.

\begin{acknowledgments}
We thank Prof.~Yoshinobu Kawahara for his valuable comments on data-driven methods. 
We also thank Dr. John J. Molina for carefully reading the manuscript and providing insightful comments. We are grateful to Prof.~Takashi Taniguchi for encouraging us to start this research. 
TS was partially supported by JST PRESTO Grant Number JPMJPR22O3.
SM was financially supported by the Kyoto University Science and Technology Innovation Creation Fellowship (FSMAT), Grant Number JPMJFS2123.
\end{acknowledgments}

\section*{Author Contributions}
T.S. and S.M. contributed equally to this work.

\bibliography{Rheo_SINDy}

\begin{thebibliography}{56}%
\makeatletter
\providecommand \@ifxundefined [1]{%
 \@ifx{#1\undefined}
}%
\providecommand \@ifnum [1]{%
 \ifnum #1\expandafter \@firstoftwo
 \else \expandafter \@secondoftwo
 \fi
}%
\providecommand \@ifx [1]{%
 \ifx #1\expandafter \@firstoftwo
 \else \expandafter \@secondoftwo
 \fi
}%
\providecommand \natexlab [1]{#1}%
\providecommand \enquote  [1]{``#1''}%
\providecommand \bibnamefont  [1]{#1}%
\providecommand \bibfnamefont [1]{#1}%
\providecommand \citenamefont [1]{#1}%
\providecommand \href@noop [0]{\@secondoftwo}%
\providecommand \href [0]{\begingroup \@sanitize@url \@href}%
\providecommand \@href[1]{\@@startlink{#1}\@@href}%
\providecommand \@@href[1]{\endgroup#1\@@endlink}%
\providecommand \@sanitize@url [0]{\catcode `\\12\catcode `\$12\catcode
  `\&12\catcode `\#12\catcode `\^12\catcode `\_12\catcode `\%12\relax}%
\providecommand \@@startlink[1]{}%
\providecommand \@@endlink[0]{}%
\providecommand \url  [0]{\begingroup\@sanitize@url \@url }%
\providecommand \@url [1]{\endgroup\@href {#1}{\urlprefix }}%
\providecommand \urlprefix  [0]{URL }%
\providecommand \Eprint [0]{\href }%
\providecommand \doibase [0]{http://dx.doi.org/}%
\providecommand \selectlanguage [0]{\@gobble}%
\providecommand \bibinfo  [0]{\@secondoftwo}%
\providecommand \bibfield  [0]{\@secondoftwo}%
\providecommand \translation [1]{[#1]}%
\providecommand \BibitemOpen [0]{}%
\providecommand \bibitemStop [0]{}%
\providecommand \bibitemNoStop [0]{.\EOS\space}%
\providecommand \EOS [0]{\spacefactor3000\relax}%
\providecommand \BibitemShut  [1]{\csname bibitem#1\endcsname}%
\let\auto@bib@innerbib\@empty
\bibitem [{\citenamefont {Brunton}\ and\ \citenamefont
  {Kutz}(2022)}]{Brunton2022}%
  \BibitemOpen
  \bibfield  {author} {\bibinfo {author} {\bibfnamefont {S.~L.}\ \bibnamefont
  {Brunton}}\ and\ \bibinfo {author} {\bibfnamefont {J.~M.}\ \bibnamefont
  {Kutz}},\ }\href@noop {} {\emph {\bibinfo {title} {Data-Driven Science and
  Engineering}}},\ \bibinfo {edition} {2nd}\ ed.\ (\bibinfo  {publisher}
  {Cambridge University Press},\ \bibinfo {year} {2022})\BibitemShut {NoStop}%
\bibitem [{\citenamefont {Brunton}, \citenamefont {Proctor},\ and\
  \citenamefont {Kutz}(2016)}]{Brunton2016}%
  \BibitemOpen
  \bibfield  {author} {\bibinfo {author} {\bibfnamefont {S.~L.}\ \bibnamefont
  {Brunton}}, \bibinfo {author} {\bibfnamefont {J.~L.}\ \bibnamefont
  {Proctor}}, \ and\ \bibinfo {author} {\bibfnamefont {J.~N.}\ \bibnamefont
  {Kutz}},\ }\bibfield  {title} {\enquote {\bibinfo {title} {Discovering
  governing equations from data by sparse identification of nonlinear dynamical
  systems},}\ }\href@noop {} {\bibfield  {journal} {\bibinfo  {journal} {Proc.
  Natl. Acad. Sci. USA}\ }\textbf {\bibinfo {volume} {113}},\ \bibinfo {pages}
  {3932--3937} (\bibinfo {year} {2016})}\BibitemShut {NoStop}%
\bibitem [{\citenamefont {de~Silva}\ \emph {et~al.}(2020)\citenamefont
  {de~Silva}, \citenamefont {Higdon}, \citenamefont {Brunton},\ and\
  \citenamefont {Kutz}}]{De_Silva2020-da}%
  \BibitemOpen
  \bibfield  {author} {\bibinfo {author} {\bibfnamefont {B.~M.}\ \bibnamefont
  {de~Silva}}, \bibinfo {author} {\bibfnamefont {D.~M.}\ \bibnamefont
  {Higdon}}, \bibinfo {author} {\bibfnamefont {S.~L.}\ \bibnamefont {Brunton}},
  \ and\ \bibinfo {author} {\bibfnamefont {J.~N.}\ \bibnamefont {Kutz}},\
  }\bibfield  {title} {\enquote {\bibinfo {title} {Discovery of physics from
  data: Universal laws and discrepancies},}\ }\href {\doibase
  10.3389/frai.2020.00025} {\bibfield  {journal} {\bibinfo  {journal} {Front.
  Artif. Intell.}\ }\textbf {\bibinfo {volume} {3}},\ \bibinfo {pages} {25}
  (\bibinfo {year} {2020})}\BibitemShut {NoStop}%
\bibitem [{\citenamefont {Kaheman}, \citenamefont {Brunton},\ and\
  \citenamefont {Nathan~Kutz}(2022)}]{Kaheman2022-jt}%
  \BibitemOpen
  \bibfield  {author} {\bibinfo {author} {\bibfnamefont {K.}~\bibnamefont
  {Kaheman}}, \bibinfo {author} {\bibfnamefont {S.~L.}\ \bibnamefont
  {Brunton}}, \ and\ \bibinfo {author} {\bibfnamefont {J.}~\bibnamefont
  {Nathan~Kutz}},\ }\bibfield  {title} {\enquote {\bibinfo {title} {Automatic
  differentiation to simultaneously identify nonlinear dynamics and extract
  noise probability distributions from data},}\ }\href {\doibase
  10.1088/2632-2153/ac567a} {\bibfield  {journal} {\bibinfo  {journal} {Mach.
  Learn.: Sci. Technol.}\ }\textbf {\bibinfo {volume} {3}},\ \bibinfo {pages}
  {015031} (\bibinfo {year} {2022})}\BibitemShut {NoStop}%
\bibitem [{\citenamefont {Fasel}\ \emph {et~al.}(2022)\citenamefont {Fasel},
  \citenamefont {Kutz}, \citenamefont {Brunton},\ and\ \citenamefont
  {Brunton}}]{Fasel2022-ae}%
  \BibitemOpen
  \bibfield  {author} {\bibinfo {author} {\bibfnamefont {U.}~\bibnamefont
  {Fasel}}, \bibinfo {author} {\bibfnamefont {J.~N.}\ \bibnamefont {Kutz}},
  \bibinfo {author} {\bibfnamefont {B.~W.}\ \bibnamefont {Brunton}}, \ and\
  \bibinfo {author} {\bibfnamefont {S.~L.}\ \bibnamefont {Brunton}},\
  }\bibfield  {title} {\enquote {\bibinfo {title} {{Ensemble-SINDy}: Robust
  sparse model discovery in the low-data, high-noise limit, with active
  learning and control},}\ }\href {\doibase 10.1098/rspa.2021.0904} {\bibfield
  {journal} {\bibinfo  {journal} {Proc. Math. Phys. Eng. Sci.}\ }\textbf
  {\bibinfo {volume} {478}},\ \bibinfo {pages} {20210904} (\bibinfo {year}
  {2022})}\BibitemShut {NoStop}%
\bibitem [{\citenamefont {Schmidt}\ and\ \citenamefont
  {Lipson}(2009)}]{Schmidt2009-zv}%
  \BibitemOpen
  \bibfield  {author} {\bibinfo {author} {\bibfnamefont {M.}~\bibnamefont
  {Schmidt}}\ and\ \bibinfo {author} {\bibfnamefont {H.}~\bibnamefont
  {Lipson}},\ }\bibfield  {title} {\enquote {\bibinfo {title} {Distilling
  {Free-Form} natural laws from experimental data},}\ }\href {\doibase
  10.1126/science.1165893} {\bibfield  {journal} {\bibinfo  {journal}
  {Science}\ }\textbf {\bibinfo {volume} {324}},\ \bibinfo {pages} {81--85}
  (\bibinfo {year} {2009})}\BibitemShut {NoStop}%
\bibitem [{\citenamefont {Bongard}\ and\ \citenamefont
  {Lipson}(2007)}]{Bongard2007-oc}%
  \BibitemOpen
  \bibfield  {author} {\bibinfo {author} {\bibfnamefont {J.}~\bibnamefont
  {Bongard}}\ and\ \bibinfo {author} {\bibfnamefont {H.}~\bibnamefont
  {Lipson}},\ }\bibfield  {title} {\enquote {\bibinfo {title} {Automated
  reverse engineering of nonlinear dynamical systems},}\ }\href {\doibase
  10.1073/pnas.0609476104} {\bibfield  {journal} {\bibinfo  {journal} {Proc.
  Natl. Acad. Sci. USA}\ }\textbf {\bibinfo {volume} {104}},\ \bibinfo {pages}
  {9943--9948} (\bibinfo {year} {2007})}\BibitemShut {NoStop}%
\bibitem [{\citenamefont {Reinbold}\ \emph {et~al.}(2021)\citenamefont
  {Reinbold}, \citenamefont {Kageorge}, \citenamefont {Schatz},\ and\
  \citenamefont {Grigoriev}}]{Reinbold2021-fa}%
  \BibitemOpen
  \bibfield  {author} {\bibinfo {author} {\bibfnamefont {P.~A.~K.}\
  \bibnamefont {Reinbold}}, \bibinfo {author} {\bibfnamefont {L.~M.}\
  \bibnamefont {Kageorge}}, \bibinfo {author} {\bibfnamefont {M.~F.}\
  \bibnamefont {Schatz}}, \ and\ \bibinfo {author} {\bibfnamefont {R.~O.}\
  \bibnamefont {Grigoriev}},\ }\bibfield  {title} {\enquote {\bibinfo {title}
  {Robust learning from noisy, incomplete, high-dimensional experimental data
  via physically constrained symbolic regression},}\ }\href {\doibase
  10.1038/s41467-021-23479-0} {\bibfield  {journal} {\bibinfo  {journal} {Nat.
  Commun.}\ }\textbf {\bibinfo {volume} {12}},\ \bibinfo {pages} {3219}
  (\bibinfo {year} {2021})}\BibitemShut {NoStop}%
\bibitem [{\citenamefont {Udrescu}\ and\ \citenamefont
  {Tegmark}(2020)}]{Udrescu2020-ny}%
  \BibitemOpen
  \bibfield  {author} {\bibinfo {author} {\bibfnamefont {S.-M.}\ \bibnamefont
  {Udrescu}}\ and\ \bibinfo {author} {\bibfnamefont {M.}~\bibnamefont
  {Tegmark}},\ }\bibfield  {title} {\enquote {\bibinfo {title} {{AI} feynman: A
  physics-inspired method for symbolic regression},}\ }\href {\doibase
  10.1126/sciadv.aay2631} {\bibfield  {journal} {\bibinfo  {journal} {Sci.
  Adv.}\ }\textbf {\bibinfo {volume} {6}},\ \bibinfo {pages} {eaay2631}
  (\bibinfo {year} {2020})}\BibitemShut {NoStop}%
\bibitem [{\citenamefont {Cranmer}\ \emph {et~al.}(2020)\citenamefont
  {Cranmer}, \citenamefont {Sanchez~Gonzalez}, \citenamefont {Battaglia},
  \citenamefont {Xu}, \citenamefont {Cranmer}, \citenamefont {Spergel},\ and\
  \citenamefont {Ho}}]{Cranmer2020}%
  \BibitemOpen
  \bibfield  {author} {\bibinfo {author} {\bibfnamefont {M.}~\bibnamefont
  {Cranmer}}, \bibinfo {author} {\bibfnamefont {A.}~\bibnamefont
  {Sanchez~Gonzalez}}, \bibinfo {author} {\bibfnamefont {P.}~\bibnamefont
  {Battaglia}}, \bibinfo {author} {\bibfnamefont {R.}~\bibnamefont {Xu}},
  \bibinfo {author} {\bibfnamefont {K.}~\bibnamefont {Cranmer}}, \bibinfo
  {author} {\bibfnamefont {D.}~\bibnamefont {Spergel}}, \ and\ \bibinfo
  {author} {\bibfnamefont {S.}~\bibnamefont {Ho}},\ }\bibfield  {title}
  {\enquote {\bibinfo {title} {Discovering symbolic models from deep learning
  with inductive biases},}\ }in\ \href@noop {} {\emph {\bibinfo {booktitle}
  {NeurIPS}}},\ Vol.~\bibinfo {volume} {33},\ \bibinfo {editor} {edited by\
  \bibinfo {editor} {\bibfnamefont {H.}~\bibnamefont {Larochelle}}, \bibinfo
  {editor} {\bibfnamefont {M.}~\bibnamefont {Ranzato}}, \bibinfo {editor}
  {\bibfnamefont {R.}~\bibnamefont {Hadsell}}, \bibinfo {editor} {\bibfnamefont
  {M.}~\bibnamefont {Balcan}}, \ and\ \bibinfo {editor} {\bibfnamefont
  {H.}~\bibnamefont {Lin}}}\ (\bibinfo {year} {2020})\ pp.\ \bibinfo {pages}
  {17429--17442}\BibitemShut {NoStop}%
\bibitem [{\citenamefont {Lemos}\ \emph {et~al.}(2023)\citenamefont {Lemos},
  \citenamefont {Jeffrey}, \citenamefont {Cranmer}, \citenamefont {Ho},\ and\
  \citenamefont {Battaglia}}]{Lemos2023}%
  \BibitemOpen
  \bibfield  {author} {\bibinfo {author} {\bibfnamefont {P.}~\bibnamefont
  {Lemos}}, \bibinfo {author} {\bibfnamefont {N.}~\bibnamefont {Jeffrey}},
  \bibinfo {author} {\bibfnamefont {M.}~\bibnamefont {Cranmer}}, \bibinfo
  {author} {\bibfnamefont {S.}~\bibnamefont {Ho}}, \ and\ \bibinfo {author}
  {\bibfnamefont {P.}~\bibnamefont {Battaglia}},\ }\bibfield  {title} {\enquote
  {\bibinfo {title} {Rediscovering orbital mechanics with machine learning},}\
  }\href {\doibase 10.1088/2632-2153/acfa63} {\bibfield  {journal} {\bibinfo
  {journal} {Mach. Learn.: Sci. Technol.}\ }\textbf {\bibinfo {volume} {4}},\
  \bibinfo {pages} {045002} (\bibinfo {year} {2023})}\BibitemShut {NoStop}%
\bibitem [{\citenamefont {Karniadakis}\ \emph {et~al.}(2021)\citenamefont
  {Karniadakis}, \citenamefont {Kevrekidis}, \citenamefont {Lu}, \citenamefont
  {Perdikaris}, \citenamefont {Wang},\ and\ \citenamefont
  {Yang}}]{Karniadakis2021-rk}%
  \BibitemOpen
  \bibfield  {author} {\bibinfo {author} {\bibfnamefont {G.~E.}\ \bibnamefont
  {Karniadakis}}, \bibinfo {author} {\bibfnamefont {I.~G.}\ \bibnamefont
  {Kevrekidis}}, \bibinfo {author} {\bibfnamefont {L.}~\bibnamefont {Lu}},
  \bibinfo {author} {\bibfnamefont {P.}~\bibnamefont {Perdikaris}}, \bibinfo
  {author} {\bibfnamefont {S.}~\bibnamefont {Wang}}, \ and\ \bibinfo {author}
  {\bibfnamefont {L.}~\bibnamefont {Yang}},\ }\bibfield  {title} {\enquote
  {\bibinfo {title} {Physics-informed machine learning},}\ }\href {\doibase
  10.1038/s42254-021-00314-5} {\bibfield  {journal} {\bibinfo  {journal} {Nat.
  Rev. Phys.}\ }\textbf {\bibinfo {volume} {3}},\ \bibinfo {pages} {422--440}
  (\bibinfo {year} {2021})}\BibitemShut {NoStop}%
\bibitem [{\citenamefont {Raissi}, \citenamefont {Perdikaris},\ and\
  \citenamefont {Karniadakis}(2019)}]{Raissi2019-bu}%
  \BibitemOpen
  \bibfield  {author} {\bibinfo {author} {\bibfnamefont {M.}~\bibnamefont
  {Raissi}}, \bibinfo {author} {\bibfnamefont {P.}~\bibnamefont {Perdikaris}},
  \ and\ \bibinfo {author} {\bibfnamefont {G.~E.}\ \bibnamefont
  {Karniadakis}},\ }\bibfield  {title} {\enquote {\bibinfo {title}
  {Physics-informed neural networks: A deep learning framework for solving
  forward and inverse problems involving nonlinear partial differential
  equations},}\ }\href {\doibase 10.1016/j.jcp.2018.10.045} {\bibfield
  {journal} {\bibinfo  {journal} {Journal of computational physics}\ }\textbf
  {\bibinfo {volume} {378}},\ \bibinfo {pages} {686--707} (\bibinfo {year}
  {2019})}\BibitemShut {NoStop}%
\bibitem [{\citenamefont {Jia}\ \emph {et~al.}(2021)\citenamefont {Jia},
  \citenamefont {Willard}, \citenamefont {Karpatne}, \citenamefont {Read},
  \citenamefont {Zwart}, \citenamefont {Steinbach},\ and\ \citenamefont
  {Kumar}}]{Jia2021-mn}%
  \BibitemOpen
  \bibfield  {author} {\bibinfo {author} {\bibfnamefont {X.}~\bibnamefont
  {Jia}}, \bibinfo {author} {\bibfnamefont {J.}~\bibnamefont {Willard}},
  \bibinfo {author} {\bibfnamefont {A.}~\bibnamefont {Karpatne}}, \bibinfo
  {author} {\bibfnamefont {J.~S.}\ \bibnamefont {Read}}, \bibinfo {author}
  {\bibfnamefont {J.~A.}\ \bibnamefont {Zwart}}, \bibinfo {author}
  {\bibfnamefont {M.}~\bibnamefont {Steinbach}}, \ and\ \bibinfo {author}
  {\bibfnamefont {V.}~\bibnamefont {Kumar}},\ }\bibfield  {title} {\enquote
  {\bibinfo {title} {{Physics-Guided} machine learning for scientific
  discovery: An application in simulating lake temperature profiles},}\ }\href
  {\doibase 10.1145/3447814} {\bibfield  {journal} {\bibinfo  {journal}
  {ACM/IMS Trans. Data Sci.}\ }\textbf {\bibinfo {volume} {2}},\ \bibinfo
  {pages} {1--26} (\bibinfo {year} {2021})}\BibitemShut {NoStop}%
\bibitem [{\citenamefont {Rosofsky}, \citenamefont {Al~Majed},\ and\
  \citenamefont {Huerta}(2023)}]{Rosofsky2023-ae}%
  \BibitemOpen
  \bibfield  {author} {\bibinfo {author} {\bibfnamefont {S.~G.}\ \bibnamefont
  {Rosofsky}}, \bibinfo {author} {\bibfnamefont {H.}~\bibnamefont {Al~Majed}},
  \ and\ \bibinfo {author} {\bibfnamefont {E.~A.}\ \bibnamefont {Huerta}},\
  }\bibfield  {title} {\enquote {\bibinfo {title} {Applications of physics
  informed neural operators},}\ }\href {\doibase 10.1088/2632-2153/acd168}
  {\bibfield  {journal} {\bibinfo  {journal} {Machine Learning: Science and
  Technology}\ }\textbf {\bibinfo {volume} {4}},\ \bibinfo {pages} {025022}
  (\bibinfo {year} {2023})}\BibitemShut {NoStop}%
\bibitem [{\citenamefont {Larson}(1988)}]{Larson1988}%
  \BibitemOpen
  \bibfield  {author} {\bibinfo {author} {\bibfnamefont {R.~G.}\ \bibnamefont
  {Larson}},\ }\href@noop {} {\emph {\bibinfo {title} {Constitutive Equations
  for Polymer Melts and Solutions}}}\ (\bibinfo  {publisher} {Butterworths
  Series in Chemical Engineering},\ \bibinfo {year} {1988})\BibitemShut
  {NoStop}%
\bibitem [{\citenamefont {Ilg}\ and\ \citenamefont
  {Kr{\"o}ger}(2011)}]{Ilg2011}%
  \BibitemOpen
  \bibfield  {author} {\bibinfo {author} {\bibfnamefont {P.}~\bibnamefont
  {Ilg}}\ and\ \bibinfo {author} {\bibfnamefont {M.}~\bibnamefont
  {Kr{\"o}ger}},\ }\bibfield  {title} {\enquote {\bibinfo {title} {{Molecularly
  derived constitutive equation for low-molecular polymer melts from
  thermodynamically guided simulation}},}\ }\href@noop {} {\bibfield  {journal}
  {\bibinfo  {journal} {J. Rheol.}\ }\textbf {\bibinfo {volume} {55}},\
  \bibinfo {pages} {69--93} (\bibinfo {year} {2011})}\BibitemShut {NoStop}%
\bibitem [{\citenamefont {Bird}, \citenamefont {Armstrong},\ and\ \citenamefont
  {Hassager}(1987)}]{Bird1987}%
  \BibitemOpen
  \bibfield  {author} {\bibinfo {author} {\bibfnamefont {R.~B.}\ \bibnamefont
  {Bird}}, \bibinfo {author} {\bibfnamefont {R.~C.}\ \bibnamefont {Armstrong}},
  \ and\ \bibinfo {author} {\bibfnamefont {O.}~\bibnamefont {Hassager}},\
  }\href@noop {} {\emph {\bibinfo {title} {Dynamics of Polymeric Liquids}}},\
  \bibinfo {edition} {2nd}\ ed.,\ Vol.~\bibinfo {volume} {2}\ (\bibinfo
  {publisher} {Oxford University Press},\ \bibinfo {year} {1987})\BibitemShut
  {NoStop}%
\bibitem [{\citenamefont {Rouse}(1953)}]{Rouse1953}%
  \BibitemOpen
  \bibfield  {author} {\bibinfo {author} {\bibfnamefont {P.~E.}\ \bibnamefont
  {Rouse}},\ }\bibfield  {title} {\enquote {\bibinfo {title} {{A Theory of the
  Linear Viscoelastic Properties of Dilute Solutions of Coiling Polymers}},}\
  }\href@noop {} {\bibfield  {journal} {\bibinfo  {journal} {J. Chem. Phys.}\
  }\textbf {\bibinfo {volume} {21}},\ \bibinfo {pages} {1272--1280} (\bibinfo
  {year} {1953})}\BibitemShut {NoStop}%
\bibitem [{\citenamefont {Doi}\ and\ \citenamefont {Edwards}(1986)}]{Doi1986}%
  \BibitemOpen
  \bibfield  {author} {\bibinfo {author} {\bibfnamefont {M.}~\bibnamefont
  {Doi}}\ and\ \bibinfo {author} {\bibfnamefont {S.~F.}\ \bibnamefont
  {Edwards}},\ }\href@noop {} {\emph {\bibinfo {title} {The Theory of Polymer
  Dynamics}}}\ (\bibinfo  {publisher} {Oxford University Press},\ \bibinfo
  {year} {1986})\BibitemShut {NoStop}%
\bibitem [{\citenamefont {Masubuchi}\ \emph {et~al.}(2001)\citenamefont
  {Masubuchi}, \citenamefont {Takimoto}, \citenamefont {Koyama}, \citenamefont
  {Ianniruberto}, \citenamefont {Marrucci},\ and\ \citenamefont
  {Greco}}]{Masubuchi2001}%
  \BibitemOpen
  \bibfield  {author} {\bibinfo {author} {\bibfnamefont {Y.}~\bibnamefont
  {Masubuchi}}, \bibinfo {author} {\bibfnamefont {J.-I.}\ \bibnamefont
  {Takimoto}}, \bibinfo {author} {\bibfnamefont {K.}~\bibnamefont {Koyama}},
  \bibinfo {author} {\bibfnamefont {G.}~\bibnamefont {Ianniruberto}}, \bibinfo
  {author} {\bibfnamefont {G.}~\bibnamefont {Marrucci}}, \ and\ \bibinfo
  {author} {\bibfnamefont {F.}~\bibnamefont {Greco}},\ }\bibfield  {title}
  {\enquote {\bibinfo {title} {{Brownian simulations of a network of reptating
  primitive chains}},}\ }\href@noop {} {\bibfield  {journal} {\bibinfo
  {journal} {J. Chem. Phys.}\ }\textbf {\bibinfo {volume} {115}},\ \bibinfo
  {pages} {4387--4394} (\bibinfo {year} {2001})}\BibitemShut {NoStop}%
\bibitem [{\citenamefont {Doi}\ and\ \citenamefont {Takimoto}(2003)}]{Doi2003}%
  \BibitemOpen
  \bibfield  {author} {\bibinfo {author} {\bibfnamefont {M.}~\bibnamefont
  {Doi}}\ and\ \bibinfo {author} {\bibfnamefont {J.}~\bibnamefont {Takimoto}},\
  }\bibfield  {title} {\enquote {\bibinfo {title} {Molecular modelling of
  entanglement},}\ }\href@noop {} {\bibfield  {journal} {\bibinfo  {journal}
  {Phil. Trans. R. Soc. A.}\ }\textbf {\bibinfo {volume} {361}},\ \bibinfo
  {pages} {641--652} (\bibinfo {year} {2003})}\BibitemShut {NoStop}%
\bibitem [{\citenamefont {Likhtman}(2005)}]{Likhtman2005}%
  \BibitemOpen
  \bibfield  {author} {\bibinfo {author} {\bibfnamefont {A.~E.}\ \bibnamefont
  {Likhtman}},\ }\bibfield  {title} {\enquote {\bibinfo {title} {Single-chain
  slip-link model of entangled polymers: Simultaneous description of neutron
  spin-echo, rheology, and diffusion},}\ }\href@noop {} {\bibfield  {journal}
  {\bibinfo  {journal} {Macromolecules}\ }\textbf {\bibinfo {volume} {38}},\
  \bibinfo {pages} {6128--6139} (\bibinfo {year} {2005})}\BibitemShut {NoStop}%
\bibitem [{\citenamefont {Jamali}(2023)}]{Jamali2023-review}%
  \BibitemOpen
  \bibfield  {author} {\bibinfo {author} {\bibfnamefont {S.}~\bibnamefont
  {Jamali}},\ }\bibfield  {title} {\enquote {\bibinfo {title} {Data-driven
  rheology: could be a new paradigm?}}\ }\href@noop {} {\bibfield  {journal}
  {\bibinfo  {journal} {Rheology Bulletin}\ }\textbf {\bibinfo {volume} {92}},\
  \bibinfo {pages} {20--24} (\bibinfo {year} {2023})}\BibitemShut {NoStop}%
\bibitem [{\citenamefont {Miyamoto}(2024)}]{Miyamoto2024-review}%
  \BibitemOpen
  \bibfield  {author} {\bibinfo {author} {\bibfnamefont {S.}~\bibnamefont
  {Miyamoto}},\ }\bibfield  {title} {\enquote {\bibinfo {title} {Short review
  on machine learning-based multi-scale simulation in rheology},}\ }\href@noop
  {} {\bibfield  {journal} {\bibinfo  {journal} {Nihon Reoroji Gakkaishi (J.
  Soc. Rheol. Jpn.)}\ }\textbf {\bibinfo {volume} {52}},\ \bibinfo {pages}
  {15--19} (\bibinfo {year} {2024})}\BibitemShut {NoStop}%
\bibitem [{\citenamefont {Fang}\ \emph {et~al.}(2022)\citenamefont {Fang},
  \citenamefont {Ge}, \citenamefont {Zhang}, \citenamefont {E},\ and\
  \citenamefont {Lei}}]{Fang2022-DNN}%
  \BibitemOpen
  \bibfield  {author} {\bibinfo {author} {\bibfnamefont {L.}~\bibnamefont
  {Fang}}, \bibinfo {author} {\bibfnamefont {P.}~\bibnamefont {Ge}}, \bibinfo
  {author} {\bibfnamefont {L.}~\bibnamefont {Zhang}}, \bibinfo {author}
  {\bibfnamefont {W.}~\bibnamefont {E}}, \ and\ \bibinfo {author}
  {\bibfnamefont {H.}~\bibnamefont {Lei}},\ }\bibfield  {title} {\enquote
  {\bibinfo {title} {{DeePN$^2$:} a deep learning-based non-newtonian
  hydrodynamic model},}\ }\href {\doibase 10.4208/jml.220115} {\bibfield
  {journal} {\bibinfo  {journal} {J. Mach. Learn.}\ }\textbf {\bibinfo {volume}
  {1}},\ \bibinfo {pages} {114--140} (\bibinfo {year} {2022})}\BibitemShut
  {NoStop}%
\bibitem [{\citenamefont {Mahmoudabadbozchelou}\ \emph
  {et~al.}(2022)\citenamefont {Mahmoudabadbozchelou}, \citenamefont {Kamani},
  \citenamefont {Rogers},\ and\ \citenamefont {Jamali}}]{Mohammadamin2022-GNN}%
  \BibitemOpen
  \bibfield  {author} {\bibinfo {author} {\bibfnamefont {M.}~\bibnamefont
  {Mahmoudabadbozchelou}}, \bibinfo {author} {\bibfnamefont {K.~M.}\
  \bibnamefont {Kamani}}, \bibinfo {author} {\bibfnamefont {S.~A.}\
  \bibnamefont {Rogers}}, \ and\ \bibinfo {author} {\bibfnamefont
  {S.}~\bibnamefont {Jamali}},\ }\bibfield  {title} {\enquote {\bibinfo {title}
  {Digital rheometer twins: Learning the hidden rheology of complex fluids
  through rheology-informed graph neural networks},}\ }\href {\doibase
  10.1073/pnas.2202234119} {\bibfield  {journal} {\bibinfo  {journal} {Proc.
  Natl. Acad. Sci. USA}\ }\textbf {\bibinfo {volume} {119}},\ \bibinfo {pages}
  {e2202234119} (\bibinfo {year} {2022})}\BibitemShut {NoStop}%
\bibitem [{\citenamefont {Jin}\ \emph {et~al.}(2023)\citenamefont {Jin},
  \citenamefont {Yoon}, \citenamefont {Park},\ and\ \citenamefont
  {Ahn}}]{Jin2023-RNN}%
  \BibitemOpen
  \bibfield  {author} {\bibinfo {author} {\bibfnamefont {H.}~\bibnamefont
  {Jin}}, \bibinfo {author} {\bibfnamefont {S.}~\bibnamefont {Yoon}}, \bibinfo
  {author} {\bibfnamefont {F.~C.}\ \bibnamefont {Park}}, \ and\ \bibinfo
  {author} {\bibfnamefont {K.~H.}\ \bibnamefont {Ahn}},\ }\bibfield  {title}
  {\enquote {\bibinfo {title} {Data-driven constitutive model of complex fluids
  using recurrent neural networks},}\ }\href {\doibase
  10.1007/s00397-023-01405-z} {\bibfield  {journal} {\bibinfo  {journal}
  {Rheol. Acta}\ }\textbf {\bibinfo {volume} {62}},\ \bibinfo {pages}
  {569--586} (\bibinfo {year} {2023})}\BibitemShut {NoStop}%
\bibitem [{\citenamefont {Mahmoudabadbozchelou}\ and\ \citenamefont
  {Jamali}(2021)}]{Mahmoudabad2021-RhINN}%
  \BibitemOpen
  \bibfield  {author} {\bibinfo {author} {\bibfnamefont {M.}~\bibnamefont
  {Mahmoudabadbozchelou}}\ and\ \bibinfo {author} {\bibfnamefont
  {S.}~\bibnamefont {Jamali}},\ }\bibfield  {title} {\enquote {\bibinfo {title}
  {Rheology-informed neural networks {(RhINNs)} for forward and inverse
  metamodelling of complex fluids},}\ }\href {\doibase
  10.1038/s41598-021-91518-3} {\bibfield  {journal} {\bibinfo  {journal} {Sci.
  Rep.}\ }\textbf {\bibinfo {volume} {11}},\ \bibinfo {pages} {12015} (\bibinfo
  {year} {2021})}\BibitemShut {NoStop}%
\bibitem [{\citenamefont {Mahmoudabadbozchelou}, \citenamefont {Karniadakis},\
  and\ \citenamefont {Jamali}(2022)}]{Jamali2021-PINN}%
  \BibitemOpen
  \bibfield  {author} {\bibinfo {author} {\bibfnamefont {M.}~\bibnamefont
  {Mahmoudabadbozchelou}}, \bibinfo {author} {\bibfnamefont {G.~E.}\
  \bibnamefont {Karniadakis}}, \ and\ \bibinfo {author} {\bibfnamefont
  {S.}~\bibnamefont {Jamali}},\ }\bibfield  {title} {\enquote {\bibinfo {title}
  {{nn-PINNs:} non-newtonian physics-informed neural networks for complex fluid
  modeling},}\ }\href {\doibase 10.1039/D1SM01298C} {\bibfield  {journal}
  {\bibinfo  {journal} {Soft Matter}\ }\textbf {\bibinfo {volume} {18}},\
  \bibinfo {pages} {172--185} (\bibinfo {year} {2022})}\BibitemShut {NoStop}%
\bibitem [{\citenamefont {Saadat}, \citenamefont {Mahmoudabadbozchelou},\ and\
  \citenamefont {Jamali}(2022)}]{Sadat2022-RhINN}%
  \BibitemOpen
  \bibfield  {author} {\bibinfo {author} {\bibfnamefont {M.}~\bibnamefont
  {Saadat}}, \bibinfo {author} {\bibfnamefont {M.}~\bibnamefont
  {Mahmoudabadbozchelou}}, \ and\ \bibinfo {author} {\bibfnamefont
  {S.}~\bibnamefont {Jamali}},\ }\bibfield  {title} {\enquote {\bibinfo {title}
  {Data-driven selection of constitutive models via rheology-informed neural
  networks {(RhINNs)}},}\ }\href {\doibase 10.1007/s00397-022-01357-w}
  {\bibfield  {journal} {\bibinfo  {journal} {Rheol. Acta}\ }\textbf {\bibinfo
  {volume} {61}},\ \bibinfo {pages} {721--732} (\bibinfo {year}
  {2022})}\BibitemShut {NoStop}%
\bibitem [{\citenamefont {Mahmoudabadbozchelou}\ \emph
  {et~al.}(2021)\citenamefont {Mahmoudabadbozchelou}, \citenamefont {Caggioni},
  \citenamefont {Shahsavari}, \citenamefont {Hartt}, \citenamefont
  {Em~Karniadakis},\ and\ \citenamefont {Jamali}}]{Mohammadamin2021-MFNN}%
  \BibitemOpen
  \bibfield  {author} {\bibinfo {author} {\bibfnamefont {M.}~\bibnamefont
  {Mahmoudabadbozchelou}}, \bibinfo {author} {\bibfnamefont {M.}~\bibnamefont
  {Caggioni}}, \bibinfo {author} {\bibfnamefont {S.}~\bibnamefont
  {Shahsavari}}, \bibinfo {author} {\bibfnamefont {W.~H.}\ \bibnamefont
  {Hartt}}, \bibinfo {author} {\bibfnamefont {G.}~\bibnamefont
  {Em~Karniadakis}}, \ and\ \bibinfo {author} {\bibfnamefont {S.}~\bibnamefont
  {Jamali}},\ }\bibfield  {title} {\enquote {\bibinfo {title} {{Data-driven
  physics-informed constitutive metamodeling of complex fluids: A multifidelity
  neural network {(MFNN)} framework}},}\ }\href {\doibase 10.1122/8.0000138}
  {\bibfield  {journal} {\bibinfo  {journal} {J. Rheol.}\ }\textbf {\bibinfo
  {volume} {65}},\ \bibinfo {pages} {179--198} (\bibinfo {year}
  {2021})}\BibitemShut {NoStop}%
\bibitem [{\citenamefont {Lennon}, \citenamefont {McKinley},\ and\
  \citenamefont {Swan}(2023)}]{Lennon2023-RUDE}%
  \BibitemOpen
  \bibfield  {author} {\bibinfo {author} {\bibfnamefont {K.~R.}\ \bibnamefont
  {Lennon}}, \bibinfo {author} {\bibfnamefont {G.~H.}\ \bibnamefont
  {McKinley}}, \ and\ \bibinfo {author} {\bibfnamefont {J.~W.}\ \bibnamefont
  {Swan}},\ }\bibfield  {title} {\enquote {\bibinfo {title} {Scientific machine
  learning for modeling and simulating complex fluids},}\ }\href {\doibase
  10.1073/pnas.2304669120} {\bibfield  {journal} {\bibinfo  {journal} {Proc.
  Natl. Acad. Sci. USA}\ }\textbf {\bibinfo {volume} {120}},\ \bibinfo {pages}
  {e2304669120} (\bibinfo {year} {2023})}\BibitemShut {NoStop}%
\bibitem [{\citenamefont {Zhao}\ \emph {et~al.}(2018)\citenamefont {Zhao},
  \citenamefont {Li}, \citenamefont {Caswell}, \citenamefont {Ouyang},\ and\
  \citenamefont {Karniadakis}}]{Zhao2018}%
  \BibitemOpen
  \bibfield  {author} {\bibinfo {author} {\bibfnamefont {L.}~\bibnamefont
  {Zhao}}, \bibinfo {author} {\bibfnamefont {Z.}~\bibnamefont {Li}}, \bibinfo
  {author} {\bibfnamefont {B.}~\bibnamefont {Caswell}}, \bibinfo {author}
  {\bibfnamefont {J.}~\bibnamefont {Ouyang}}, \ and\ \bibinfo {author}
  {\bibfnamefont {G.~E.}\ \bibnamefont {Karniadakis}},\ }\bibfield  {title}
  {\enquote {\bibinfo {title} {Active learning of constitutive relation from
  mesoscopic dynamics for macroscopic modeling of non-newtonian flows},}\
  }\href {\doibase 10.1016/j.jcp.2018.02.039} {\bibfield  {journal} {\bibinfo
  {journal} {J. Comput. Phys.}\ }\textbf {\bibinfo {volume} {363}},\ \bibinfo
  {pages} {116--127} (\bibinfo {year} {2018})}\BibitemShut {NoStop}%
\bibitem [{\citenamefont {Zhao}\ \emph {et~al.}(2021)\citenamefont {Zhao},
  \citenamefont {Li}, \citenamefont {Wang}, \citenamefont {Caswell},
  \citenamefont {Ouyang},\ and\ \citenamefont {Karniadakis}}]{Zhao2021}%
  \BibitemOpen
  \bibfield  {author} {\bibinfo {author} {\bibfnamefont {L.}~\bibnamefont
  {Zhao}}, \bibinfo {author} {\bibfnamefont {Z.}~\bibnamefont {Li}}, \bibinfo
  {author} {\bibfnamefont {Z.}~\bibnamefont {Wang}}, \bibinfo {author}
  {\bibfnamefont {B.}~\bibnamefont {Caswell}}, \bibinfo {author} {\bibfnamefont
  {J.}~\bibnamefont {Ouyang}}, \ and\ \bibinfo {author} {\bibfnamefont {G.~E.}\
  \bibnamefont {Karniadakis}},\ }\bibfield  {title} {\enquote {\bibinfo {title}
  {Active- and transfer-learning applied to microscale-macroscale coupling to
  simulate viscoelastic flows},}\ }\href {\doibase 10.1016/j.jcp.2020.110069}
  {\bibfield  {journal} {\bibinfo  {journal} {J. Comput. Phys.}\ }\textbf
  {\bibinfo {volume} {427}},\ \bibinfo {pages} {110069} (\bibinfo {year}
  {2021})}\BibitemShut {NoStop}%
\bibitem [{\citenamefont {Seryo}\ \emph {et~al.}(2020)\citenamefont {Seryo},
  \citenamefont {Sato}, \citenamefont {Molina},\ and\ \citenamefont
  {Taniguchi}}]{Seryo2020-hp}%
  \BibitemOpen
  \bibfield  {author} {\bibinfo {author} {\bibfnamefont {N.}~\bibnamefont
  {Seryo}}, \bibinfo {author} {\bibfnamefont {T.}~\bibnamefont {Sato}},
  \bibinfo {author} {\bibfnamefont {J.~J.}\ \bibnamefont {Molina}}, \ and\
  \bibinfo {author} {\bibfnamefont {T.}~\bibnamefont {Taniguchi}},\ }\bibfield
  {title} {\enquote {\bibinfo {title} {Learning the constitutive relation of
  polymeric flows with memory},}\ }\href {\doibase
  10.1103/PhysRevResearch.2.033107} {\bibfield  {journal} {\bibinfo  {journal}
  {Phys. Rev. Res.}\ }\textbf {\bibinfo {volume} {2}},\ \bibinfo {pages}
  {033107} (\bibinfo {year} {2020})}\BibitemShut {NoStop}%
\bibitem [{\citenamefont {Seryo}, \citenamefont {Molina},\ and\ \citenamefont
  {Taniguchi}(2021)}]{Seryo2021-ug}%
  \BibitemOpen
  \bibfield  {author} {\bibinfo {author} {\bibfnamefont {N.}~\bibnamefont
  {Seryo}}, \bibinfo {author} {\bibfnamefont {J.~J.}\ \bibnamefont {Molina}}, \
  and\ \bibinfo {author} {\bibfnamefont {T.}~\bibnamefont {Taniguchi}},\
  }\bibfield  {title} {\enquote {\bibinfo {title} {Select applications of
  bayesian data analysis and machine learning to flow problems},}\ }\href
  {\doibase 10.1678/rheology.49.97} {\bibfield  {journal} {\bibinfo  {journal}
  {Nihon Reoroji Gakkaishi (J. Soc. Rheol. Jpn.)}\ }\textbf {\bibinfo {volume}
  {49}},\ \bibinfo {pages} {97--113} (\bibinfo {year} {2021})}\BibitemShut
  {NoStop}%
\bibitem [{\citenamefont {Miyamoto}, \citenamefont {Molina},\ and\
  \citenamefont {Taniguchi}(2023)}]{Miyamoto2023-rq}%
  \BibitemOpen
  \bibfield  {author} {\bibinfo {author} {\bibfnamefont {S.}~\bibnamefont
  {Miyamoto}}, \bibinfo {author} {\bibfnamefont {J.~J.}\ \bibnamefont
  {Molina}}, \ and\ \bibinfo {author} {\bibfnamefont {T.}~\bibnamefont
  {Taniguchi}},\ }\bibfield  {title} {\enquote {\bibinfo {title}
  {Machine-learned constitutive relations for multi-scale simulations of
  well-entangled polymer melts},}\ }\href {\doibase 10.1063/5.0156272}
  {\bibfield  {journal} {\bibinfo  {journal} {Phys. Fluids}\ }\textbf {\bibinfo
  {volume} {35}},\ \bibinfo {pages} {063113} (\bibinfo {year}
  {2023})}\BibitemShut {NoStop}%
\bibitem [{\citenamefont {Fukami}\ \emph {et~al.}(2021)\citenamefont {Fukami},
  \citenamefont {Murata}, \citenamefont {Zhang},\ and\ \citenamefont
  {Fukagata}}]{Fukami2021-zx}%
  \BibitemOpen
  \bibfield  {author} {\bibinfo {author} {\bibfnamefont {K.}~\bibnamefont
  {Fukami}}, \bibinfo {author} {\bibfnamefont {T.}~\bibnamefont {Murata}},
  \bibinfo {author} {\bibfnamefont {K.}~\bibnamefont {Zhang}}, \ and\ \bibinfo
  {author} {\bibfnamefont {K.}~\bibnamefont {Fukagata}},\ }\bibfield  {title}
  {\enquote {\bibinfo {title} {Sparse identification of nonlinear dynamics with
  low-dimensionalized flow representations},}\ }\href {\doibase
  10.1017/jfm.2021.697} {\bibfield  {journal} {\bibinfo  {journal} {J. Fluid
  Mech.}\ }\textbf {\bibinfo {volume} {926}},\ \bibinfo {pages} {A10} (\bibinfo
  {year} {2021})}\BibitemShut {NoStop}%
\bibitem [{\citenamefont {Mahmoudabadbozchelou}\ \emph
  {et~al.}(2024)\citenamefont {Mahmoudabadbozchelou}, \citenamefont {Kamani},
  \citenamefont {Rogers},\ and\ \citenamefont
  {Jamali}}]{Mohammadamin2024-SINDy}%
  \BibitemOpen
  \bibfield  {author} {\bibinfo {author} {\bibfnamefont {M.}~\bibnamefont
  {Mahmoudabadbozchelou}}, \bibinfo {author} {\bibfnamefont {K.~M.}\
  \bibnamefont {Kamani}}, \bibinfo {author} {\bibfnamefont {S.~A.}\
  \bibnamefont {Rogers}}, \ and\ \bibinfo {author} {\bibfnamefont
  {S.}~\bibnamefont {Jamali}},\ }\bibfield  {title} {\enquote {\bibinfo {title}
  {Unbiased construction of constitutive relations for soft materials from
  experiments via rheology-informed neural networks},}\ }\href {\doibase
  10.1073/pnas.2313658121} {\bibfield  {journal} {\bibinfo  {journal} {Proc.
  Natl. Acad. Sci. USA}\ }\textbf {\bibinfo {volume} {121}},\ \bibinfo {pages}
  {e2313658121} (\bibinfo {year} {2024})}\BibitemShut {NoStop}%
\bibitem [{\citenamefont {Chartrand}(2011)}]{Chartrand2011}%
  \BibitemOpen
  \bibfield  {author} {\bibinfo {author} {\bibfnamefont {R.}~\bibnamefont
  {Chartrand}},\ }\bibfield  {title} {\enquote {\bibinfo {title} {Numerical
  differentiation of noisy, nonsmooth data},}\ }\href@noop {} {\bibfield
  {journal} {\bibinfo  {journal} {ISRN Applied Mathematics}\ }\textbf {\bibinfo
  {volume} {2011}},\ \bibinfo {pages} {164564} (\bibinfo {year}
  {2011})}\BibitemShut {NoStop}%
\bibitem [{\citenamefont {Rudy}\ \emph {et~al.}(2017)\citenamefont {Rudy},
  \citenamefont {Brunton}, \citenamefont {Proctor},\ and\ \citenamefont
  {Kutz}}]{Rudy2017}%
  \BibitemOpen
  \bibfield  {author} {\bibinfo {author} {\bibfnamefont {S.~H.}\ \bibnamefont
  {Rudy}}, \bibinfo {author} {\bibfnamefont {S.~L.}\ \bibnamefont {Brunton}},
  \bibinfo {author} {\bibfnamefont {J.~L.}\ \bibnamefont {Proctor}}, \ and\
  \bibinfo {author} {\bibfnamefont {J.~N.}\ \bibnamefont {Kutz}},\ }\bibfield
  {title} {\enquote {\bibinfo {title} {Data-driven discovery of partial
  differential equations},}\ }\href@noop {} {\bibfield  {journal} {\bibinfo
  {journal} {Sci. Adv.}\ }\textbf {\bibinfo {volume} {3}},\ \bibinfo {pages}
  {e1602614} (\bibinfo {year} {2017})}\BibitemShut {NoStop}%
\bibitem [{\citenamefont {Pedregosa}\ \emph {et~al.}(2011)\citenamefont
  {Pedregosa}, \citenamefont {Varoquaux}, \citenamefont {Gramfort},
  \citenamefont {Michel}, \citenamefont {Thirion}, \citenamefont {Grisel},
  \citenamefont {Blondel}, \citenamefont {Prettenhofer}, \citenamefont {Weiss},
  \citenamefont {Dubourg}, \citenamefont {Vanderplas}, \citenamefont {Passos},
  \citenamefont {Cournapeau}, \citenamefont {Brucher}, \citenamefont {Perrot},\
  and\ \citenamefont {Duchesnay}}]{Pedregosa2011}%
  \BibitemOpen
  \bibfield  {author} {\bibinfo {author} {\bibfnamefont {F.}~\bibnamefont
  {Pedregosa}}, \bibinfo {author} {\bibfnamefont {G.}~\bibnamefont
  {Varoquaux}}, \bibinfo {author} {\bibfnamefont {A.}~\bibnamefont {Gramfort}},
  \bibinfo {author} {\bibfnamefont {V.}~\bibnamefont {Michel}}, \bibinfo
  {author} {\bibfnamefont {B.}~\bibnamefont {Thirion}}, \bibinfo {author}
  {\bibfnamefont {O.}~\bibnamefont {Grisel}}, \bibinfo {author} {\bibfnamefont
  {M.}~\bibnamefont {Blondel}}, \bibinfo {author} {\bibfnamefont
  {P.}~\bibnamefont {Prettenhofer}}, \bibinfo {author} {\bibfnamefont
  {R.}~\bibnamefont {Weiss}}, \bibinfo {author} {\bibfnamefont
  {V.}~\bibnamefont {Dubourg}}, \bibinfo {author} {\bibfnamefont
  {J.}~\bibnamefont {Vanderplas}}, \bibinfo {author} {\bibfnamefont
  {A.}~\bibnamefont {Passos}}, \bibinfo {author} {\bibfnamefont
  {D.}~\bibnamefont {Cournapeau}}, \bibinfo {author} {\bibfnamefont
  {M.}~\bibnamefont {Brucher}}, \bibinfo {author} {\bibfnamefont
  {M.}~\bibnamefont {Perrot}}, \ and\ \bibinfo {author} {\bibfnamefont
  {E.}~\bibnamefont {Duchesnay}},\ }\bibfield  {title} {\enquote {\bibinfo
  {title} {Scikit-learn: Machine learning in {P}ython},}\ }\href@noop {}
  {\bibfield  {journal} {\bibinfo  {journal} {J. Mach. Learn. Res.}\ }\textbf
  {\bibinfo {volume} {12}},\ \bibinfo {pages} {2825--2830} (\bibinfo {year}
  {2011})}\BibitemShut {NoStop}%
\bibitem [{\citenamefont {Zou}(2006)}]{Zou2006}%
  \BibitemOpen
  \bibfield  {author} {\bibinfo {author} {\bibfnamefont {H.}~\bibnamefont
  {Zou}},\ }\bibfield  {title} {\enquote {\bibinfo {title} {The adaptive lasso
  and its oracle properties},}\ }\href {\doibase 10.1198/016214506000000735}
  {\bibfield  {journal} {\bibinfo  {journal} {J. Am. Stat. Assoc.}\ }\textbf
  {\bibinfo {volume} {101}},\ \bibinfo {pages} {1418--1429} (\bibinfo {year}
  {2006})}\BibitemShut {NoStop}%
\bibitem [{\citenamefont {Giesekus}(1982)}]{Giesekus1982}%
  \BibitemOpen
  \bibfield  {author} {\bibinfo {author} {\bibfnamefont {H.}~\bibnamefont
  {Giesekus}},\ }\bibfield  {title} {\enquote {\bibinfo {title} {A simple
  constitutive equation for polymer fluids based on the concept of
  deformation-dependent tensorial mobility},}\ }\href@noop {} {\bibfield
  {journal} {\bibinfo  {journal} {J. Non-Newtonian Fluid Mech.}\ }\textbf
  {\bibinfo {volume} {11}},\ \bibinfo {pages} {69--109} (\bibinfo {year}
  {1982})}\BibitemShut {NoStop}%
\bibitem [{\citenamefont {Laso}\ and\ \citenamefont
  {{\"O}ttinger}(1993)}]{Laso1993}%
  \BibitemOpen
  \bibfield  {author} {\bibinfo {author} {\bibfnamefont {M.}~\bibnamefont
  {Laso}}\ and\ \bibinfo {author} {\bibfnamefont {H.}~\bibnamefont
  {{\"O}ttinger}},\ }\bibfield  {title} {\enquote {\bibinfo {title}
  {Calculation of viscoelastic flow using molecular models: the connffessit
  approach},}\ }\href@noop {} {\bibfield  {journal} {\bibinfo  {journal} {J.
  Non-Newtonian Fluid Mech.}\ }\textbf {\bibinfo {volume} {47}},\ \bibinfo
  {pages} {1--20} (\bibinfo {year} {1993})}\BibitemShut {NoStop}%
\bibitem [{\citenamefont {Graham}(2014)}]{Graham2014}%
  \BibitemOpen
  \bibfield  {author} {\bibinfo {author} {\bibfnamefont {M.~D.}\ \bibnamefont
  {Graham}},\ }\bibfield  {title} {\enquote {\bibinfo {title} {Drag reduction
  and the dynamics of turbulence in simple and complex fluids},}\ }\href@noop
  {} {\bibfield  {journal} {\bibinfo  {journal} {Phys. Fluids}\ }\textbf
  {\bibinfo {volume} {26}},\ \bibinfo {pages} {101301} (\bibinfo {year}
  {2014})}\BibitemShut {NoStop}%
\bibitem [{\citenamefont {Mochimaru}(1983)}]{Mochimaru1983}%
  \BibitemOpen
  \bibfield  {author} {\bibinfo {author} {\bibfnamefont {Y.}~\bibnamefont
  {Mochimaru}},\ }\bibfield  {title} {\enquote {\bibinfo {title}
  {Unsteady-state development of plane couette flow for viscoelastic fluids},}\
  }\href@noop {} {\bibfield  {journal} {\bibinfo  {journal} {J. Non-Newtonian
  Fluid Mech.}\ }\textbf {\bibinfo {volume} {12}},\ \bibinfo {pages} {135--152}
  (\bibinfo {year} {1983})}\BibitemShut {NoStop}%
\bibitem [{\citenamefont {Du}, \citenamefont {Liu},\ and\ \citenamefont
  {Yu}(2005)}]{Du2005}%
  \BibitemOpen
  \bibfield  {author} {\bibinfo {author} {\bibfnamefont {Q.}~\bibnamefont
  {Du}}, \bibinfo {author} {\bibfnamefont {C.}~\bibnamefont {Liu}}, \ and\
  \bibinfo {author} {\bibfnamefont {P.}~\bibnamefont {Yu}},\ }\bibfield
  {title} {\enquote {\bibinfo {title} {{FENE Dumbbell Model and Its Several
  Linear and Nonlinear Closure Approximations}},}\ }\href@noop {} {\bibfield
  {journal} {\bibinfo  {journal} {Multiscale Model. Simul.}\ }\textbf {\bibinfo
  {volume} {4}},\ \bibinfo {pages} {709--731} (\bibinfo {year}
  {2005})}\BibitemShut {NoStop}%
\bibitem [{\citenamefont {Young}\ \emph {et~al.}(2023)\citenamefont {Young},
  \citenamefont {Corona}, \citenamefont {Datta}, \citenamefont {Helgeson},\
  and\ \citenamefont {Graham}}]{Young2023}%
  \BibitemOpen
  \bibfield  {author} {\bibinfo {author} {\bibfnamefont {C.~D.}\ \bibnamefont
  {Young}}, \bibinfo {author} {\bibfnamefont {P.~T.}\ \bibnamefont {Corona}},
  \bibinfo {author} {\bibfnamefont {A.}~\bibnamefont {Datta}}, \bibinfo
  {author} {\bibfnamefont {M.~E.}\ \bibnamefont {Helgeson}}, \ and\ \bibinfo
  {author} {\bibfnamefont {M.~D.}\ \bibnamefont {Graham}},\ }\bibfield  {title}
  {\enquote {\bibinfo {title} {{Scattering-Informed Microstructure Prediction
  during Lagrangian Evolution (SIMPLE)―a data-driven framework for modeling
  complex fluids in flow}},}\ }\href@noop {} {\bibfield  {journal} {\bibinfo
  {journal} {Rheol. Acta}\ }\textbf {\bibinfo {volume} {62}},\ \bibinfo {pages}
  {587--604} (\bibinfo {year} {2023})}\BibitemShut {NoStop}%
\bibitem [{\citenamefont {Loiseau}\ and\ \citenamefont
  {Brunton}(2018)}]{Loiseau2018}%
  \BibitemOpen
  \bibfield  {author} {\bibinfo {author} {\bibfnamefont {J.-C.}\ \bibnamefont
  {Loiseau}}\ and\ \bibinfo {author} {\bibfnamefont {S.~L.}\ \bibnamefont
  {Brunton}},\ }\bibfield  {title} {\enquote {\bibinfo {title} {Constrained
  sparse galerkin regression},}\ }\href@noop {} {\bibfield  {journal} {\bibinfo
   {journal} {J. Fluid Mech.}\ }\textbf {\bibinfo {volume} {838}},\ \bibinfo
  {pages} {42--67} (\bibinfo {year} {2018})}\BibitemShut {NoStop}%
\bibitem [{\citenamefont {Sato}\ and\ \citenamefont
  {Taniguchi}(2019)}]{Sato2019}%
  \BibitemOpen
  \bibfield  {author} {\bibinfo {author} {\bibfnamefont {T.}~\bibnamefont
  {Sato}}\ and\ \bibinfo {author} {\bibfnamefont {T.}~\bibnamefont
  {Taniguchi}},\ }\bibfield  {title} {\enquote {\bibinfo {title} {Rheology and
  entanglement structure of well-entangled polymer melts: A slip-link
  simulation study},}\ }\href@noop {} {\bibfield  {journal} {\bibinfo
  {journal} {Macromolecules}\ }\textbf {\bibinfo {volume} {52}},\ \bibinfo
  {pages} {3951--3964} (\bibinfo {year} {2019})}\BibitemShut {NoStop}%
\bibitem [{\citenamefont {Miyamoto}, \citenamefont {Sato},\ and\ \citenamefont
  {Taniguchi}(2023)}]{Miyamoto2023}%
  \BibitemOpen
  \bibfield  {author} {\bibinfo {author} {\bibfnamefont {S.}~\bibnamefont
  {Miyamoto}}, \bibinfo {author} {\bibfnamefont {T.}~\bibnamefont {Sato}}, \
  and\ \bibinfo {author} {\bibfnamefont {T.}~\bibnamefont {Taniguchi}},\
  }\bibfield  {title} {\enquote {\bibinfo {title} {{Stretch-orientation-induced
  reduction of friction in well-entangled bidisperse blends: a dual slip-link
  simulation study}},}\ }\href@noop {} {\bibfield  {journal} {\bibinfo
  {journal} {Rheol. Acta}\ }\textbf {\bibinfo {volume} {62}},\ \bibinfo {pages}
  {57--70} (\bibinfo {year} {2023})}\BibitemShut {NoStop}%
\bibitem [{\citenamefont {Sato}\ \emph {et~al.}(2020)\citenamefont {Sato},
  \citenamefont {Moghadam}, \citenamefont {Tan},\ and\ \citenamefont
  {Larson}}]{Sato2020}%
  \BibitemOpen
  \bibfield  {author} {\bibinfo {author} {\bibfnamefont {T.}~\bibnamefont
  {Sato}}, \bibinfo {author} {\bibfnamefont {S.}~\bibnamefont {Moghadam}},
  \bibinfo {author} {\bibfnamefont {G.}~\bibnamefont {Tan}}, \ and\ \bibinfo
  {author} {\bibfnamefont {R.~G.}\ \bibnamefont {Larson}},\ }\bibfield  {title}
  {\enquote {\bibinfo {title} {{A slip-spring simulation model for predicting
  linear and nonlinear rheology of entangled wormlike micellar solutions}},}\
  }\href@noop {} {\bibfield  {journal} {\bibinfo  {journal} {J. Rheol.}\
  }\textbf {\bibinfo {volume} {64}},\ \bibinfo {pages} {1045--1061} (\bibinfo
  {year} {2020})}\BibitemShut {NoStop}%
\bibitem [{\citenamefont {Sato}\ and\ \citenamefont {Larson}(2022)}]{Sato2022}%
  \BibitemOpen
  \bibfield  {author} {\bibinfo {author} {\bibfnamefont {T.}~\bibnamefont
  {Sato}}\ and\ \bibinfo {author} {\bibfnamefont {R.~G.}\ \bibnamefont
  {Larson}},\ }\bibfield  {title} {\enquote {\bibinfo {title} {{Nonlinear
  rheology of entangled wormlike micellar solutions predicted by a
  micelle-slip-spring model}},}\ }\href@noop {} {\bibfield  {journal} {\bibinfo
   {journal} {J. Rheol.}\ }\textbf {\bibinfo {volume} {66}},\ \bibinfo {pages}
  {639--656} (\bibinfo {year} {2022})}\BibitemShut {NoStop}%
\bibitem [{\citenamefont {Hyun}\ \emph {et~al.}(2011)\citenamefont {Hyun},
  \citenamefont {Wilhelm}, \citenamefont {Klein}, \citenamefont {Cho},
  \citenamefont {Nam}, \citenamefont {Ahn}, \citenamefont {Lee}, \citenamefont
  {Ewoldt},\ and\ \citenamefont {McKinley}}]{Hyun2011}%
  \BibitemOpen
  \bibfield  {author} {\bibinfo {author} {\bibfnamefont {K.}~\bibnamefont
  {Hyun}}, \bibinfo {author} {\bibfnamefont {M.}~\bibnamefont {Wilhelm}},
  \bibinfo {author} {\bibfnamefont {C.~O.}\ \bibnamefont {Klein}}, \bibinfo
  {author} {\bibfnamefont {K.~S.}\ \bibnamefont {Cho}}, \bibinfo {author}
  {\bibfnamefont {J.~G.}\ \bibnamefont {Nam}}, \bibinfo {author} {\bibfnamefont
  {K.~H.}\ \bibnamefont {Ahn}}, \bibinfo {author} {\bibfnamefont {S.~J.}\
  \bibnamefont {Lee}}, \bibinfo {author} {\bibfnamefont {R.~H.}\ \bibnamefont
  {Ewoldt}}, \ and\ \bibinfo {author} {\bibfnamefont {G.~H.}\ \bibnamefont
  {McKinley}},\ }\bibfield  {title} {\enquote {\bibinfo {title} {A review of
  nonlinear oscillatory shear tests: Analysis and application of large
  amplitude oscillatory shear {(LAOS)}},}\ }\href@noop {} {\bibfield  {journal}
  {\bibinfo  {journal} {Prog. Polym. Sci.}\ }\textbf {\bibinfo {volume} {36}},\
  \bibinfo {pages} {1697--1753} (\bibinfo {year} {2011})}\BibitemShut {NoStop}%
\end{thebibliography}%
\end{document}


\preprint{Supplementary Material for {\it Journal of Rheology}}

\title[Supplementary Material]{Rheo-SINDy: Finding a constitutive model from rheological data for complex fluids \\using sparse identification for nonlinear dynamics}

\author{Takeshi Sato}%
\email{takeshis@se.kanazawa-u.ac.jp}
 \affiliation{
Advanced Manufacturing Technology Institute, Kanazawa University, Kanazawa 920-1192, Japan
 }
 
\author{Souta Miyamoto}%
\affiliation{Department of Chemical Engineering, Graduate School of Engineering, Kyoto University, Kyoto 615-8510, Japan}


\author{Shota Kato}
\affiliation{Graduate School of Informatics, Kyoto University, Kyoto 606-8501, Japan}

\date{\today}

\maketitle
\section{\label{a_Lasso_Hyper_Param}Hyperparameter of the Adaptive Lasso}
\begin{figure}[t]
 \centering
  \includegraphics[width=3in]{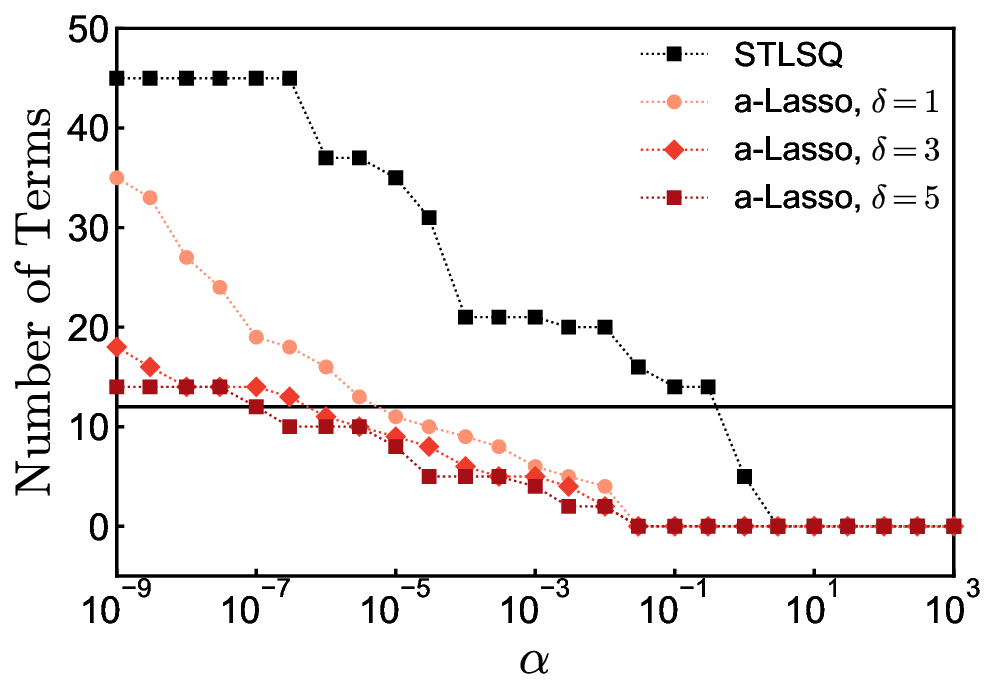}
  \caption{The total number of terms obtained by {\it Rheo}-SINDy with the STLSQ (black symbols) and a-Lasso (red symbols) for the Giesekus model. Here, the circles, diamonds, and squares in the red series indicate the results with $\delta = 1$, $3$, and $5$ for the adaptive weight $w_{\mu\nu,j}$, respectively.}
  \label{FigS01}
\end{figure}
Here, we shortly note the effect of changing the hyperparameter $\delta$ of the a-Lasso, which determines the adaptive weight.
Figure~\ref{FigS01} compares the total number of terms for the Giesekus model obtained by the a-Lasso with three different $\delta$ values.
The training data include the multiple trajectories, which are the same as those in Fig.~4 in the main text.
Here, the results for the STLSQ are also shown for comparison.
As shown in Fig.~\ref{FigS01}, the solutions obtained by the a-Lasso with $\delta = 1$, $3$, and $5$ are sparser than those obtained by the STLSQ.
Due to the increased effects of weights, the solutions for $\delta = 3$ and $5$ are sparser compared to the solutions for $\delta = 1$.
Moreover, the results with $\delta = 3$ are almost the same as those with $\delta=5$,
although the a-Lasso with $\delta=5$ provides slightly sparser solutions.
Thus, the hyperparameter $\delta=3$ can be considered sufficiently large to obtain sparse solutions.
We note, in general, that a sparser solution is superior from the perspective of overfitting and helps prevent unexpected divergence during test simulations.
From these discussions, in this study, we used $\delta = 3$ as the adaptive weight in the a-Lasso.
\newpage

\section{\label{A_FENE_P}Stress Expressions for the FENE-P Dumbbell Model}
As noted in Sec.~IIIC2 in the main text, the constitutive equation for the FENE-P dumbbell model can be expressed in terms of the stress (cf. Eqs.~(39)--(41)).
We here show the derivation of the constitutive equation for the FENE-P model~\cite{Mochimaru1983}.

To improve clarity,
let us rewrite the stress $\bm \tau$ in Eq.~(28) in the main text as follows:
\begin{equation}
{\bm \tau} (t) = \rho h_{\rm eq} Z^{-1}_{\rm eq} Z \langle {\bm R}(t){\bm R}(t) \rangle - G {\bm I},
\label{stress_mod}
\end{equation}
where $Z$ has already been defined in Eq.~(38) in the main text and $Z_{\rm eq}$ indicates $Z$ at equilibrium.
In what follows, we express all variables in dimensionless forms by using the unit time $\lambda$ and the unit stress $\rho k_{\rm B} T$. Additionally, for simplicity, we omit the tilde representing dimensionless quantities.
Taking the trace of both sides of Eq.~\eqref{stress_mod} and using the relation $\langle {\bm R}^2(t)\rangle = R_{\rm max}^2 (1 - Z^{-1})$, we can rewrite $Z$ as a function of $\bm \tau$:
\begin{equation}
Z = 1 + \frac{1}{3 n_{\rm K} Z^{-1}_{\rm eq}} ( {{\rm tr}} {\bm \tau} + 3).
\label{Z_tau}
\end{equation}
Taking the convected derivative of $\bm \tau /Z$, the time evolution of stress can be expressed as
\begin{equation}
\frac{{\rm d}  {\bm \tau}}{{\rm d}t} -  {\bm \tau} \cdot  {\bm \kappa}^{\rm T} -  {\bm \kappa} \cdot  {\bm \tau} = - Z^{-1}_{\rm eq} Z  {\bm \tau}  + 2{\bm D} + \frac{{\rm D} \ln Z}{{\rm D}t} ( {\bm \tau} + {\bm I}),
\label{FENE_P_stress}
\end{equation}
which is the same as in Eq.~(37) in the main text.
Since we do not address the spatial gradient in rheological calculations,
${\rm D}(\cdots)/{\rm D}t$ simply reduces ${\rm d}(\cdots)/{\rm d}t$.
To obtain Eq.~\eqref{FENE_P_stress}, we used the following relation that can be obtained by Eqs.~(32) and (33) in the main text:
\begin{equation}
\frac{{\rm d} {\bm C}}{{\rm d} t} - {\bm C} \cdot {\bm \kappa}^{\rm T} - {\bm \kappa} \cdot {\bm C} = - \frac{n_{\rm K}}{3}  {\bm \tau}.
\end{equation}
From Eq.~\eqref{Z_tau}, the time evolution of $\ln Z$ can be expressed in terms of ${\rm tr} \bm \tau$ as
\begin{equation}
\frac{{\rm d}}{{\rm d} t} {\rm tr}  {\bm \tau} = \left \{ 3 n_{\rm K} Z^{-1}_{\rm eq} + ( {\rm tr}  {\bm \tau} + 3) \right \} \frac{{\rm d}\ln Z}{{\rm d} t}.
\label{Z_tr_tau}
\end{equation}
Furthermore, taking trace of Eq.~\eqref{FENE_P_stress} and using Eq.~\eqref{Z_tr_tau}, we can have
\begin{equation}
\frac{{\rm d}\ln Z}{{\rm d} t}
= \frac{1}{3 n_{\rm K} Z^{-1}_{\rm eq}} \left \{  - Z^{-1}_{\rm eq} Z {\rm tr}  {\bm \tau}  + 2 {\rm tr} {\bm D} + {\rm tr} \left (  {\bm \tau} \cdot  {\bm \kappa}^{\rm T} +  {\bm \kappa} \cdot  {\bm \tau} \right ) \right \}.
\label{lnZ}
\end{equation}
Combining Eqs.~\eqref{FENE_P_stress} and \eqref{lnZ},
we can express the time evolution of $\bm \tau$ (i.e., $\dot {\bm \tau}$) as a function of $\bm \tau$ and $\bm \kappa$.
Specifically,
Eqs.~\eqref{FENE_P_stress} and \eqref{lnZ} reduce to Eqs.~(39)--(41) in the main text under shear flow.

\section{\label{STRidge_FENE}STRidge Regressions for the FENE Dumbbell Model}
\begin{figure}[t]
 \centering
  \includegraphics[width=3in]{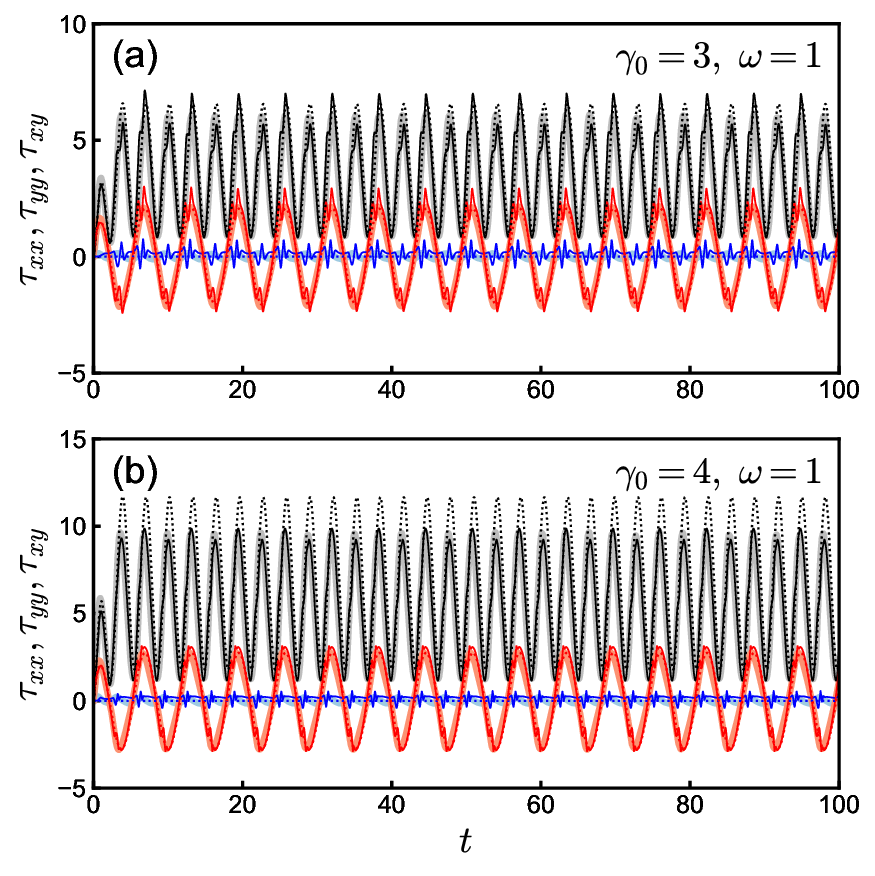}
  \caption{Test simulation results using equations obtained by {\it Rheo}-SINDy with the STRidge for the library shown in Eq.~(43) in the main text. The test simulations were conducted under the oscillatory shear flows with (a) $\gamma_0 = 3$ and $\omega=1$ and (b) $\gamma_0 = 4$ and $\omega=1$. The bold lines show the exact solutions, and the thin solid and short-dashed lines show the results with the smaller $\alpha$ value ($\alpha = 1 \times 10^{-1}$) and the larger $\alpha$ value ($\alpha = 1$).}
  \label{FigS02}
\end{figure}
\begin{figure}[t]
 \centering
  \includegraphics[width=3in]{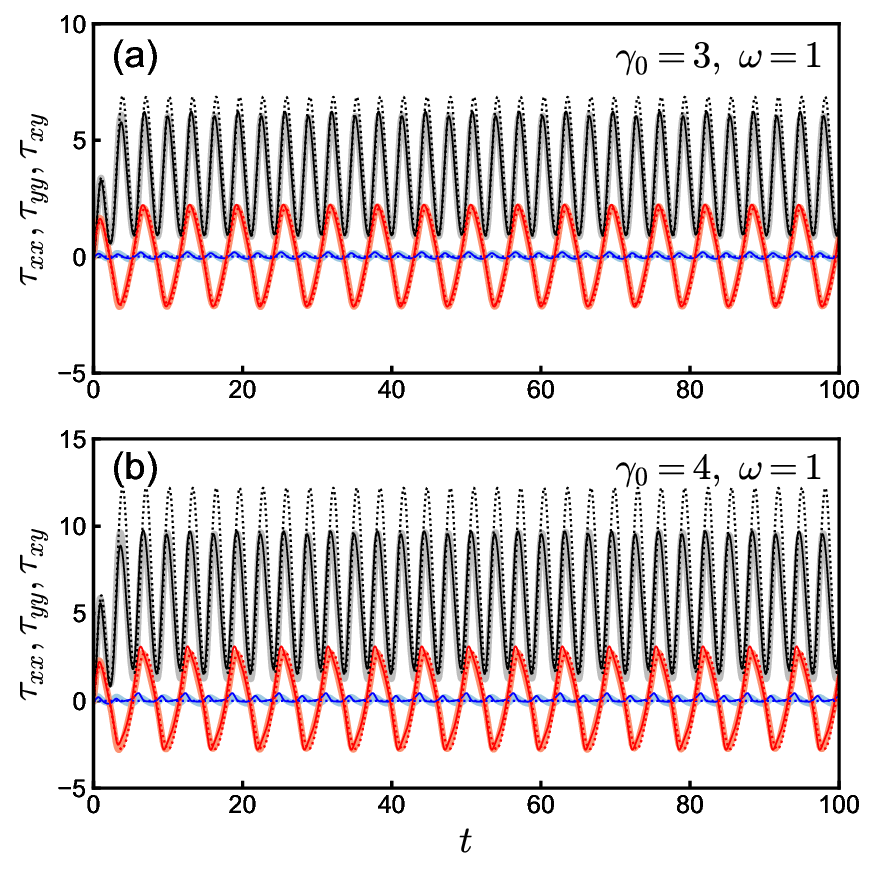}
  \caption{Test simulation results using equations obtained by {\it Rheo}-SINDy with the STRidge for the library including polynomial terms up to the third order of $\{\tau_{xx}, \tau_{yy}, \tau_{zz}, \tau_{xy}, \kappa_{xy}\}$. The flow parameters for the test simulations are the same as those in Fig.~\ref{FigS02}. The bold lines show the exact solutions, and the thin solid and short-dashed lines show the results with the smaller $\alpha$ value ($\alpha = 3 \times 10^{-2}$) and the larger $\alpha$ value ($\alpha = 1$).}
  \label{FigS03}
\end{figure}
Figures~\ref{FigS02} and \ref{FigS03} show the test simulation results for the FENE dumbbell model using the approximate constitutive equations obtained by {\it Rheo}-SINDy with the STRidge.
Here, in Fig.~\ref{FigS02}, we employed the custom library shown in Eq.~(43) in the main text ($N_{\bm \Theta} = 29$), while in Fig.~\ref{FigS03}, we utilized a polynomial library including polynomial terms up to the third order of $\{\tau_{xx}, \tau_{yy}, \tau_{zz}, \tau_{xy}, \kappa_{xy}\}$ ($N_{\bm \Theta} = 56$).
From the thin solid lines in Fig.~\ref{FigS02}, which show the results with the smaller $\alpha = 1 \times 10^{-1}$,
while the magnitudes of the predicted stress components almost match the results of the BD simulation,
spike-like predictions are occasionally observed.
When using the third-order polynomial library, the solutions for the small $\alpha$, indicated by thin solid lines in Fig.~\ref{FigS03}, closely resemble the results of the BD simulations.
This is likely attributed to the fact that the larger number of candidate terms included in the library improves the predictive ability of the model.
Nevertheless, we note that increasing the number of terms in the library without careful consideration does not necessarily lead to an improvement in the model performance.
By increasing the number of terms in the library, overfitting issues may arise.
For example, when {\it Rheo}-SINDy chooses terms that are likely to be significantly large under shear flow, such as $\tau_{xx} \kappa_{xy}^2$, there is an increased possibility that the differential equations may fail to be solved when conducting test simulations for the parameters outside of the training data.

\newpage
\section{\label{FENE_Large_Strain}Regressions using the training data with the large $\gamma_0$}
\subsection{FENE dumbbell model}
\begin{figure*}[t]
 \centering
  \includegraphics[width=6.25in]{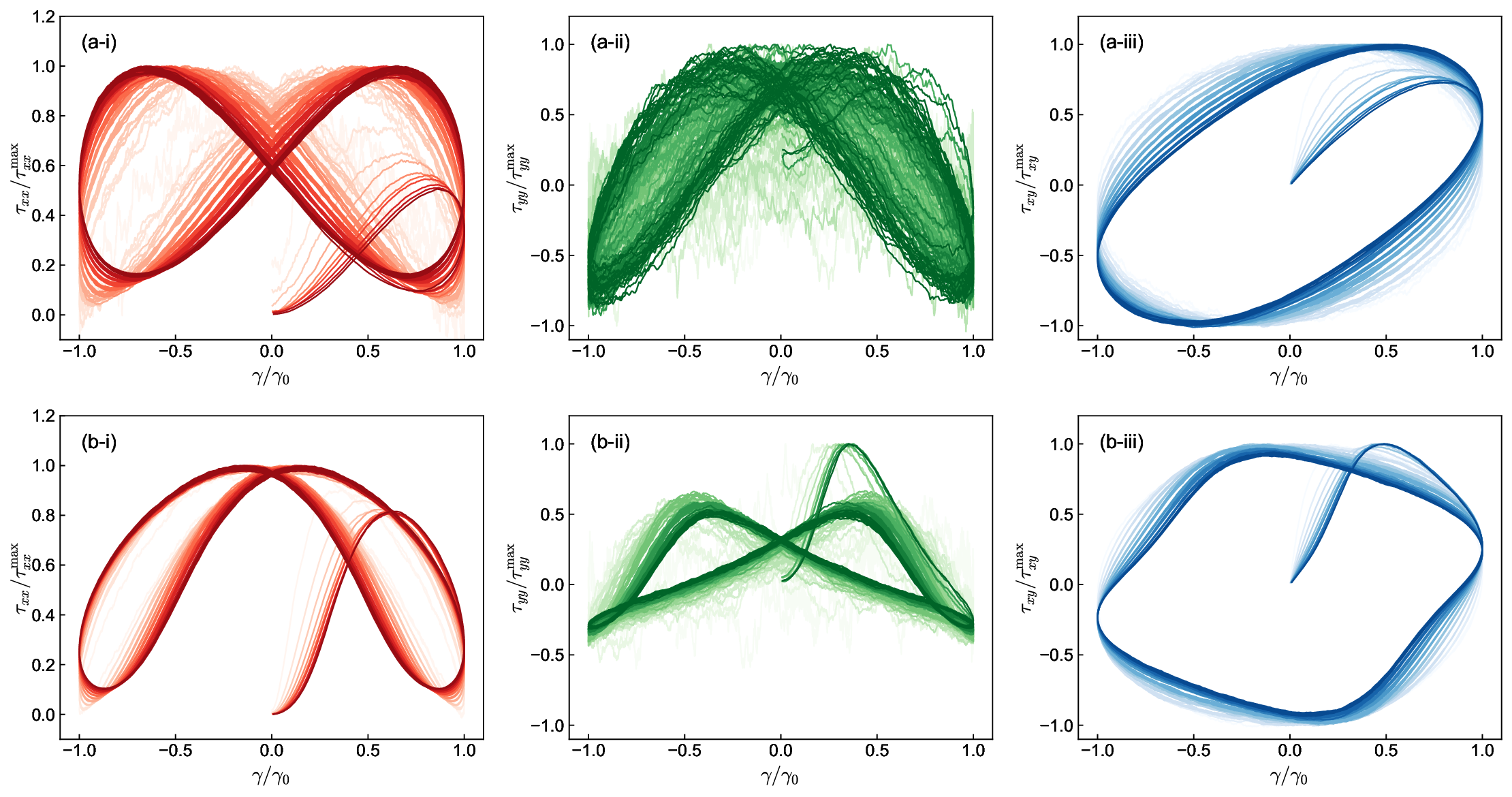}
  \caption{The normalized training data (a) with the standard $\gamma_0$ ($\gamma_0 = 2$) and (b) with the larger $\gamma_0$ ($\gamma_0 = 8$). The different colors in each panel distinguish the different $\omega$ values ($\omega \in \{0.1, 0.2, \ldots, 1\}$).}
  \label{FigS04}
\end{figure*}
\begin{figure}[t]
 \centering
  \includegraphics[width=3in]{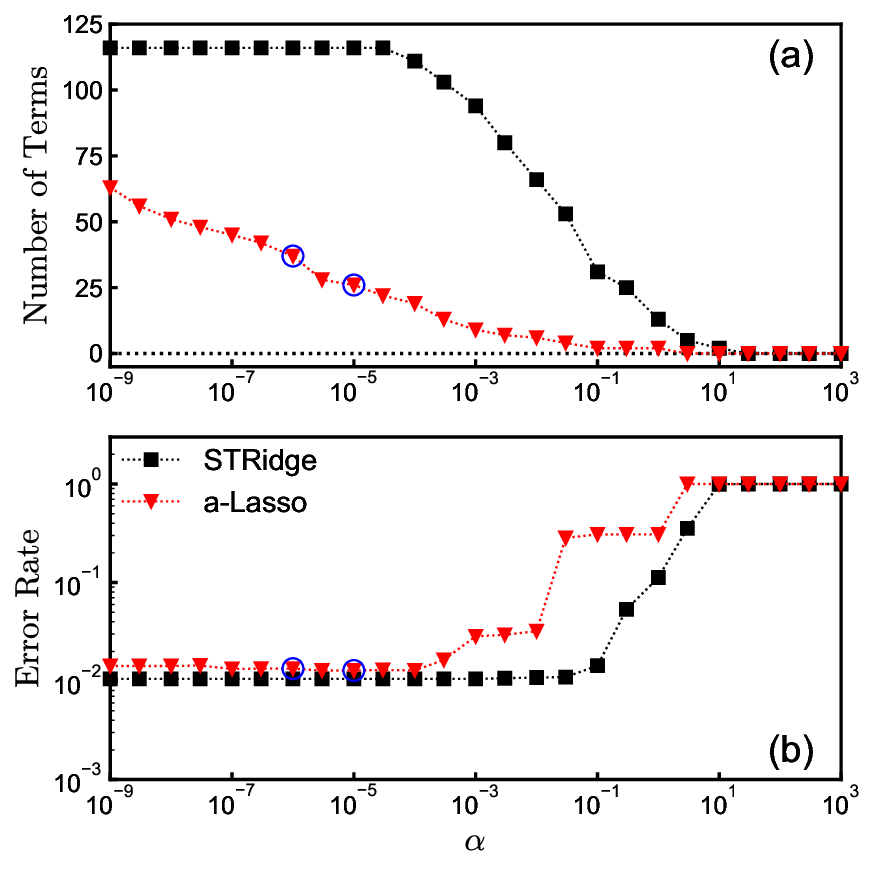}
  \caption{(a) The total number of terms and (b) the error rate of the constitutive model obtained by {\it Rheo}-SINDy with the STRidge (black squares) and the a-Lasso (red reverse triangles) for the FENE dumbbell model. To obtain constitutive models, we used the training data with $\gamma_0 = 8$. The horizontal short-dashed line in (a) indicates that the number of terms is zero. The blue circles represent the $\alpha$ values selected for test simulations.}
  \label{FigS05}
\end{figure}
\begin{figure*}[t]
 \centering
  \includegraphics[width=6.25in]{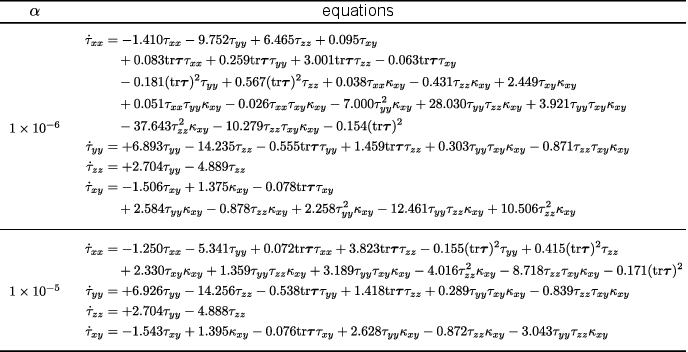}
  \caption{The constitutive equations of the FENE dumbbell model obtained by {\it Rheo}-SINDy with the a-Lasso for the training data with $\gamma_0 = 8$.}
  \label{FigS06}
\end{figure*}
Here, we conducted the {\it Rheo}-SINDy regressions for the FENE dumbbell model using training data with a larger $\gamma_0$ value ($\gamma_0 = 8$) than the standard $\gamma_0$ value ($\gamma_0 = 2$). 
As shown in Fig.~\ref{FigS04}, the training data with $\gamma_0 = 8$ include the nonlinear shear responses.

Figure~\ref{FigS05} shows the total number of terms and the error rate obtained by {\it Rheo}-SINDy for the training data with $\gamma_0 = 8$. 
Although the $\alpha$ dependence of the solutions shows a similar trend to Fig.~13 in the main text, the total number of terms at the same $\alpha$ is larger for the {\it Rheo}-SINDy regression using the training data with $\gamma_0 = 8$ than with $\gamma_0 = 2$.
In Fig.~\ref{FigS06}, we show the equations obtained by {\it Rheo}-SINDy for the two $\alpha$ values indicated in blue circles in Fig.~\ref{FigS05}. 
As shown in Fig.~\ref{FigS06}, the regression results using the training data generated with $\gamma_0 = 8$ include nonlinear terms that are not present in Fig.~14 of the main text.

\clearpage
\subsection{FENE-P dumbbell model}

\begin{figure}[t]
   \centering
    \includegraphics[width=6.25in]{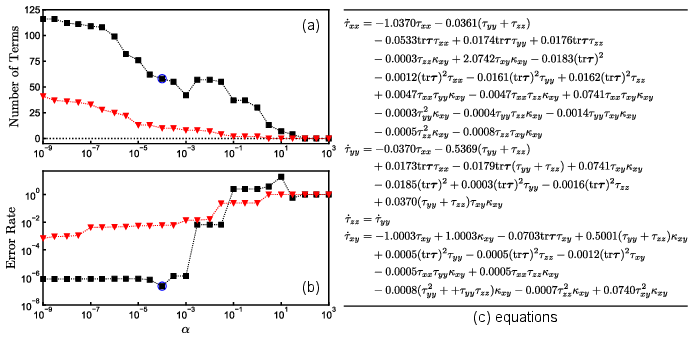}
    \caption{(a) The total number of terms and (b) the error rate of the constitutive equations obtained by {\it Rheo}-SINDy with the STRidge (black squares) and a-Lasso (red reverse triangles) for the FENE-P dumbbell model in the stress expression (Eqs.~(42)--(44) with $n_{\rm K} = 10$ in the main text). 
    Here, we conducted the {\it Rheo}-SINDy regressions using training data with a larger $\gamma_0$ value ($\gamma_0 = 8$) than the standard $\gamma_0$ value ($\gamma_0 = 2$). 
    The blue circles represent the $\alpha$ value selected for test simulations. Based on our criterion, $\alpha=1\times10^{-4}$ for the STRidge was picked. In (c), the obtained equations using the STRidge with $\alpha=1\times10^{-4}$ are shown.}
    \label{FigS07}
\end{figure}

Figure~\ref{FigS07} indicates the total number of terms and the error rate obtained by {\it Rheo}-SINDy for the FENE-P dumbbell model using training data with $\gamma_0 = 8$. Similar to Fig.~10 in the main text, the STRidge gives the smaller error rate than the a-Lasso. Within the range of $0.1 \le \alpha \le 10$, the error rate obtained by the STRidge exceeds unity, indicating that the regressions with the STRidge were not successful. This failure is presumably due to our current algorithm for the STRidge, which is based on the textbook by Brunton and coworkers~\cite{Brunton2022}, and should be addressed in the future. Nevertheless, outside that range, we can obtain the appropriate $\alpha$ value according to our criterion as $\alpha = 1 \times 10^{-4}$. The equations we obtained from {\it Rheo}-SINDy are shown in Fig.~\ref{FigS07}(c). These expressions were used for test simulations shown in Fig.~16 in the main text. 
\newpage

\section{Material frame indifference}
Here, we use the UCM and Giesekus models as examples to demonstrate the transformation that {\it Rheo}-SINDy has made to address the principle of material frame indifference.
To account for this principle, the velocity gradient tensor $\bm \kappa$ in the laboratory frame (LF) is decomposed into the deformation rate tensor $\bm D$ and the rotation rate tensor $\bm W$ as $\bm \kappa = \bm D + \bm W$, where $\bm D = (\bm \kappa + \bm \kappa^{\rm T})/2$ and $\bm W = (\bm \kappa - \bm \kappa^{\rm T})/2$. 
As explained in the main text (and the study of Young and coworkers~\cite{Young2023}), in the reference frame (RF) rotated by $\bm W$, the stress tensor $\bm \tau'$ and strain rate tensor $\bm D'$ can be expressed as $\bm \tau' = \bm R^{\rm T} (\theta_{\bm W}) \bm \tau \bm R (\theta_{\bm W})$ and $\bm D' = \bm R^{\rm T} (\theta_{\bm W}) \bm D \bm R (\theta_{\bm W})$, respectively, where $\theta_{\bm W}$ is the rotation angle. 
$\theta_{\bm W}$ at time $t$ can be determined by the time evolution equation of $\bm R (\theta_{\bm W})$ as $\dot {\bm R} (\theta_{\bm W}) = {\bm W} \cdot {\bm R} (\theta_{\bm W})$. 
In the RF, the UCM and Giesekus models can be expressed as
\begin{equation}
\dot {\bm \tau}' - {\bm \tau}' \cdot {\bm D}' - {\bm D}' \cdot {\bm \tau}' = -\frac{1}{\lambda} {\bm \tau}' + 2G {\bm D}'
\label{rot_UCM}
\end{equation}
and 
\begin{equation}
\dot {\bm \tau}' - {\bm \tau}' \cdot {\bm D}' - {\bm D}' \cdot {\bm \tau}' = -\frac{1}{\lambda} {\bm \tau}' -\frac{\alpha_{\rm G}}{G \lambda} {\bm \tau}' \cdot {\bm \tau}' + 2G {\bm D}',
\label{rot_Giesekus}
\end{equation}
respectively. 

To make a constitutive equation with quantities that are independent of a particular direction of orthogonal coordinate system, $\bm \tau'$ and $\bm D'$ in the RF are diagonalized using rotation matrices around an axis $\bm P(\theta^{\ast}_{\bm \tau})$ and $\bm P(\theta^{\ast}_{\bm D})$ as $\bm \tau^{\ast} = \bm P^{\rm T} (\theta^{\ast}_{\bm \tau}) \bm \tau' \bm P (\theta^{\ast}_{\bm \tau})$ and $\bm D^{\ast} = \bm P^{\rm T} (\theta^{\ast}_{\bm D}) \bm \tau' \bm P (\theta^{\ast}_{\bm D})$, respectively. Here, $\theta^{\ast}_{\bm \tau}$ and $\theta^{\ast}_{\bm D}$ are the rotation angles in a plane under deformations to obtain $\bm \tau^{\ast}$ and $\bm D^{\ast}$, respectively. 
These angles can be expressed as 
\begin{equation}
\theta_{\bm \tau}^{\ast} = \frac{1}{2} \arctan \left ( -\frac{2 \tau'_{xy}}{\tau'_{xx} - \tau'_{yy}} \right )
\end{equation}
and
\begin{equation}
\theta_{\bm D}^{\ast} = \frac{1}{2} \arctan \left ( -\frac{D'_{xy}}{D'_{xx}} \right ). 
\end{equation}

Substituting $\bm \tau' = \bm P (\theta^{\ast}_{\bm \tau}) \bm \tau^{\ast} \bm P^{\rm T} (\theta^{\ast}_{\bm \tau})$ and $\bm D' = \bm P (\theta^{\ast}_{\bm D}) \bm \tau^{\ast} \bm P^{\rm T} (\theta^{\ast}_{\bm D})$ into Eq.~\eqref{rot_UCM} and left- and right-multiplying the equation by $\bm P^{\rm T} (\theta^{\ast}_{\bm \tau})$ and $\bm P (\theta^{\ast}_{\bm \tau})$, we can obtain 
\begin{align}
&\dot {\bm \tau}^\ast + \left \{ {\bm P} \left (+ \frac{\pi}{2} \right ) {\bm \tau}^\ast - {\bm \tau}^\ast {\bm  P} \left (+\frac{\pi}{2} \right ) \right \} \dot \theta_{\bm \tau}^{\ast} \nonumber \\
&= \bm \tau^\ast \bm P \{-(\theta_{\bm \tau}^{\ast} - \theta_{\bm D}^{\ast})\} \bm D^\ast \bm P \{ +(\theta_{\bm \tau}^{\ast} -\theta_{\bm D}^{\ast}) \} + \bm P \{ -(\theta_{\bm \tau}^{\ast} - \theta_{\bm D}^{\ast})\} \bm D^\ast \bm P \{ +(\theta_{\bm \tau}^{\ast} - \theta_{\bm D}^{\ast})\} \bm \tau^\ast \nonumber \\
&\quad - \frac{1}{\lambda} \bm \tau^\ast + 2G \bm P \{ -(\theta_{\bm \tau}^{\ast} - \theta_{\bm D}^{\ast})\} \bm D^\ast \bm P \{ +(\theta_{\bm \tau}^{\ast} - \theta_{\bm D}^{\ast})\} \nonumber \\
&= \bm \tau^\ast \bm P (-\phi^{\ast}) \bm D^\ast \bm P (+\phi^{\ast}) + \bm P (-\phi^{\ast}) \bm D^\ast \bm P (+\phi^{\ast}) \bm \tau^\ast - \frac{1}{\lambda} \bm \tau^\ast + 2G \bm P (-\phi^{\ast}) \bm D^\ast \bm P (+\phi^{\ast}), 
\label{objective_UCM}
\end{align}
where $\phi^{\ast} \equiv \theta_{\bm \tau}^{\ast} - \theta_{\bm D}^{\ast}$. 
Similarly, the Giesekus model shown in Eq.~\eqref{rot_Giesekus} can be transformed into 
\begin{align}
&\dot {\bm \tau}^\ast + \left \{ {\bm P} \left (+ \frac{\pi}{2} \right ) {\bm \tau}^\ast - {\bm \tau}^\ast {\bm  P} \left (+\frac{\pi}{2} \right ) \right \} \dot \theta_{\bm \tau}^{\ast} \nonumber \\
&= \bm \tau^\ast \bm P (-\phi^{\ast}) \bm D^\ast \bm P (+\phi^{\ast}) + \bm P (-\phi^{\ast}) \bm D^\ast \bm P (+\phi^{\ast}) \bm \tau^\ast - \frac{1}{\lambda} \bm \tau^\ast \nonumber \\
&\quad -\frac{\alpha_{\rm G}}{G \lambda} {\bm \tau}^{\ast} \cdot {\bm \tau}^{\ast} + 2G \bm P (-\phi^{\ast}) \bm D^\ast \bm P (+\phi^{\ast}). 
\label{objective_Giesekus}
\end{align}
Here, we used the following relations: 
\begin{equation}
\dot {\bm P} (\theta) = {\bm P} \left ( \theta + \frac{\pi}{2} \right )
\end{equation}
and
\begin{equation}
\dot {\bm P} (-\theta) = {\bm P} \left ( -\theta + \frac{\pi}{2} \right ). 
\end{equation}

Under deformations in a plane, the time evolutions of the three stress components and the angle of the UCM and Giesekus models are expressed as 
\begin{align}
\dot{\tau}^{\ast}_{xx} &= +2\tau^{\ast}_{xx} D_{xx}^{\ast} \cos 2\phi^{\ast}-\frac{1}{\lambda} {\tau}^{\ast}_{xx} + 2 G D_{xx}^{\ast} \cos 2\phi^{\ast}, \label{eq_diag_01} \\
\dot{\tau}^{\ast}_{yy} &= -2\tau^{\ast}_{yy} D_{xx}^{\ast} \cos 2\phi^{\ast}-\frac{1}{\lambda} {\tau}^{\ast}_{yy} - 2 G D_{xx}^{\ast} \cos 2\phi^{\ast}, \label{eq_diag_02} \\
\dot{\tau}^{\ast}_{zz} &= -\frac{1}{\lambda} {\tau}^{\ast}_{zz} = 0, \label{eq_diag_03} \\
\dot {\theta}_{\bm \tau}^{\ast} &= - \left ( \frac{\tau^{\ast}_{xx}}{\tau^{\ast}_{xx} - \tau^{\ast}_{yy}} + \frac{\tau^{\ast}_{yy}}{\tau^{\ast}_{xx} - \tau^{\ast}_{yy}} + \frac{2G}{\tau^{\ast}_{xx} - \tau^{\ast}_{yy}}  \right ) D_{xx}^{\ast} \sin 2\phi^{\ast}, \label{eq_diag_04}
\end{align}
and
\begin{align}
\dot{\tau}^{\ast}_{xx} &= +2\tau^{\ast}_{xx} D_{xx}^{\ast} \cos 2\phi^{\ast}-\frac{1}{\lambda} {\tau}^{\ast}_{xx} - \frac{\alpha_{\rm G}}{G\lambda} ({\tau}^{\ast}_{xx})^2 + 2 G D_{xx}^{\ast} \cos 2\phi^{\ast}, \label{G_eq_diag_01} \\
\dot{\tau}^{\ast}_{yy} &= -2\tau^{\ast}_{yy} D_{xx}^{\ast} \cos 2\phi^{\ast}-\frac{1}{\lambda} {\tau}^{\ast}_{yy} - \frac{\alpha_{\rm G}}{G\lambda} ({\tau}^{\ast}_{yy})^2 - 2 G D_{xx}^{\ast} \cos 2\phi^{\ast}, \label{G_eq_diag_02} \\
\dot{\tau}^{\ast}_{zz} &= -\frac{1}{\lambda} {\tau}^{\ast}_{zz} - \frac{\alpha_{\rm G}}{G\lambda} ({\tau}^{\ast}_{zz})^2 = 0, \label{G_eq_diag_03} \\
\dot {\theta}_{\bm \tau}^{\ast} &= - \left ( \frac{\tau^{\ast}_{xx}}{\tau^{\ast}_{xx} - \tau^{\ast}_{yy}} + \frac{\tau^{\ast}_{yy}}{\tau^{\ast}_{xx} - \tau^{\ast}_{yy}} + \frac{2G}{\tau^{\ast}_{xx} - \tau^{\ast}_{yy}}  \right ) D_{xx}^{\ast} \sin 2\phi^{\ast}, \label{G_eq_diag_04}
\end{align}
respectively.